\documentclass[12pt]{article}
\pdfoutput=1

\usepackage[square,comma]{natbib}
\setcitestyle{numbers,citesep={,}}

\usepackage{url}

\usepackage{euscript}
\usepackage{amssymb}
\usepackage{amsfonts}
\usepackage{amsbsy}
\usepackage{amsmath}
\usepackage{epsfig}
\usepackage{amsthm}
\usepackage{amscd}
\usepackage{amstext}

\usepackage{color}
\definecolor{darkgreen}{rgb}{0,0.5,0}
\definecolor{darkblue}{rgb}{0,0,0.85}
\definecolor{purple}{rgb}{0.4,.2,0.7}
\usepackage[colorlinks=true,citecolor=darkgreen,linkcolor=darkblue,urlcolor=purple]{hyperref}

\def\IC{\mathbb{C}}

\def\IZ{{\mathbb{Z}}}
\def\IR{{\mathbb{R}}}
\def\IP{\mathbb{P}}

\def\ICP{{\mathbb{CP}}}

\def\CM {{\cal M}}

\def\CN {{\cal N}}
\def\CR {{\cal R}}
\def\CD {{\cal D}}
\def\CF {{\cal F}}

\def\CP {{\cal P }}
\def\CL {{\cal L}}

\def\CO {{\cal O}}
\def\CZ {{\cal Z}}

\def\CH {{\cal H}}

\def\CS {{\cal S}}
\def\CA{{\cal A}}

\def\CZ{{\cal Z}}

\numberwithin{equation}{section}

\def\one{{\hbox{ 1\kern-.8mm l}}}

\newcommand{\be}{\begin{equation}}
\newcommand{\ee}{\end{equation}}

\begin{document}

\begin{center}

\vspace{1cm} { \LARGE {\bf TASI lectures on complex structures}}

\vspace{1cm}

{\large Frederik Denef}

\vspace{0.8cm}

{\it Center for the Fundamental Laws of Nature, Harvard University, \\
17 Oxford Street, Cambridge, MA 02138.
}
\vspace{0.5cm}

{\it Simons Center for Geometry and Physics, \\
Stony Brook, NY 11794-3636.
} 
\vspace{0.5cm}

{\it Institute for Theoretical Physics, University of Leuven, \\
Celestijnenlaan 200D, B-3001 Heverlee, Belgium.
 }

\vspace{0.6cm}

{\tt  denef physics.harvard.edu} \\

\vspace{2cm}

\end{center}

\begin{abstract}

These lecture notes give an introduction to a number of ideas and methods that have been useful in the study of complex systems ranging from spin glasses to D-branes on Calabi-Yau manifolds. Topics include the replica formalism, Parisi's solution of the Sherrington-Kirkpatrick model, overlap order parameters, supersymmetric quantum mechanics, D-brane landscapes and their black hole duals.

\end{abstract}
\pagebreak
\setcounter{page}{1}

\tableofcontents

\newpage

\section{Introduction}

\subsection{The case for complexity}

If you are a string theorist, chances are that you think of a complex structure as something squaring to minus one, rather than as something consisting of many intricately interacting degrees of freedom. These lecture notes, however, are concerned with the latter.

There is a deeply rooted belief in the natural sciences that the more fundamental a theory gets, the less important complexity becomes. In particular there has traditionally been an almost unquestioned assumption that physics at subatomic scales must be intrinsically simple, and that complexity is only relevant in the macroscopic or living world. Certainly there is a decrease in complexity when going from cells to proteins and from there to amino acids, atoms and finally elementary particles. It would seem logical that any step further down this reductive tree --- for example explaining the origin of elementary particles and their properties, accounting for the entropy of black holes or understanding the initial conditions of the universe --- should involve only structures and concepts of the utmost simplicity and elegance. 

By now considerable theoretical and experimental evidence has accumulated suggesting this is wrong. In particular,  whenever quantum mechanics and gravity pair up --- as in cosmological dynamics on the largest space and time scales, the determination of effective low energy parameters in theories with compact extra dimensions, and the stringy microphysics of black holes --- complexity appears not only unavoidable, but seems to some extent essential. On the theoretical side I am referring here more specifically to the highly complex, fractal-like iterated structures arising in eternal inflation, the perplexing complexity of the string theory landscape, and the closely related complexity of wrapped D-brane systems that, among many other applications, have given us the first quantitative explanation of the entropy of black holes.\footnote{Except for very specific items, references will be given at the end of this section and in subsequent sections.} On the experimental side, the most influential development has been the gathering of abundant cosmological precision data: detailed cosmic microwave spectroscopy providing convincing evidence for the slow roll inflation hypothesis, and the measurement of an impossibly tiny yet nonzero vacuum energy density, of just the right magnitude to allow self-reproducing resonances to crawl out of gravitationally collapsed dust at precisely the right time to see the onset of cosmic acceleration. Whether we like it or not, these developments, and the absence of plausible alternative explanations for these and other fine-tuning conundrums, have added significant credibility to the idea that some fine-tunings may have no other explanation than being tautological post-selection effects in a vast, eternally inflating multiverse, scanning a huge ensemble of low energy effective field theories, only a tiny fraction of which will allow structures to evolve of sufficient complexity to be capable of organizing summer schools. This idea requires no fundamentally new concepts of any kind, works by exactly the same semiclassical mechanisms as ordinary inflation, eliminates some of the worst fine-tuning problems in physics, provides in principle a precise, quantitative definition of naturalness, and appears to fit naturally into a UV complete, unifying framework incorporating all known principles of physics, that has no external parameters yet spawns all the necessary complexity, out of simple equations following unambiguously from an even simpler idea: the quantization of strings.  

The price to pay for this remarkable feat is that it forces us to consider structures and concepts far more diverse and complex than those it seeks to explain, without immediate prospects of directly observing them. The Standard Models of elementary particles and cosmology is arguably a lot simpler than any of the string compactifications that could produce it; for example it could be F-theory compactified on the Calabi-Yau fourfold hypersurface in weighted projective space $\ICP^{1,1,84,516,1204,1806}$, which boasts about three hundred thousand deformation moduli, up to seventy-five thousand D3-branes, and a flux-induced superpotential specified by a choice of almost two million integers multiplying an equal amount of independent period integrals \cite{Lynker1999,Kreuzer}. According to simple estimates, this compactification, assuming all moduli can be stabilized, gives rise to more than $10^{10^5}$ intricately interconnected flux compactifications. Similarly, a macroscopic extremal Reissner-Nordstrom black hole looks considerably simpler than the corresponding weakly coupled D-brane systems one is led to consider in microscopic computations of the black hole entropy. For example a wrapped D4 producing a modest extremal black hole of say the mass of the sun easily involves D4-branes with $10^{100}$ moduli subject to superpotentials specified by $10^{100}$ flux quanta. Finally, even in simple toy model landscapes, the large scale structure generated by eternal inflation is infinitely more complex than what we observe in our universe. 


Although the importance and urgency of a better understanding of these matters is clear to many, there has been widespread reluctance to face this kind of stringy complexity directly. Instead, the dominant approach has been to try to isolate particular phenomena of interest, for example by 
studying simplified compactification models like local Calabi-Yau spaces, effective field theory models, non-disordered toy models of the landscape, D-brane models for black holes restricted to charge regimes where the dominant contributions to the entropy are highly structured by symmetry (Cardy regime), or coarse grained asymptotic parameter distributions in ensembles of compactifications allowing similar complexity-minimizing limits. There are of course obvious excellent reasons to follow this reductionist approach. However, there are also excellent reasons to try to probe complexity itself:
\begin{enumerate}
\item In contrast to supersymmetric AdS compactifications, a specific choice of compactification data does in general \emph{not} correspond to a superselection sector in cosmologies with a positive vacuum energy, as quantum and thermal fluctuations will force changes in geometric moduli, fluxes and topology. Thus, any complete, nonperturbative description of a sector of string theory rich enough to describe our own universe should be able to encode not just one compactification, but the full space of internal space configurations that can be dynamically reached in one way or another.\footnote{It is not necessary that such configurations also support metastable ``vacua''.} Since various topological transitions between for example Calabi-Yau manifolds are known to be perfectly sensible physical processes in string theory, the  space of such interconnected configurations is likely to be huge. If we imagine for a moment that there exists a complete holographic description of eternally inflating cosmologies in string theory, say in the form of a field theory living at future infinity, then this field theory would somehow have to encode the full complexity of those googols of geometries --- it would literally be a theory of everything, and it couldn't be anything less.
\item There is often striking organization, universality and elegance emerging in disordered systems, as has become clear over the past decades in studies of spin and structural glasses, neural networks and other complex systems. Conventional notions of symmetry are largely irrelevant in such systems, but other, equally powerful structures and their associated order parameters appear in their place. This includes hierarchical cluster organization of the state space, replica symmetry breaking and overlap order parameters. Uncovering these led to highly nontrivial exact solutions of various models of complex, disordered systems. Much like symmetries in ordered systems, these structures also determine to a large extent the dynamics and other physical properties. They have a wide range of applicability in fields as diverse as condensed matter physics, neuroscience, biology and computer science, and have led to practical applications such as new efficient algorithms for optimization,  data mining and artificial intelligence. Thus, rather than an annoyance to be avoided, complexity can be the essence, and the key feature to focus on. The lessons learned from these investigations are bound to be useful in the context of complex systems in string theory and cosmology.
\item Conversely, string dualities such as holography may provide a new and useful approach to the general theory of complex systems. What makes this particularly promising compared to other instances of applied holography is that the natural large $N$ limit is also the limit one is a priori interested in here. Concrete examples are given by supersymmetric branes wrapped on compact cycles in Calabi-Yau manifolds, which exhibit many of the typical characteristics of mean field models of glasses.
\end{enumerate}

\subsection{Contents and goals of these notes}

In these lecture notes, I will give an introduction to some of the concepts and techniques which have been useful in studies of complex systems of many degrees of freedom, with intricate, disordered interactions. Although I will discuss ideas and techniques developed in the theory of glasses as well as ideas and techniques developed in string theory, there will be little or no discussion of applications of one to the other. For this I refer to the work that will appear in \cite{Barandes,ADnew,AABDG}, and in which any new idea that might be present in these notes originated. The focus will be on the basics, meaning  material usually assumed to be known in most of the recent specialized literature. Rather than to give a comprehensive review, I will treat a number of specific topics in a more or less self-contained and hopefully pedagogical way, to avoid variants of step 7 of
\cite{Goose}, and to allow the reader to learn how to actually compute a number of things rather than to just get a flavor of the ideas. This comes at a price of having to leave out many interesting and important topics (including some that were discussed to a certain extent in my actual lectures at TASI). In particular, despite its importance in the motivation given above, I will say almost nothing directly here about the landscape of string compactifications, and focus instead on space localized D-brane systems arising as the weak coupling description of charged black holes. This has three additional reasons: $(i)$ I have already written extensive lecture notes on string vacua \cite{Denefb}, and wanted to avoid overlap altogether. $(ii)$ The internal space geometry of the space-localized D-brane systems describing black holes is identical to that of space-filling D-brane systems describing compactifications, and the techniques used to analyze the former can directly be transported to the latter. The advantages of looking first at space-localized branes are furthermore numerous: They are conceptually and technically easier, allow arbitrarily large charges and hence a proper thermodynamic limit, and have holographic dual descriptions as black holes, providing effective ``experimental measurements'' of thermodynamic quantities like the entropy. $(iii)$ It prevents, for the time being, potential diffusion of confusion from partially unresolved conceptual problems of quantum gravity in a cosmological setting (as reviewed by Tom Banks at this school \cite{Banks2010}) into what should be an exposition of well-understood and diversely applicable methods. 

More specifically, I will cover the following topics:
\begin{enumerate}
 \item An introduction to the theory of spin glasses, in particular Parisi's solution of the Sherrington-Kirkpatrick model, using the replica formalism. This was the first nontrivial energy landscape to be studied and understood in detail in physics. Special attention is given to Parisi's overlap order parameters, which allow to detect a nontrivial equilibrium state space structure without having to be able to explicitly know those equilibrium states. A generalization of this order parameter for arbitrary quantum systems is proposed at the end of the section. Some other complex systems such as the Hopfield model for memory and learning are briefly discussed. Other, non-replica approaches such as the cavity method and Langevin dynamics are important but not treated in any detail in these notes.
 \item An introduction to supersymmetric quantum mechanics, where simple but powerful concepts such as the Witten index make it possible to compute exact quantum ground state degeneracies of highly complex systems. The computation of nonperturbative lifting effects due to landscape barrier tunneling as well as relations to Morse theory are discussed in some detail. Finally, the map between nonsupersymmetric Langevin dynamics and supersymmetric quantum mechanics is introduced, and its appearance in glass theory is outlined.
 \item An introduction to the low energy quantum mechanics description of D-branes wrapped on compact cycles in Calabi-Yau manifolds. These are the prime examples of complex systems in string theory, and excellent models to introduce techniques ubiquitous in the field of string compactification. Again I start from scratch, assuming only some background knowledge in elementary differential geometry and the basics of D-branes in string theory. I describe in detail the low energy reduction of the D4-brane wrapped on a high degree 4-cycle. This results in a supersymmetric quantum mechanics of high complexity, which nevertheless is manageable thanks to the underlying supersymmetric geometry. If the D4 charge is of order $N$, the brane allows of order $N^3$ worldvolume magnetic fluxes, inducing an extremely complicated superpotential for its order $N^3$ moduli, and leading to an enormous energy landscape with exponentially many minima. Nevertheless it is possible to explicitly construct vast numbers of exact supersymmetric critical points, by giving the critical point condition the interpretation of ``capturing'' holomorphic 2-cycles. At large D0-charge, this method becomes inadequate, but at the same time it becomes easier to give good estimates for the number of ground states. A formula for the index is derived by directly applying the machinery of Lagrangian supersymmetric quantum mechanics, treating the worldvolume fluxes as momenta canonically conjugate to angular variables. This reproduces the well known continuum approximation at leading order, but also provides all corrections in terms of differential euler characteristics of critical point loci. Finally, bound states with mobile D0-branes are introduced and counted in a rudimentary fashion, a microscopic formula for the D4-D0 is derived, and some open ends and generalizations are summed up to conclude. 
 \item An overview of the construction of the zoo of multicentered supersymmetric black hole bound state solutions that are the holographic duals of the D-brane energy landscapes discussed in the previous section. 
\end{enumerate}

\subsection{Some general references}

For the topics covered in these lectures, I will give references along the way. For the topics mentioned above but which will not be treated in any detail in what follows, here are a few general references
that may be useful: \cite{Starobinsky1982,Linde1983,Linde1994,Winitzki2002,Strominger2001,Strominger2001a,Spradlin2001a,Witten2001a,Maldacena2003b,Anninos2010b,Bousso2010,Bousso2010a,Bousso2008a,Bousso2007,Garriga,Vilenkin2011,
multiverse,Everett1957,Giddings1988,Coleman1988,Arkani-hamed,Weinberg1987,Bousso2007,Strominger1986,Bousso2000,Kachru2003,Susskind2003,Schellekens2006,Douglas2003,GRANA2006,Acharya2006,Douglas2007,Denef2007new,Denefb,Douglas2010b,Nabutovsky2003}.

\section{Spin glasses and beyond} \label{sec:spinglass}

\subsection{Introduction}

A spin glass \cite{Mezard1987new,Binder1986,Fisher1992} is a system of localized spins with disordered interactions. An example is copper sprinkled with manganese atoms at random positions. The sign of the spin-spin interaction potential between the manganese atoms oscillates as a function of distance and is relatively long range,
so we get effectively randomized mixed ferromagnetic and antiferromagnetic interactions. Some spin pairs will want to align, while others will wish to anti-align. If among three given spins, two pairs want to align and one pair wants to anti-align, or if all three pairs want to anti-align, one pair will not get what it wants and the triangle is called frustrated. The presence of many frustrated triangles typically leads to exponentially many local minima of the energy, and thus to a landscape.

This landscape is highly complex; the problem of finding the ground state in even the simplest spin glass models is effectively intractable. Nevertheless, as we shall see, some models of spin glasses have exactly solvable thermodynamics, thanks to a remarkable underlying mathematical structure which greatly simplifies the analysis but nevertheless leads to a very rich phase space structure. Its most striking consequence is \emph{ultrametricity} of the state space. An ultrametric space is a metric space with a distance function $d$ that satisfies something stronger than the standard triangle inequality, namely
\begin{equation}
 d(a,b) \leq \max\{ d(b,c),d(c,a) \} \, .
\end{equation}
What this means is that all triangles are isosceles, with the unequal side the shortest of the three.  Ultrametric spaces appear in various branches of mathematics; for example the $p$-adic distance\footnote{The $p$-adic norm of a rational number is $|q|_p=p^{-k}$ when $q$ can be written as $q=p^k m/n$ where $m$ and $n$ contain no powers of $p$. The $p$-adic distance is then $d_p(a,b) \equiv |a-b|_p$.} between rational numbers is ultrametric. For nice introductions to the mathematics of ultrametric spaces, see chapter 9 of \cite{Cartier1992}, and \cite{Brekke1988new}. Ultrametric spaces are also familiar in biology: If we define the distance between two current species as how far back in time one has to go to find a common ancestor, this distance is trivially ultrametric. Indeed, picking a turkey, a monkey and a donkey for example, the distance turkey-donkey and turkey-monkey is the same, while the distance monkey-donkey is the shorter one of the three. Alternatively, one can define a species distance as the degree of difference of DNA. Since this will be roughly proportional to how far back a common ancestor is found, this distance will again be essentially ultrametric. The same is true for any representative set of proteins. This is used with great success to reconstruct evolutionary trees (for a recent account see e.g.\ \cite{Roch2010}). The lesson here for us is that ultrametricity is equivalent to the possibility of clustering objects and organizing them in a \emph{hierarchical tree}, similar to the evolution tree of species. 

\begin{figure}
\begin{center}
\includegraphics[height=4cm]{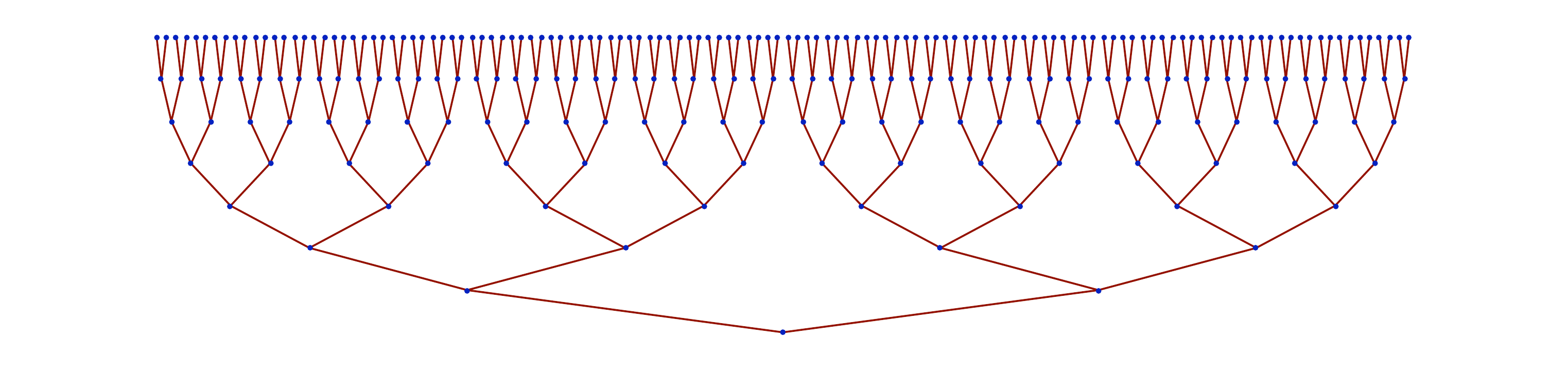}
\end{center}
\caption{\footnotesize An ultrametric tree. Distances between the points in the top layer are set by the vertical distance to the first common ancestor node.
 \label{tree}}
\end{figure}

\begin{figure}
\begin{center}
\includegraphics[height=5.8cm]{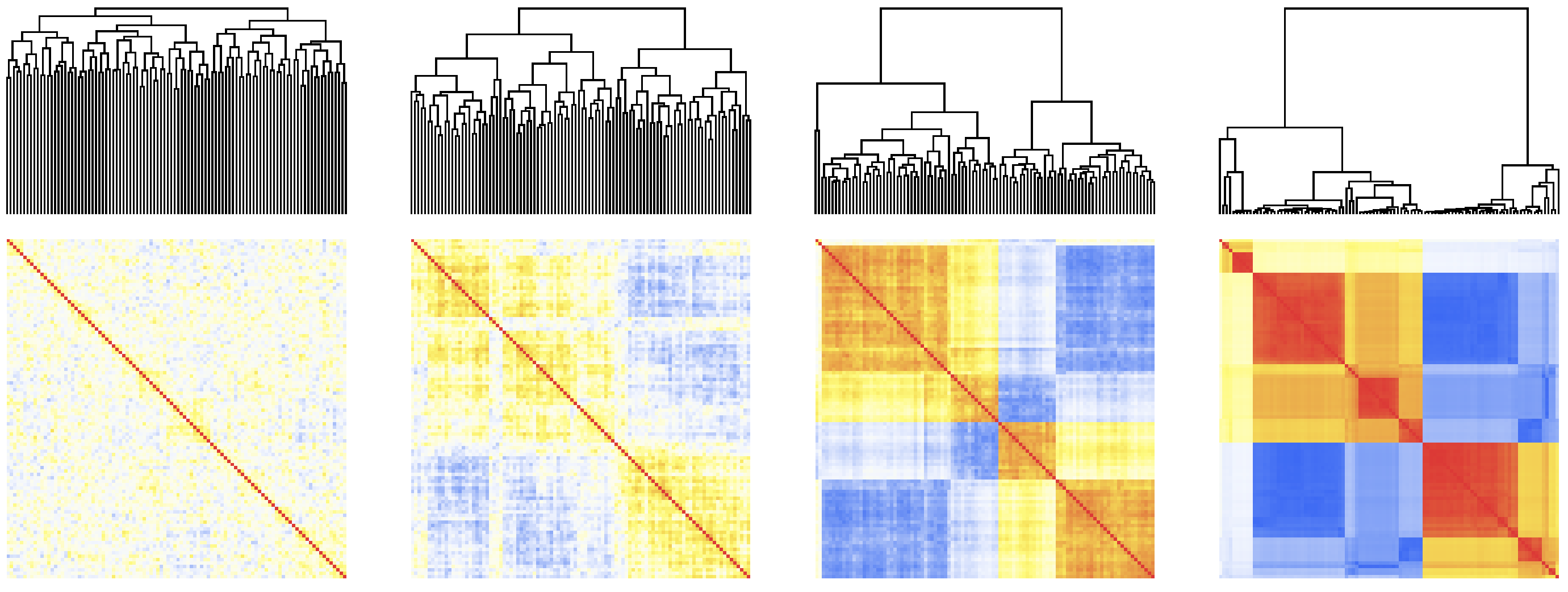}
\end{center}
\caption{\footnotesize \cite{AABDG} Dendrogram plots and overlap matrices for the SK model with $N=800$ spins. 
We used parallel tempering Monte Carlo \cite{Marinari1998,Earl2005} to reach thermal equilibrium, with 50 replicas at equally spaced temperatures between $T=0.1$ and $T=1.2$. Based on overlaps of 100 configurations sampled with separation of 100 sweeps, clustered using Mathematica. Results are shown for $T/T_c=1.2, 0.86, 0.55, 0.12$. Red = maximal positive overlap $q=+1$ (i.e. minimal distance), white = overlap $q=0$, dark blue = maximal negative overlap $q=-1$. Although the data is too limited to draw firm conclusions, the plots are suggestive of hierarchical clustering.  
 \label{SKplots}}
\end{figure}

In the case of spin glasses, the points in the metric space are the different equilibrium states, that is the different ergodic components or superselection sectors in which the standard Gibbs measure breaks up in the glass phase. We will describe this in detail in the subsequent sections, here we give a qualitative picture. The distance between different equilibrium states is given by some measure of microscopic dissimilarity, for example the sum of local magnetization differences squared. Remarkably, in  mean-field models such as the Sherrington-Kirkpatrick model \cite{Sherrington1975}, this space turns out to be ultrametric \cite{Parisi1979,Parisi1980,Mezard1984}: In the thermodynamic limit, with probability 1, triples of magnetization distances are isosceles. Equilibrium states get hierarchically organized in clusters. The cluster division is independent of the distance measure used, as long as it is statistically representative, meaning zero distance implies equal states. Figure \ref{SKplots} shows the results of a simulation illustrating this phenomenon.

What is remarkable is that such a nontrivial tree structure appears in a purely static setting, without any apparent underlying evolution process creating it.\footnote{To be precise, the remarkable thing is the existence of a \emph{nontrivial} ultrametric structure, with multiple branchings. Trivial ultrametricity, meaning all triangles are equilateral, is very easy to realize in infinite dimensional spaces. For example all pairs of randomly chosen points on an infinite dimensional sphere are with probability 1 at the same distance from each other.} Instead, the role of evolutionary time is played by temperature or more generally energy scale: At high temperatures there is only one ergodic component of phase space. When the temperature is lowered below the spin glass phase transition, the phase space starts breaking up into distinct parts that no longer talk to each other, separated by effectively unsurpassable free energy barriers. Initially these valleys in the free energy landscape will all still look very similar, but upon further cooling, their mutual magnetization distance grows, and eventually they in turn will start breaking up in components separated by free energy barriers. This branching process proceeds all the way down to zero temperature, generating an ultrametric tree of states. Any pair of ground states defines a ``time'' on this tree, proportional to their ultrametric distance. 

As we will see, ultrametricity plays an important role in the exact solution of mean field spin glass models, and determines many of their static and dynamic properties. It plays in this sense a role similar to that of symmetries and symmetry breaking in physics. In fact, its emergence is closely related to the breaking of an auxiliary symmetry, called replica symmetry, which naturally arises in the description of disordered systems. The order parameter (or rather the order \emph{function}) capturing the symmetry breaking pattern exhibits ultrametricity as one of its characteristic features. We will discuss this extensively in what follows.

For more realistic, local (short range) spin glass models in three dimensions, for which the mean field approximation is poor, there is no consensus on the physical presence or relevance of ultrametricity, and different schools exist with different favorite models and results, the most widespread being the droplet model of \cite{Fisher1986}. On the other hand, in systems where little or no locality is present, such as neural networks or combinatorial optimization problems, mean field methods and ultrametricity are ubiquitous \cite{Mezard1987new}. A recent numerical study of ultrametricity in simple Ising type models of different degrees of locality can be found in \cite{Katzgraber2009}. 

For a review of ultrametricity in physics, see \cite{Rammal1986b}. General criteria for the appearance ultrametricity were formulated in \cite{Parisi2000}. Introductory texts on spin glass theory include \cite{Mezard1987new,DeDominicis2006new,Castellani2005,Crisanti1992a} 

In the following we will make the above qualitative description more precise, introduce the new order parameters and explain the mysterious but powerful replica trick and the phenomenon of replica symmetry breaking associated with ultrametricity.

\subsection{The Parisi order parameter}

\subsubsection{Spin glass models} \label{sec:spinglassmodels}

The most basic models for spin glasses are formally similar to the classical Ising model, with $N$ spins $s_i = \pm 1$ interacting according to a Hamiltonian of the form\footnote{In the statistical mechanics literature, the common convention is to put a minus sign in front of the sum over the spins, so ferromagnetic couplings are positive. We will not do this in these lecture notes.}
\begin{equation}
 H = \sum_{ij} J_{ij} s_i s_j \, .
\end{equation}
In the case of the Ising model, we have $J_{ij} = -J$ for nearest neighbors $(ij)$ and zero otherwise. At high temperatures the magnetization $M \equiv \frac{1}{N} \sum_i m_i$ is zero ($m_i$ is the average value of $s_i$). In dimension 2 or higher, below a critical temperature $T_c$, the magnetization acquires a nonzero value and the $\IZ_2$ spin flip symmetry is spontaneously broken. In the mean field approximation, $T_c = J$ and $M$ is a solution of the mean field consistency equation $M = \tanh(M J/T)$.

In 1975, to model spin glasses, Edwards and Anderson \cite{Edwards1975} proposed to study the above Hamiltonian, still with nearest neighbor couplings $J_{ij}$, but now drawn randomly and independently from a Gaussian distribution with zero mean and standard deviation $J$. Although at finite $N$ everything in this model will depend on the actual values of $J_{ij}$, in the thermodynamic limit $N \to \infty$, intensive quantities such as the (spatially averaged) magnetization or free energy density are self-averaging: they become equal to their $J_{ij}$-averaged value with unit probability. This makes computations possible. Due to the disorder, the magnetization $M$ vanishes at all temperatures in this model. Nevertheless, there exists a critical temperature below which the spins freeze, in the sense that they acquire locally preferred directions. This disordered frozen phase is the spin glass phase. The magnetization is clearly not a good order parameter to detect this situation. Instead Edwards and Anderson introduced a new order parameter:
\begin{equation} 
\label{qEA}
 q_{EA} \equiv \frac{1}{N} \sum_i m_i^2 \, .
\end{equation}
Here the $m_i$ are the local magnetizations in the equilibrium state the system finds itself in, which in an actual physical setting can be thought of as the time averaged value of the spin $s_i$. For the Ising model $m_i$ is independent of $i$, so $q_{EA} = M^2$. The Edwards-Anderson parameter is related to the experimentally measured magnetic susceptibility: $\chi = (1-q_{EA})/T$.

Though the meaning of time averaging is physically clear, we have not specified any dynamics for this model, so we need a purely static definition for $m_i$. This is subtle. It is \emph{not} the canonical ensemble average $\langle s_i \rangle$ obtained from the Gibbs probability measure $p(s) \propto e^{-\beta H(s)}$, as this average is trivially zero due to the $\IZ_2$ symmetry of $H$. The same issue arises already for the Ising model, but there it is clear what is going on: below $T_c$ the Gibbs probability measure splits in two ``superselection'' sectors, characterized by opposite values of the magnetization. The two sectors can be separated by switching on a small  background magnetic field $h$, which shifts $H \to H + h \sum_i s_i$, lifting the $\IZ_2$ degeneracy and eliminating one of the two sectors, depending on the sign of $h$. We can then define the magnetization for each sector as $m_{i {\pm}} \equiv \lim_{h \to 0 \pm} \lim_{N \to \infty} \langle s_i \rangle_h$. 

For spin glasses this does not work; since $M=0$, switching on a constant $h$ will not lift the degeneracy. Switching on an inhomogeneous field $h_i$ tailored to the $m_i$ of the frozen equilibrium state we want to single out would of course do the job, but it is impossible to know in advance what the required profile is going to be; it could be anything, and it will be different for different values of $J_{ij}$. To make things worse, there may be many distinct equilibrium states, all with different values of the $m_i$, and again there is no way of telling without actually solving the system. There is no standard symmetry and no order apparent in the system at any temperature. What ``order'' parameter could possibly distinguish between those phases? 

These questions were considerably sharpened and then answered through the study of an even more simplified spin glass model, proposed by Sherrington and Kirkpatrick \cite{Sherrington1975} in 1975. The Hamiltonian is as above except that now the couplings $J_{ij}$ are nonzero for \emph{all} pairs of spins, not just nearest neighbors; they are all independent Gaussian random variables with zero mean and variance $J^2/2N$,\footnote{The variance is scaled with $N$ such that the typical size of the interaction potential at each lattice site remains finite in the thermodynamic limit: $\overline{(\sum_j J_{ij} s^j)^2}=J^2/2$. The factor of 2 is added for consistency with standard conventions, in which usually $J_{ij}=J_{ji}$ is imposed, which we will not do.}
\begin{equation}
 \overline{J_{ij} J_{kl}} = \tfrac{J^2}{2N} \delta_{ik} \delta_{jl} \, . 
\end{equation} 
Overlines will denote averages over the $J_{ij}$ throughout these notes. The model is completely nonlocal; spatial geometry has no meaning in this setup. It is exactly solvable, but the solution turned out much more interesting than Sherrington and Kirkpatrick originally imagined. The simple solution they originally proposed was correct at high temperatures, but manifestly wrong near $T=0$, where it predicted for instance negative entropy densities. After considerable effort, the correct solution was finally found in 1979, in a seminal breakthrough by Parisi \cite{Parisi1979,Parisi1980}. (It took another 25 years before the solution was rigorously proven to be correct \cite{Guerra2008,Talagrand2006}.) In the course of the process, Parisi uncovered the proper order parameters to fully describe the spin glass phase, as well as a hidden ``statistical'' symmetry group whose breaking they parametrize. We will describe the solution in section \ref{sec:RSsolSK}, but first we describe the physical meaning of the order parameter.\footnote{Historically it was first obtained as a formal mathematical object in the replica formalism. The physical interpretation was given later in {\cite{Parisi1983}}.} To do this we must introduce the notion of thermodynamic pure states.

\subsubsection{Pure states}

The probability measure characterizing the Gibbs state is
\begin{equation}
 p_G(s) = \frac{1}{Z} \, e^{-\beta H(s)} \, , \quad Z = \sum_s e^{-\beta H(s)} \, .
\end{equation}
For a given probability measure $p$, we denote the expectation value of an observable $A(s)$ by $\langle A \rangle_p$. For the Gibbs state we usually just write $\langle A \rangle_G = \langle A \rangle$. We say $p$ satisfies cluster decomposition if correlation functions of local observables factorize in the thermodynamic limit for almost all points. That is, for any finite $r$, in the limit $N \to \infty$,
\begin{equation}
 \langle A_{i_1} B_{i_2} \cdots C_{i_r} \rangle_p =  \langle A_{i_1} \rangle_p  \langle B_{i_2} \rangle_p \, \cdots \langle C_{i_r} \rangle_p \, + \, R_{i_1 i_2 \cdots i_r} ,
  \label{clusterproperty}
\end{equation} 
where the remainder $R$ is negligible on average: $\lim_{N \to \infty} \frac{1}{N^r} \sum_{i_1 \cdots i_r} |R_{i_1 \cdots i_r}| = 0$.
Here $A_i$, $B_i$, \ldots, $C_i$ are local observables, like for example $A_i = s_i$ or $B_i = s_{i-1} s_i s_{i+3}$. In particular the clustering property implies that intensive quantities like the free energy density have definite, non-fluctuating values. In local theories (\ref{clusterproperty}) is equivalent to the property that correlation functions factorize in the limit of infinite spatial separation.

The Gibbs measure for the Ising model satisfies cluster decomposition above the critical temperature, but not below: at high temperatures we have $\langle s_i s_j \rangle = 0 = \langle s_i \rangle \langle s_j \rangle$ when $|i-j| \to \infty$, but at low temperatures we have instead $\langle s_i s_j \rangle = M^2 \neq \langle s_i \rangle \langle s_j \rangle = 0 \cdot 0$. 
However we can canonically decompose the Gibbs measure into a direct sum of probability measures which do satisfy cluster decomposition. Thus, for the Ising example, we have below the critical temperature
\begin{equation}
 p_G(s) = \tfrac{1}{2} p_+(s) + \tfrac{1}{2} p_-(s) \, ,
\end{equation}
where $p_\pm(s) = \lim_{h \to \pm 0} \lim_{N \to \infty} \frac{1}{Z} e^{-\beta (H(s) + h \sum_i s_i)}$. Sets of finite measure according to $p_+$ have zero measure according to $p_-$ and vice versa. The measures $p_\pm$ do satisfy cluster decomposition: $\langle s_i s_j \rangle_+ = M^2 = \langle s_i \rangle_+ \langle s_j \rangle_+$. The superselection sectors described by $p_{\pm}$ are called ``pure states'' in the statistical mechanics literature. 

Rigorously defining pure states for general systems is subtle. For a discussion in the context of spin glasses we refer to page 89 (appendix 1) of \cite{Young1997}. What is known rigorously is that any probability measure $p_G$ defined for a system of \emph{infinite} size, which locally behaves like the Gibbs measure (i.e.\ it gives relative probabilities proportional to $e^{-\beta \Delta H}$ for finite size fluctuations), can always be uniquely decomposed into pure states $p_\alpha$ as
\begin{equation} \label{decomp}
 p_G(s) = \sum w_\alpha \, p_\alpha(s) \, .
\end{equation}
Here $w_\alpha>0$ is the Gibbs probability of being in the pure state $\alpha$. The pure states satisfy the clustering property and cannot be further decomposed. This gives them an implicit definition.\footnote{More intuitively, in physical systems, pure states correspond to distinct ergodic components, individually invariant under time evolution and not further decomposable into smaller time invariant components. In ergodic components the ergodic theorem implies that the time average of any observable equals its ensemble (phase space) average.}

\subsubsection{Overlap distributions} \label{sec:overlap}

As mentioned earlier, for disordered systems, it is in general impossible to explicitly find the actual decomposition into pure states. But granting the decomposition exists, one can formally define the \emph{overlap} $q_{\alpha \beta}$ between pure states as
\begin{equation} \label{defoverlap}
 q_{\alpha \beta} \equiv \frac{1}{N} \sum_i \langle s_i \rangle_\alpha \langle s_i \rangle_\beta \, .
\end{equation}
As a special case, notice that $q_{\alpha \alpha}$ is nothing but the Edwards-Andersen order parameter $q_{EA}$ defined in (\ref{qEA}). Although not at all obvious, it turns out \cite{Mezard1985new} that $q_{\alpha \alpha}$ is independent of the pure state $\alpha$ (and of the disorder realization $J_{ij}$), so $q_{EA}$ is actually an invariant of the system, depending only on $T/J$.\footnote{Since the SK model is completely nonlocal and does not a priori distinguish any site $j$ relative to a given site $i$, we can think of the permutation symmetry acting on the site indices $i$ as a gauge symmetry, similar to the diffeomorphism group in gravity (where $J_{ij}$ is the analog of the spacetime metric). Then the local magnetization distribution $\CP(m) \equiv \frac{1}{N} \sum_i \delta(m-m_{i\alpha}) = \overline{\delta(m - m_{i\alpha})}$ encodes all gauge invariant information based on the ``vevs'' $m_{i\alpha}$. It is shown in \cite{Mezard1985new} that again this quantity is independent of the state $\alpha$. So we can say that there is no gauge invariant distinction between different pure states based on just the magnetizations; all equilibrium states look the same as far as non-fluctuating quantities are concerned.} Using this, we see that the overlap is closely related to the Euclidean distance between states:
\begin{equation}
 d(\alpha,\beta)^2 = \frac{1}{N} \sum_i \left( m_{i \alpha} - m_{i \beta} \right)^2 
 = 2(q_{EA} - q_{\alpha \beta}) \, .
\end{equation}
At this point the overlap matrix may still seem like an abstract, incomputable quantity. However, consider the \emph{overlap probability distribution}:
\begin{equation} \label{Pqdef}
 P(q) = \sum_{\alpha \beta} w_\alpha w_\beta \, \delta(q-q_{\alpha \beta}) \, . 
\end{equation}
This is the probability of finding an overlap $q_{\alpha \beta} = q$ when one samples the Gibbs state. The wonderful thing is that this quantity is actually computable without any knowledge of the actual decomposition into pure states. It can be rewritten purely in terms of the Gibbs state as the overlap distribution for two identical replicas of the system, with spins $s^{(1)}$ and $s^{(2)}$:
\begin{equation} \label{Pqrep}
 P(q) = \left\langle \delta \left( q - \mbox{$\frac{1}{N} \sum_i s_i^{(1)} s_i^{(2)}$}\right) \right\rangle_{n=2} \, ,
\end{equation}
where $n=2$ means we are considering two replicas:
\begin{equation}
 \left\langle A(s^{(1)},s^{(2)}) \right\rangle_{n=2} \equiv \frac{1}{Z^2} \sum_s e^{-\beta H(s^{(1)})-\beta H(s^{(2)})} \, A(s^{(1)},s^{(2)}) \, . 
\end{equation}
To prove this, one shows the moments are equal. Consider for example its second moment:
\begin{eqnarray*}
 \langle q^2 \rangle &=& \int dq \, P(q) \, q^2  \\
 &=& \sum_{\alpha \beta} w_\alpha w_\beta \, \frac{1}{N} \sum_i 
 \langle s_{i} \rangle_\alpha  \langle s_{i} \rangle_\beta \, \frac{1}{N} \sum_j
 \langle s_{j} \rangle_\alpha \langle s_{j} \rangle_\beta \\
 &=& \frac{1}{N^2} \sum_{\alpha \beta} w_\alpha w_\beta  \sum_{i j} 
 \langle s_{i} s_{j} \rangle_\alpha \langle s_{i} s_{j} \rangle_\beta \\
 &=& \frac{1}{N^2} \sum_{i j} 
 \langle s_{i} s_{j} \rangle \langle s_{i} s_{j} \rangle \, \\
 &=& 
 \frac{1}{N^2} \left\langle \left( \sum_i s^{(1)}_{i} s^{(2)}_{i} \right)^2 \right\rangle_{n=2} \, . 
\end{eqnarray*}
To get to the third line we used the clustering property (\ref{clusterproperty}), and for the fourth line we used (\ref{decomp}) in reverse to express everything in terms of Gibbs state expectation values. The higher moments are treated analogously.

Equation (\ref{Pqrep}) in principle allows us to compute or approximate $P(q)$ by standard methods, for example by using Monte Carlo simulations sampling the Boltzmann-Gibbs distribution. It captures the presence and properties of pure states very well and is therefore a good order parameter for the spin glass phase. If only one pure state is present, namely the Gibbs state itself, as is the case at high temperatures, then $q_{11}=q_{EA}=0$ due to the $\IZ_2$ symmetry, and $P(q)=\delta(q)$. If the system freezes and splits into one $\IZ_2$ pair of pure states, with local magnetizations $m_i$ and $-m_i$, then for both states the self-overlap is $q_{EA}=\frac{1}{N} \sum_i m_i^2$ and $P(q)=\frac{1}{2} \delta(q-q_{EA}) + \frac{1}{2} \delta(q+q_{EA})$. If there are many pure states, there will be many delta-functions, and if there is an infinite number of them, $P(q)$ becomes continuous.

\begin{figure}
\begin{center}
\includegraphics[height=6.2cm]{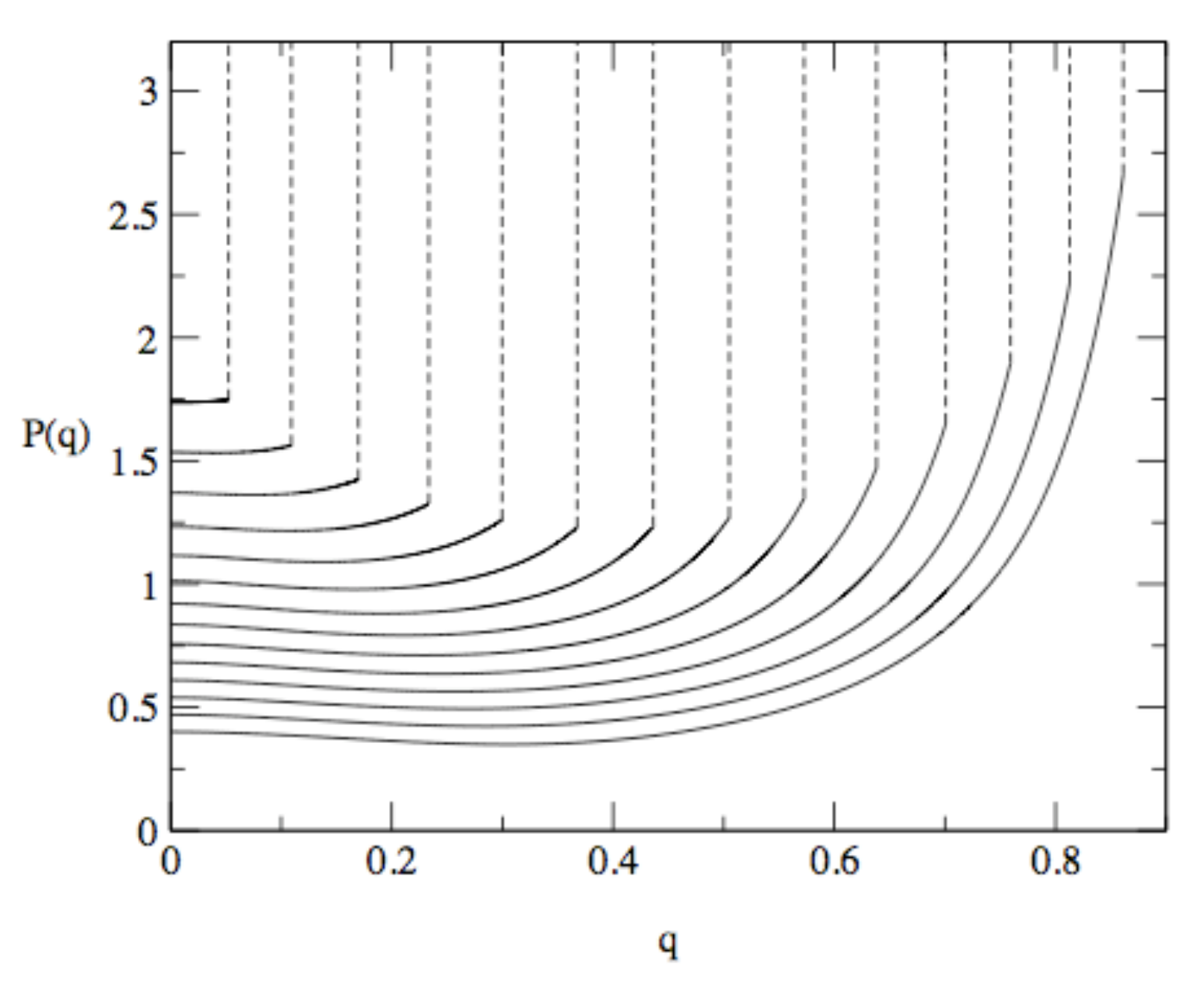}
\end{center}
\vskip-2mm\caption{\footnotesize From \cite{Crisanti2002a}. Disorder-averaged overlap distribution $\overline{P(q)}$ for the SK model at various temperatures, from $T = 0.95$ on right to $T = 0.30$ on left in steps of 0.05 ($T_c = 1$). The dotted lines represent delta-functions, localized at $q_{\alpha\alpha}=q_{EA}$.  \label{Pofq}
}
\end{figure}

Computing $P(q)$ analytically for a given coupling matrix $J_{ij}$ is not possible. It is however possible to compute its average $\overline{P(q)}$, by a variety of methods that can also be used to compute more basic thermodynamic quantities; the most prominent ones are the replica, cavity (TAP), Langevin dynamics and supersymmetry methods \cite{Mezard1987new,DeDominicis2006new,Thouless1977,Parisi1979b,Kurchan1992,Kurchan2002a}. In these lecture notes we will focus on the replica method. To already get an idea of where all this is heading, a result of such computations is shown in fig.\ \ref{Pofq}, displaying the disorder-averaged $\overline{P(q)}$ for various temperatures below $T_c$. The function is smooth except for a delta-function peak at the highest value of $q$, which is the self-overlap $q_{\alpha\alpha}=q_{EA}$. Its presence is a consequence of the state and disorder independence of $q_{EA}$, leading to a term $(\sum_\alpha w_\alpha^2) \delta(q-q_{EA})$ in (\ref{Pqdef}). The nonvanishing of $\sum_\alpha w_\alpha^2$ in turn indicates the presence of states of finite weight $w_\alpha$. The continuity of $\overline{P(q)}$ shows that there are in general multiple pure states, but it does not imply $P_J(q)$ is continuous at fixed $J$; in fact because some $w_\alpha$ are finite, $P_J(q)$ must have multiple delta peaks, but with $J$-dependent locations, so integrating over $J$ smooths them out. Indeed, unlike the magnetization or the free energy, $P_J(q)$ is not a self-averaging quantity; it fluctuates between different realizations of the disorder.

\subsubsection{Ultrametricity}

One can also define more refined overlap distributions, such as the  overlap triangle distribution
\begin{eqnarray}
 P(q_1,q_2,q_3) &\equiv& \sum_{\alpha \beta \gamma} w_\alpha w_\beta w_\gamma \, \delta(q_1 - q_{\beta\gamma})
 \delta(q_2 - q_{\gamma\alpha}) \, \delta(q_3-q_{\alpha\beta}) \, ,
\end{eqnarray}
which similar to (\ref{Pqrep}) can be written in terms of the overlap distribution of three independent replicas in the Gibbs state. Restricting to triples with positive overlaps,\footnote{This is necessary because without this restriction, the $\IZ_2$ symmetry trivially destroys the ultrametric structure: A sign flip of the spins in state $\alpha$ will flip the sign of two out of three overlaps of an $(\alpha \beta \gamma)$ triangle, and clearly this does not preserve isoscelesness. This also explains the structure in the blue regions of fig.\ \ref{SKplots}.} one finds \cite{Mezard1984}
\begin{eqnarray}
 P(q_1,q_2,q_3) &=& \frac{1}{2} \int_{0}^{q_1} dq' P(q') P(q_1)  \, \delta(q_1-q_2) \, \delta(q_2-q_3) \nonumber \\
 && +\frac{1}{2} \biggl( P(q_1) P(q_2) \, \theta(q_1-q_2) \, \delta(q_2-q_3)  \, + \, \mbox{permutations} \biggr) \, ,
\end{eqnarray}
where $\theta$ is the step function and $P(q)$ is as in (\ref{Pqdef}). This manifestly exhibits ultrametricity, with the first term encoding equilateral and the last three terms more general isosceles triangles.

\subsection{Replica solution of the SK model} \label{sec:RSsolSK}

\subsubsection{The replica trick}

We now turn to the actual computations in the SK model, first using the replica formalism \cite{Edwards1975,Sherrington1975,Parisi1979,Parisi1980}. The Hamiltonian is, as stated earlier:
\begin{equation} \label{SKHam}
 H_J[s] = \sum_{i \neq j} J_{ij} s_i s_j \, ,
\end{equation}
where $i,j = 1, \ldots, N$, $s_i = \pm 1$ and $J_{ij}$ is drawn randomly out of a Gaussian distribution with mean 0 and variance $\frac{J^2}{2N}$, i.e.\
\begin{equation}
 p(J_{ij}) \propto e^{-\frac{N}{J^2} J_{ij}^2} \, .
\end{equation}
Other distributions would give equivalent results; only the first two moments matter.
It is impossible to solve this system exactly for arbitrary given couplings $J_{ij}$ --- just finding the absolute minimum energy configuration is already an NP-hard problem \cite{Barahona1982}. 
Fortunately the usual intensive thermodynamic quantities of interest are \emph{self-averaging}, that is they are independent of the random matrix $J_{ij}$ with probability going to 1 in the thermodynamic limit. Thus we can compute them by computing averages. For example the free energy density is
 \begin{equation}
 F = \overline{F_{J}} \equiv \int {\scriptsize{\prod_{ij}}} dJ_{ij} \, p(J_{ij}) \, F_{J} \, , \qquad
 F_{J} \equiv - \frac{1}{\beta N} \log Z_{J} \, , \qquad Z_{J} \equiv \sum_{s \in \{ \pm 1 \}^N } e^{-\beta H_{J}[s]} \, .
\end{equation}
Crucially, the average $\overline{\log Z_{J}}$ has to be computed \emph{after} taking the logarithm. Such an average is called \emph{quenched}, the disorder represented by the $J_{ij}$ is called quenched disorder, etc. Computing the average first, i.e.\ on the partition function itself, is called \emph{annealed} averaging. Physically, this corresponds to a situation in which the couplings themselves are fluctuating variables. The annealed average is much easier to compute than the quenched average, but it is not the situation we are in; in a real spin glass for example the couplings are determined by the positions of the impurity atoms in the host crystal, which vary randomly over space but do not fluctuate in time, or if they do, on much longer time scales than the spins fluctuate.

To deal with the quenched average, one can use the replica trick. This is based on the observation that
\begin{equation} \label{replicaLog}
 \log Z = \lim_{n \to 0} \frac{Z^n - 1}{n}  = \frac{\partial}{\partial n} Z^n |_{n=0} \, ,
\end{equation}
together with the fact that for positive integers $n$, the average of $Z^n$ is just an annealed average of the disorder coupled to $n$ replicas of the original system, and therefore easy to compute. Of course, the subtle part is the ``analytic continuation'' to $n = 0$, which is hard to make rigorous.\footnote{A perhaps somewhat more precise version starts from the observation that $Z^n$ is an entire function of $n$, with Taylor expansion $Z^n = e^{n \log Z} =\sum_{k=0}^\infty \frac{n^k}{k!} (\log Z)^k$ valid for all $n$, so $\overline{Z^n} = 1 + n \, \overline{\log Z} + \cdots$ for all $n$. The idea is then that the term linear in $n$ can be extracted by considering arbitrary integers $n$ and computing $\overline{Z^n}$ as an expansion in $n$. However this is in principle ambiguous: for example $1 + \sin (\pi n)$ is entire and evaluates to 1 for all integers, yet it has a nonvanishing first order term in its Taylor expansion. For a deeper analysis see e.g.\ \cite{Kondor1983,Verbaarschot1985}. \label{replicamoreprec}} 
However, its effective power in a large range of applications is undeniable, so the lack of a rigorous framework is no doubt more a failure of our arsenal of rigorous frameworks than of the idea itself.

Introducing replica indices $a,b=1,\ldots,n$, the coupling-averaged $n$-fold replicated partition function is, putting $J = 1$,   
\begin{eqnarray}
 \overline{Z^n} &=& \int dJ_{ij} \, e^{-N J_{ij}^2} \, \sum_{s \in \{\pm 1 \}^{nN}}  \, e^{-\beta J_{ij} s^i_a s^j_a} \, .
\end{eqnarray}
Repeated indices are summed over and we absorb the Gaussian normalization factors in the measure $dJ_{ij}$.  At this point, the replicas do not interact with each other. However, because they all couple to the same $J_{ij}$, integrating out the disorder induces effective interactions between the replicas:
\begin{eqnarray} \label{avreplZ}
 \overline{Z^n} &=& \sum_{s \in \{\pm 1 \}^{nN}}  \, e^{\frac{\beta^2}{4N}  s^i_a s^j_a s^i_b s^j_b} \, .
\end{eqnarray}
The positive sign in the exponent means the interaction is attractive. Now, reverse-mimicking how replicas got coupled by integrating out the site coupling matrix $J_{ij}$, we \emph{de}couple the lattice sites by integrating \emph{in} a ``replica coupling matrix'' $Q_{ab}$:
\begin{eqnarray}
 \overline{Z^n} &=& \sum_{s \in \{\pm 1 \}^{nN}} \, \int dQ \, e^{-N \frac{\beta^2}{4} (Q_{ab})^2} \, e^{\frac{\beta^2}{2} Q_{ab} s^i_a s^i_b } \nonumber \\
 &=& \int dQ \, e^{-N \frac{\beta^2}{4} (Q_{ab})^2 } \biggl( \sum_{S \in \{\pm 1 \}^{n}} e^{\frac{\beta^2}{2} Q_{ab} S_a S_b }  \biggr)^N \, . \label{replicresF}
\end{eqnarray}
The $S$-sum here runs over the $n$ replica copies only; lattice indices no longer occur. The sum over the lattice sites has simply produced an overall power of $N$. This is awesome because it means that in the thermodynamic limit $N \to \infty$, we can evaluate the integral over $Q$ in a saddle point approximation. It all boils down now to finding the critical points of
\begin{equation}
 \CF(Q) \equiv \frac{\beta}{4} Q_{ab}^2 - \frac{1}{\beta} \log \CZ(Q) \, , \qquad
 \CZ(Q) \equiv \sum_{S \in \{\pm 1 \}^{n}} e^{\frac{\beta^2}{2} Q_{ab} S_a S_b } \, . \label{CZCFdef}
\end{equation}
Denoting the dominant critical point(s) of $\CF(Q)$ by $Q_\star$, we thus get $\overline{Z^n}=e^{-\beta N\CF(Q_\star)}$. If we can find the saddle points for general $n$, we are done. Using the trick (\ref{replicaLog}) and noting that consistency with $Z^0=1$ requires $\CF(Q_\star)|_{n=0} = 0$, we obtain the free energy density:
\begin{equation} \label{finalZrep}
 F =  \frac{\partial}{\partial n} \CF(Q_\star)|_{n=0} \, . 
\end{equation}
To summarize, what we have done is trade summing over the lattice for summing over replicas, which allows us to do a saddle point computation of the quenched average. 
Before proceeding to find the solution, we pause to ponder the meaning of the matrix $Q_{ab}$ that apparently captures the large $N$ behavior of the model.

\subsubsection{Meaning of $Q_{ab}$}

So far the physical meaning of the matrix $Q_{ab}$ is obscure, although the conspicuous notational similarity with the overlap matrix $q_{\alpha\beta}$ introduced in (\ref{defoverlap}) suggests the two are related. We will now show that this is indeed the case, in that its structure in the limit $n \to 0$ will produce for us the Parisi order parameter $P(q)$ defined in (\ref{Pqdef}), or more precisely $\overline{P(q)}$. To see this, first observe that the saddle point equations $\CF'(Q_\star)=0$ can be written as the self-consistency equation
\begin{equation} \label{Qabsaddleeq}
 Q_{\star ab} = \langle S_a S_b \rangle_{Q_\star} \equiv \frac{1}{\CZ(Q_\star)} \sum_{S \in \{\pm 1 \}^{n}} S_a S_b \, e^{\frac{\beta^2}{2} \, Q_{\star cd} S_c S_d} \, .
\end{equation}
Hence $Q_{ab}$ equals the overlap  \emph{in replica space}. Now consider the expression (\ref{Pqrep}) for $P(q)$:
\begin{equation} \label{Pexpr2}
 P(q) = \sum_{s_1,s_2 \in \{\pm 1\}^N} \frac{e^{-\beta \left( H[s_1] + H[s_2] \right)}}{Z^2} \,
 \delta(q - \frac{1}{N} \sum_{i=1}^N s_1^i s_2^i) \, .
\end{equation}
This depends on the random couplings $J_{ij}$. We want to average it but are facing a problem similar to the problem we had when we wanted to compute the average of $\log Z$: now the problem is the explicit appearance of $J$ in the factors $1/Z$. We deploy again the replica trick, this time based on
\begin{equation} \label{Pqreplica}
 P(q) = \lim_{n \to 0} \sum_{s \in \{\pm 1\}^{nN}} e^{-\beta \sum_{a=1}^n H[s_a]} \, \,
 \delta(q - \frac{1}{N} \sum_{i=1}^N s_1^i s_2^i) \, ,
\end{equation}
where we have singled out the first two replicas to appear in the delta function. Indeed, the right hand side can for $n \geq 2$ be written as
\begin{equation}
 Z^{n-2} \sum_{s_1,s_2 \in \{\pm 1\}^N} e^{-\beta (H[s_1]+H[s_2])} \,
 \delta(q - \frac{1}{N} \sum_{i=1}^N s_1^i s_2^i) \, ,
\end{equation}
formally reducing to (\ref{Pexpr2}) when continued to $n \to 0$. Since in (\ref{Pqreplica}) there are no longer any denominators, we can compute the disorder average by simple Gaussian integration.

Using (\ref{Pqreplica}) and manipulations similar to those leading up to (\ref{replicresF}), one computes the finite moments $\overline{\langle q^ k \rangle} \equiv \int dq \, \overline{P(q)} \, q^k$. Dropping $1/N$ suppressed terms:
\begin{eqnarray}
\overline{\langle q^k \rangle} &=& \lim_{n \to 0} 
\int dQ \, e^{-\frac{\beta^2 N}{4} (Q_{ab})^2 } \biggl( \sum_{S \in \{\pm 1 \}^{n}} e^{\frac{\beta^2}{2}  Q_{ab} S_a S_b }  \biggr)^{N-k} \biggl( \sum_{S \in \{\pm 1 \}^{n}} e^{\frac{\beta^2}{2}  Q_{ab} S_a S_b } \, S_1 S_2 \biggr)^k \nonumber \\
&=&
 \lim_{n \to 0} \int dQ \, e^{- \beta N \CF(Q)} \, \biggl( \sum_{S \in \{\pm 1 \}^{n}} \frac{e^{\frac{\beta^2}{2}  Q_{ab} S_a S_b }}{\CZ(Q)} S_1 S_2 \biggr)^k \, ,
\end{eqnarray}
where we have absorbed in the measure $dQ$ a normalization factor $1/\int dQ \, e^{-\beta N \CF(Q)}$, which ensures $\overline{\langle q^0 \rangle} = 1$. The functions $\CZ(Q)$ and $\CF(Q)$ were defined in (\ref{CZCFdef}). We may replace $S_1 S_2$ by $S_a S_b$ with $a \neq b$ and average over replicas, since there is no distinction between replicas. The resulting expression has the advantage of being the same at distinct saddle points related by the permutation symmetry of the replicas. In view of (\ref{Qabsaddleeq}), the saddle point evaluation then gives
\begin{equation}
 \overline{\langle q^k \rangle} = \lim_{n \to 0} \sum_{[Q_\star]} W_{Q_\star} \frac{1}{n(n-1)} \sum_{a \neq b} Q_{\star ab}^k  \, = - \lim_{n \to 0} \sum_{[Q_\star]} W_{Q_\star} \sum_{b \neq 1} Q_{\star 1b}^k \, .
\end{equation}
The sum over $Q_\star$ is a sum over \emph{distinct} replica permutation symmetry orbits, i.e.\ it is nontrivial only if there are different saddle points \emph{not} related by the permutation symmetry of the replicas, as may be the case when there are additional symmetries beyond the replica permutation symmetry. For each such orbit $Q_\star$, $W_{Q_\star}$ denotes its relative weight. In the case at hand, we do have an additional $\IZ_2$ spin flip symmetry in the system. If this is the only other source of saddle degeneracy, we have $W_{Q_\star}=\frac{1}{2}$. In general we conclude:
\begin{equation} \label{Pqf}
 \overline{P(q)} = - \lim_{n \to 0} \sum_{[Q_\star]} W_{Q_\star} \sum_{b \neq 1} \delta(q-Q_{\star 1b}) \, ,
\end{equation}
and if there is no degeneracy besides the $\IZ_2$ one (as one would expect generically), this becomes
\begin{equation} \label{Pqf2}
 \overline{P(q)} = - \lim_{n \to 0} \sum_{b \neq 1} \tfrac{1}{2} \delta(q-Q_{\star 1b}) +
 \tfrac{1}{2} \delta(q+Q_{\star 1b}) 
 \, .
\end{equation}
Thus the pure state overlap distribution will be entirely determined by the saddle point solution $Q_\star$ continued to $n \to 0$. To obtain the probability to find an overlap $q$, all we need to do is compute the fraction of entries $Q_{\star 1b}$ that are equal to $q$. The weird looking minus sign is not a typo: it must be there because $\sum_{b \neq 1} 1 = (n-1) \to -1$ when $n \to 0$. Still, counting the number of entries $Q_{1b}$ in a $0 \times 0$ matrix may cause a feeling of unease in the reader. In the following section we will try to make sense of this, and conclude that, paradoxically, the properly continued $0 \times 0$ matrix $Q_{ab}$ in fact has infinitely many degrees of freedom!

\subsubsection{Replica symmetric solution} \label{sec:RSsol}

Finding the critical points of $\CF(Q)$ for general $n$ is still a nontrivial task. Some obvious consequences of (\ref{Qabsaddleeq}) are $Q_{\star aa} = 1$ and the fact that it is a positive definite symmetric matrix, but to make further progress, one must make an ansatz for the form of the solution. The simplest one, which was the one used by Sherrington and Kirkpatrick \cite{Sherrington1975}, is the replica symmetric (RS) ansatz, which is the unique ansatz leaving the permutation symmetry unbroken (we will drop the explicit $\star$ subscripts from here on):
\begin{equation} \label{RS}
 Q_{ab} = u \delta_{ab} + q (1-\delta_{ab}) \, .
\end{equation}
So $Q_{ab} = u$ if $a = b$ and $Q_{ab}=q$ if $a \neq b$. The relation (\ref{Pqf2}) then gives 
\begin{equation} \label{PqRSA}
 \overline{P(q')} = \frac{1}{2} \bigl( \delta(q'-q) + \delta(q'+q) \bigr) \, .
\end{equation}
Thus the RS ansatz is equivalent to assuming not more than one $\IZ_2$ pair of pure states. If $q=0$, there is just one pure state and the system is in the paramagnetic phase. 

We wish to extremize $\CF(Q) = \frac{\beta}{4} Q_{ab}^2 - \frac{1}{\beta} \log \sum e^{\frac{\beta^2}{2} Q_{ab} S_a S_b }$. From the ansatz (\ref{RS}) we obtain
\begin{equation} \label{Qab2}
 Q_{ab}^2 = n u^2 + n(n-1) q^2 \, , \qquad
e^{\frac{\beta^2}{2} Q_{ab} S_a S_b} = e^{\frac{\beta^2}{2} \left(n(u-q)  + q (\sum_a S_a)^2 \right)  } \, .
\end{equation}
Assuming without loss of generality $q \geq 0$, we linearize the term involving $S_a$ in the exponent with the transformation
\begin{eqnarray}
 \sum_{S \in \{\pm 1 \}^{n}} e^{\frac{\beta^2}{2} q (\sum_a S_a)^2 }
 &=& \frac{1}{\sqrt{2 \pi q}} \int dz \, e^{-\frac{z^2}{2 q}}  \sum_{S \in \{\pm 1 \}^{n}} e^{z \beta  \sum_a S_a}  \label{Gausstransform} \\
 &=& \frac{1}{\sqrt{2 \pi q}} \int dz \, e^{-\frac{z^2}{2 q}}  \left[ 2\cosh(\beta z) \right]^n \\
 &=& 1 \, + \, n \, \frac{1}{\sqrt{2 \pi q}} \int dz \, e^{-\frac{z^2}{2 q}} \log \left[2 \cosh(\beta z) \right] \, + \, \CO(n^2) \, ,
\end{eqnarray}
whence
\begin{eqnarray}
 \partial_n \CF|_{n=0} = 
 \frac{\beta}{4} (u^2-q^2) -\frac{\beta}{2} (u-q)
 - \frac{1}{\beta} \frac{1}{\sqrt{2 \pi q}} \int dz \, e^{-\frac{z^2}{2 q}} \log \left[ 2 \cosh(\beta z) \right] \,  . \label{Fnqexact}
\end{eqnarray}
This is to be extremized with respect to $u$ and $q$. Extremizing $u$ is trivial:
\begin{equation}
 u=1 \, ,
\end{equation}
reproducing the result $Q_{aa}=1$ we arrived at earlier directly from (\ref{Qabsaddleeq}). We set $u=1$ in what follows. There is no simple closed form solution for the saddle point value of $q$, so let us consider limiting cases. In the high temperature limit $\beta \to 0$, we expand in powers of $\beta$: 
\begin{equation} \label{CFexp}
 \partial_n \CF|_{n=0} =  - \frac{\beta }{4}  \left(1+q^2\right)-\frac{\log 2}{\beta } \, + \,\CO(\beta^2) 
  \, .
\end{equation}
The extremum is at $q=0$: As expected for high temperatures the system is in its paramagnetic phase. The free energy and entropy densities at $q=0$ are, from (\ref{finalZrep}):
\begin{equation} \label{paramagnetic}
 F = -\frac{\log 2}{\beta} -\frac{\beta}{4}   \, , \qquad  S = \beta^2 \partial_\beta F= \log 2 - \frac{\beta^2}{4} \, .
\end{equation}
The $\log 2$ corresponds the two-fold spin degeneracy at each lattice site.

Below a critical temperature $T_c$, $q=0$ ceases to be the thermodynamically stable saddle point. To find this transition point, we expand (\ref{Fnqexact}) for small $q$
\begin{equation}
 \partial_n \CF|_{n=0} = 
 -\frac{\log 2}{\beta} -\frac{\beta}{4} +
 \frac{\beta \left(\beta ^2-1\right)}{4} \, q^2 - \frac{\beta^5}{3} \, q^3 \, \,
+ \, \, \CO(q^4) \, .
\end{equation}
The coefficient of $q^2$ changes sign when $\beta=1$, and a new saddle point with $q \approx \beta-1 > 0$ becomes the thermodynamically stable one.\footnote{What ``stable'' means in the limit $n \to 0$ is actually rather subtle \cite{Almeida1978}. The reader may notice for example that in the case at hand, the stable saddle points are local \emph{maxima} in $q$-space, whereas ordinarily in physical parameter spaces, stable equilibria are those that \emph{minimize} the free energy. This ``inverted'' rule finds its origin in the formally negative dimension of the fluctuation modes of $Q_{ab}$ in the limit $n = 0$. For example, the expression $U = \frac{1}{n}\sum_{a,b} Q_{ab}^2$ is manifestly positive definite for positive integers $n$. It has a positive definite Hessian (the $n^2 \times n^2$ unit matrix) for all $n$ and the $Q=0$ extremum of $U$ is stable. Nevertheless, inserting the RS ansatz (\ref{RS}) gives $U(q)=1-q^2$. It is in general nontrivial to do a full stability analysis in the replica formalism. For more discussion see \cite{Almeida1978,DeDominicis2006new}.  \label{invertedrule}} This signals the spin glass phase transition. Recalling we chose units such that the coupling variance parameter $J = 1$, we conclude 
\begin{equation}
 T_c = J  \,  .
\end{equation}

This seems all fine, and is consistent with Monte Carlo simulations. However, in the low temperature limit $\beta \to \infty$ something awkward happens. We have 
\begin{equation}
 \partial_n \CF|_{n=0} = -\frac{(1-q)^2}{4} \, \beta - \sqrt{\frac{2 q}{\pi}} \, + \, \CO(\beta^{-2}) \, .
\end{equation}
The relevant saddle point is at $q = 1 - \sqrt{\frac{2}{\pi}} \frac{1}{\beta}$. This means the self-overlap $q_{EA}$ of the pure states approaches $1$, as expected. But the free energy and entropy density are, to leading order
\begin{equation} \label{RSBFS}
 F = - \sqrt{\frac{2}{\pi}} + \frac{1}{2 \pi \beta} \approx -0.798 \, , \qquad
 \qquad
 S = -\frac{1}{2 \pi} \approx -0.159 \, .
\end{equation}
The free energy disagrees with numerical simulations, which indicate $F \approx -0.76$ at $T=0$. More dramatically, the negative entropy clearly does not make any sense. Therefore the solution must be wrong at low temperatures. The replica symmetric ansatz (\ref{RS}) is apparently too restrictive. The replica symmetry must be broken.

\subsubsection{Replica symmetry breaking and Parisi matrices}

\begin{figure}
\begin{center}
\includegraphics[height=5.8cm]{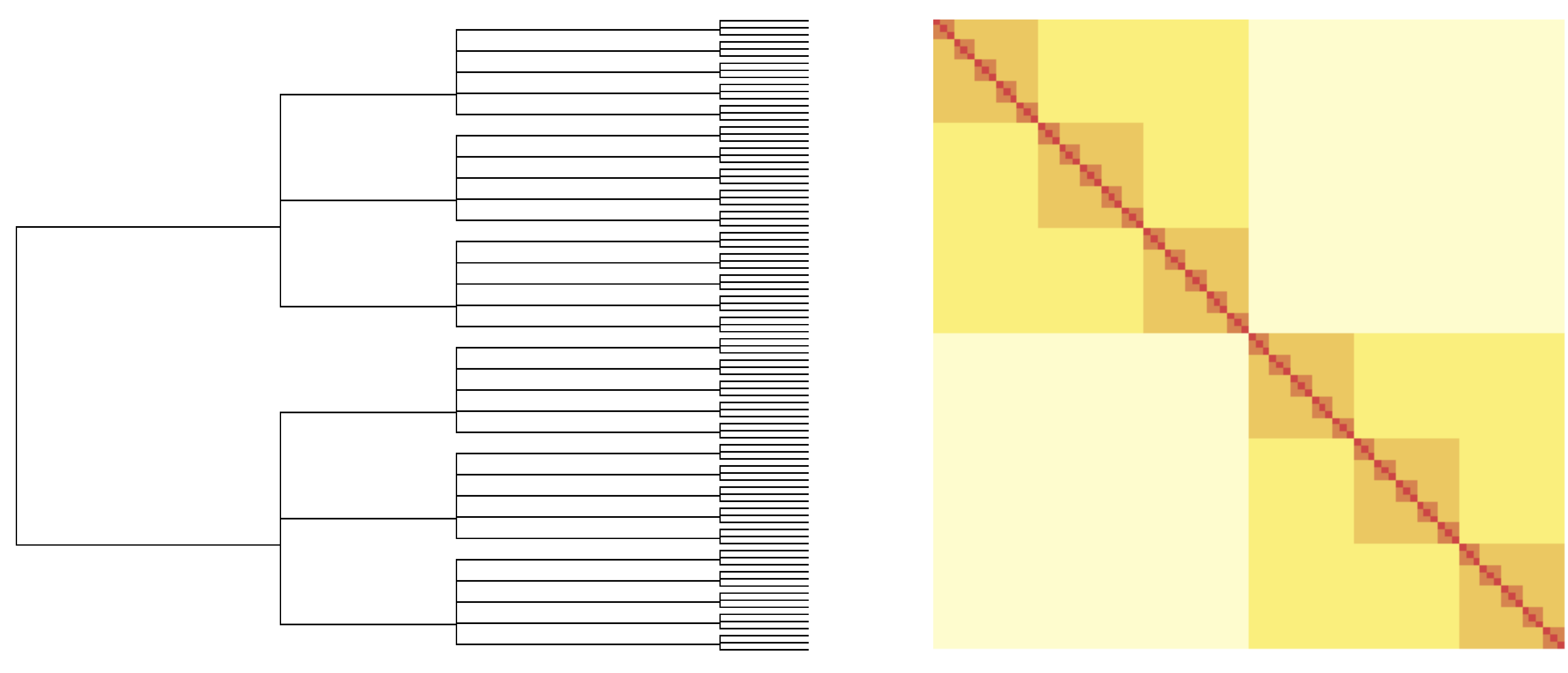}
\end{center}
\caption{\footnotesize Tree and matrix representation of a Parisi matrix $Q_{ab}$. Different colors and different tree connection heights correspond to different values of $Q_{ab}$, with darker red representing larger values. In this example, $n=90$, $K=3$, $\{m_0,m_1,m_2,m_3,m_4\}=\{1,3,15,45,90\}$, $\{u,q_0,q_1,q_2,q_3,q_4\} = \{ 1, 0.9, 0.6, 0.4, 0.1,0 \}$.
 \label{ParisiMatrix}}
\end{figure}

We want to relax the replica symmetric ansatz (\ref{RS}), which was $Q_{ab} = \delta_{ab} + q (1-\delta_{ab})$.\footnote{We put the diagonal part of $Q_{ab}$ equal to $\delta_{ab}$ here and it what follows. This is justified by observing that the diagonal part of $Q_{ab}$ decouples from the off-diagonal part in (\ref{Qabsaddleeq}), and that for any saddle point $Q_{aa}=1$.}

The effective replica Hamiltonian $H_R = - \frac{\beta N}{4}  \sum_{ab} \bigl( \frac{1}{N} \sum_i s^i_a s^i_b \bigr)^2$ appearing in (\ref{avreplZ}) describes an attractive interaction between the replicas: It is energetically favorable if the replicas line up. Competing with this is the fact that lining up means less phase space. Below $T_c$, the energy gain is more important than the entropy loss and we get a nonzero overlap expectation value between the replicas. In the RS ansatz the degree of correlation is assumed to be the same for all pairs of replicas. Relaxing this assumption means considering the possibility that groups of replicas form clusters that have larger overlaps amongst themselves than with other clusters. The maximally symmetric situation in such a scenario corresponds to the case in which the clusters are indistinguishable from each other, in particular equal in size and mutual overlaps. In this case there is also no absolute distinction between the replicas, only relative to some fixed other replica. 

To find the thermodynamically relevant critical points of $\CF(Q)$ (defined in equation (\ref{CZCFdef})), it is therefore natural to consider \cite{Parisi1979,Parisi1980} a minimal permutation symmetry breaking scheme $\CS_n \to \CS_{n/m} \times \CS_m$, by splitting up\footnote{When $m$ does not divide $n$, it is of course not possible to split the replicas in equal blocks of size $m$. The idea is to consider values of $n$ and $m$ for which it is possible, and then to analytically continue the resulting expressions, treating $m$, like $n$, as a continuous parameter.  
In this section we only consider the integer $n$ case. The continuation will be discussed in section \ref{sec:rsbn0}.} the $n$ replicas in clusters of size $m$, and assuming the overlap between distinct replicas within one cluster to be $q_{0}$ and the overlap between replicas in different clusters to be $q_1 < q_0$. Refining the replica labeling by $a = a_1 a_0$ where $a_1=1,\ldots,n/m$ labels the clusters and $a_0=1,\ldots,m$ the replicas inside each cluster, this translates to
\begin{equation}
 Q_{a_1a_0,b_1b_0} = \delta_{a_1 b_1} \delta_{a_0 b_0} + q_{0} \, \delta_{a_1 b_1} \epsilon_{a_0 b_0} + q_{1} \, \epsilon_{a_1 b_1} \, , \qquad \epsilon_{ab} \equiv 1 - \delta_{ab} 
\end{equation}
(so $\epsilon_{ab} = 1$ if $a \neq b$ and $0$ if $a=b$). The idea is then to substitute this new ansatz in $\CF(Q)$, extract the $\CO(n)$ term, and look for saddle points by varying $q_{0}, q_{1}$ and $m$. 

The clustering process can be iterated, by breaking up the clusters into smaller clusters and then those in turn into even smaller ones. When there are $K$ distinct nontrivial cluster sizes, this is referred to as replica symmetry breaking at level $K$ ($K$-RSB). 

To write things out explicitly, put $m_0 \equiv 1$ and denote the size of the smallest nontrivial cluster by $m_1$, the next larger one by $m_2$ and so on, up to $m_{K+1}=n$. Let $q_i$ be the overlap between replicas within a cluster of size $m_{i+1}$ (excluding those contained in an even smaller cluster). Labeling replicas by $a=a_K a_{K-1} \cdots a_1 a_0$ where $a_i=1,\ldots,\frac{m_{i+1}}{m_i}$ labels the clusters of size $m_i$, the ansatz can be written as  
\begin{eqnarray}
 Q_{a_K \cdots a_0,b_K \cdots b_0} &=&  \delta_{ab} + \sum_{i=0}^{K} q_i \, \delta_{a_K b_K} \cdots \delta_{a_{i+1} b_{i+1}}  \, \epsilon_{a_{i} b_{i}} \, \\
 &=& \sum_i \Delta_i \, \delta_{a_K b_K} \cdots \delta_{a_{i} b_{i}}\, , \qquad \Delta_i \equiv q_{i-1} - q_i > 0 \, , \label{QRSBdiff}
\end{eqnarray}
where $q_{i} \equiv 1$ if $i<0$ and $q_{i} \equiv 0$ if $i>K$. For example for $K=2$ this becomes
\begin{eqnarray*}
 Q_{a_2 a_1 a_0,b_2 b_1 b_0} 
 &=& \delta_{a_2 b_2} \delta_{a_1 b_1} \delta_{a_0 b_0} 
 + q_0 \, \delta_{a_2 b_2} \delta_{a_1 b_1} \epsilon_{a_0 b_0} + q_1 \, \delta_{a_2 b_2} \epsilon_{a_1 b_1} + q_2 \, \epsilon_{a_2 b_2} \,  \\
 &=& (1-q_0) \delta_{a_2 b_2} \delta_{a_1 b_1} \delta_{a_0 b_0} 
 + (q_0-q_1) \delta_{a_2 b_2} \delta_{a_1 b_1} + (q_1-q_2) \delta_{a_2 b_2} + q_2 \, .
\end{eqnarray*}
Reading this from right to left corresponds to zooming in to finer structures of the overlap matrix. An example with $K=3$ is shown in fig.\ \ref{ParisiMatrix}. 

The hierarchical block structure is equivalent to having an ultrametric structure in the overlap matrix. In other words, it can be organized as a tree, as illustrated in fig.\ \ref{ParisiMatrix}. Defining an ultrametric reference distance $r_{ab}$ between two replicas to be, say, the size $m$ of the smallest cluster to which they both belong, we can reformulate the Parisi ansatz simply as the statement that the overlap only depends on the distance:
\begin{equation} \label{Qdist}
 Q_{ab} = q(r_{ab}) \, ,
\end{equation} 
where $q(r)$ is an arbitrary function (encoding the $q_i$).


We want to evaluate the replica free energy $\CF(Q) = \frac{\beta}{4} Q_{ab}^2 - \frac{1}{\beta} \log \sum_S e^{\frac{\beta^2}{2} Q_{ab} S_a S_b }$ defined in (\ref{CZCFdef}). The first term is straightforward: 
\begin{equation} \label{RSBQQexample}
  \sum_{ab} Q_{ab}^2 = n \, \biggl( 1 + \sum_{i=0}^K q_i^2 \, (m_{i+1} - m_i) \biggr) \, .
\end{equation}
The interpretation of the coefficient $(m_{i+1}-m_{i})$ multiplying each $q_i^2$ is clear: it is the number of replicas having overlap $q_i$ with some fixed reference replica. The second term in $\CF(Q)$, proportional to $\log \CZ(Q)$, where $\CZ(Q) \equiv \sum_{S} e^{\frac{\beta^2}{2} Q_{ab} S^a S^b}$, is more interesting. Using (\ref{QRSBdiff}), we have
\begin{eqnarray} 
 \CZ = \sum_S \exp \biggl[ \frac{\beta^2}{2} \sum_i \Delta_i \sum_{a_K \cdots a_{i}}  \biggl( \sum_{a_{i-1} \cdots a_0} S_{a_K \cdots a_0} \biggr)^2 \biggr] \, .
\end{eqnarray}
The squares in the exponential can be linearized by Gaussian transforms similar to (\ref{Gausstransform}). The resulting expression initially involves Gaussian integrals over many variables $z^{(i)}_{a_K \cdots a_i}$, but they can be evaluated iteratively  in clusters starting from the smallest cluster and integrating up to larger and larger distance scales (i.e.\ larger $i$ / larger $m_i$ / smaller $q_i$). The result of this little exercise is conveniently and suggestively expressed in terms of convolutions with the Green's function of the heat equation,
\begin{equation}
 G_q(z) \equiv \frac{1}{\sqrt{2 \pi q}} \exp\bigr[ -\frac{z^2}{2q} \bigl] ,
\end{equation} 
producing $\CZ$ as the outcome of the following recursion:
\begin{eqnarray}
 \CZ_0(z) &\equiv& \sum_s e^{-\beta z s} = 2 \cosh(\beta z) \, ,\\
 \CZ_{i+1}(z) &\equiv& \biggl(  \int dz' \, G_{|q_i-q_{i-1}|}\bigl(z-z' \bigr) \, \CZ_{i}(z') \biggr)^{k_i} \, , 
\qquad k_i = \frac{m_{i+1}}{m_i} \, ,
 \label{DRG} \\
 \CZ &=& \int dz \, G_{q_{K}}\bigl(z \bigr) \, \CZ_{K+1}(z) \, .
\end{eqnarray}
The powers $k_i$ arise from identical copies of Gaussian integrals, the copies corresponding to different values of the sub-index $a_i=1,\ldots,k_i$. The first step in the recursion is easy enough: $\CZ_1(z) = \left( e^{\frac{\beta^2}{2} (1-q_0)} (2 \cosh(\beta z)) \right)^{m_1}$, but past this point one has to resort to expansions or numerical evaluation. As a simple check of the above result, notice that when we remove the distinction between the clusters, i.e. $q_K = \cdots = q_1 = q_0$ (but keeping $q_{-1} \equiv 1$, $q_{K+1} \equiv 0$), we have $G_{|q_i-q_{i-1}|}(z-z') = \delta(z-z')$ for $1 \leq i \leq K$, so $\CZ= \int dz \, G_{q_0}(z) \left( e^{\frac{\beta^2}{2} (1-q_0)} (2 \cosh(\beta z)) \right)^{m_{K+1}}$. Recalling $m_{K+1}=n$, we see this correctly reproduces the replica symmetric formulae of section \ref{sec:RSsol}.

\begin{figure}
\begin{center}
\includegraphics[height=7cm]{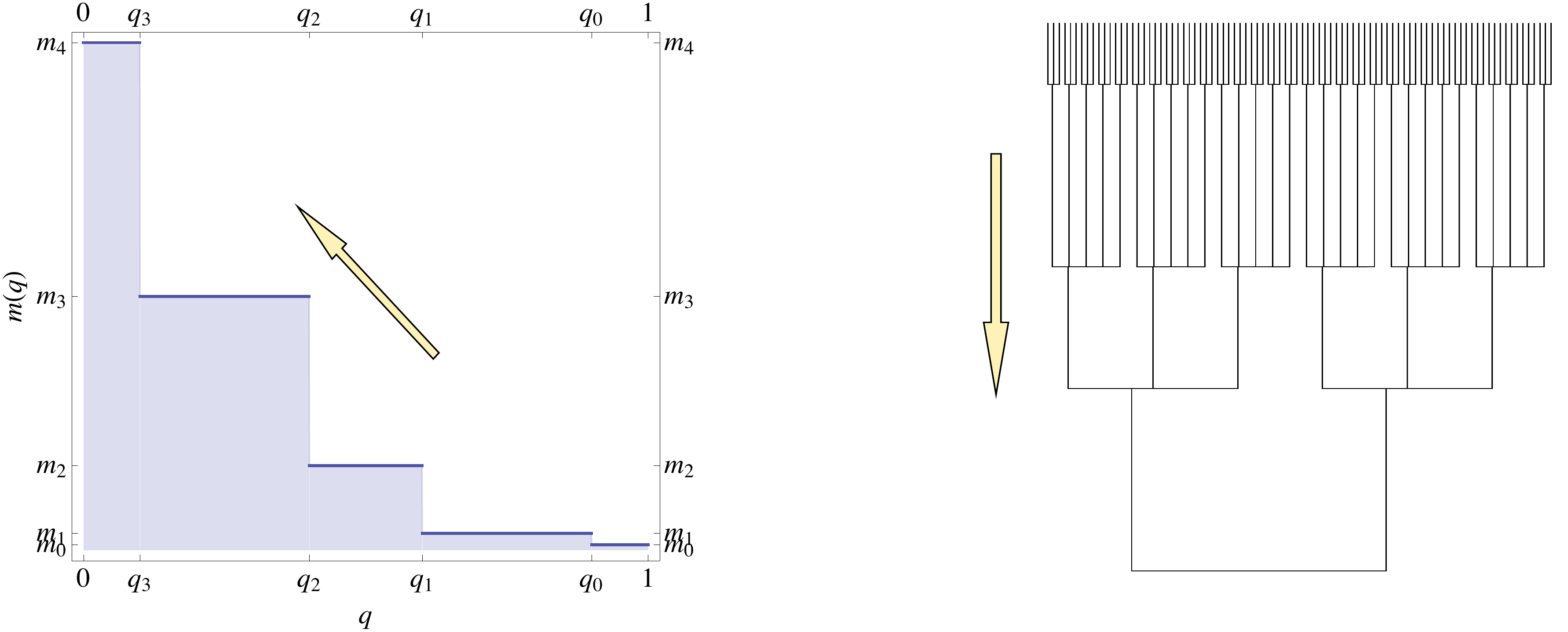}
\end{center}
\caption{\footnotesize The cluster size function $m(q)$ for the $K=3$ example of fig.\ \ref{ParisiMatrix}, and the corresponding tree. The arrow indicates the direction of the RG flow. 
 \label{mq}}
\end{figure}

The recursion (\ref{DRG}) can be thought of as an exact Wilsonian renormalization group action, evolving from the UV (small clusters) to the IR (large clusters), rescaling the number of degrees of freedom in jumps by factors $k_i=m_{i+1}/m_i$. We can make the equations more familiar looking by defining a function $\CZ(q,z)$ for all $q \in [0,1]$ such that $\CZ(q_i-\epsilon,z)=\CZ_{i+1}(z)$. The equations (\ref{DRG}) and the identification of $G$ as the Green's function of the heat equation show that we can take this function $\CZ(q,z)$ to be the single spin sum $2 \cosh(\beta z)$ at $q=1$, and then evolve it down in $q$ according to the heat equation $\partial_q \CZ = -\frac{1}{2} \partial_z^2 \CZ$, with a jump $\CZ \to \CZ^{k_i}$ whenever $q$ crosses a $q_i$. As the latter transformation is nonlinear, it is more convenient to work with $g(q,z)=\log \CZ(q,z)$, since on $g$ it acts linearly: $g \to k_i g$. (The price to pay for this is that the heat equation becomes nonlinear.) To implement this without reference to the indices $i$, we define a function $m(q)$ (illustrated in fig.\ \ref{mq}) as
\begin{equation} \label{mqdef}
 m(q) = \mbox{size of clusters with overlap $> q$} \, .
\end{equation}
So in particular $m(q_i)=m_i$ and $\partial_q m(q) = \sum_i (m_i-m_{i+1}) \, \delta(q-q_i)$. Then $g(q,z)=\log \CZ(q,z)$ is the solution to 
\begin{equation}
 \CD_q g = -\frac{1}{2} \biggl[ \left( \frac{\partial g}{\partial z} \right)^2 + \frac{\partial^2 g}{\partial z^2} \biggr] \, , \qquad
 \CD_q \equiv \frac{\partial}{\partial q} - \frac{1}{m} \frac{\partial m}{\partial q} \, ,
\end{equation}
evolving down from $q=1$ to $q=0$ with initial condition
\begin{equation}
 \lim_{q \nearrow 1} g(q,z) = \log\left(2 \cosh(\beta z)\right) \, ,
\end{equation}
and final identification
\begin{equation}
 \log \CZ = \lim_{q \searrow 0} g(q,0) \, .
\end{equation}
The ``gauge connection'' $A_q = (\partial_q m)/m$ keeps track of the increase in the number of degrees of freedom, while the nonlinear equation in $g$ is equivalent to the heat equation. 
The connection may be gauged away by redefining $f \equiv \frac{1}{m} g$, so $\partial_q f = \frac{1}{m} \CD_q g$, for which
\begin{equation}
 \frac{\partial f}{\partial q} = -\frac{1}{2} \biggl[ m(q) \biggl( \frac{\partial f}{\partial z} \biggr)^2 + \frac{\partial^2 f}{\partial z^2} \biggr] \,.
\end{equation}
We thus see that the iterative system is nothing but a particularly simple example of the exact renormalization group equations, with $q$ the analog of the scale and $f(q,z)$ the analog of the Wilsonian action.

For some purposes it is more convenient to take the scale variable to be some arbitrary auxiliary parameter $\lambda$, parametrizing the $(q,m)$ staircase  (fig.\ \ref{mq}) by \emph{continuous} functions $q(\lambda)$ and $m(\lambda)$. Letting $\lambda$ evolve from $\lambda=0$ to $\lambda=1$, with
\begin{equation} \label{bcond}
  (q,m)|_{\lambda=0}=(1,1) \, , \qquad (q,m)_{\lambda=1}=(0,n) \, ,
\end{equation} 
and denoting $\lambda$-derivatives by a dot, the RG equation becomes 
\begin{equation} \label{RGlambda}
 \dot{f} = -\frac{\dot{q}}{2}  \biggl[ m(\lambda) \biggl( \frac{\partial f}{\partial z} \biggr)^2 + \frac{\partial^2 f}{\partial z^2} \biggr] \, ,\qquad f(z)|_{\lambda=0} = \log\bigl(2 \cosh(\beta z) \bigr) \, .
\end{equation}
Obviously this introduces a reparametrization gauge symmetry. Putting everything together, we get 
\begin{equation} \label{CFlambda}
 \frac{1}{n} \CF[q,m] = \frac{\beta}{4} \biggl( 1 + \int_{\lambda=0}^1 dm \, q^2 \biggr) - \frac{1}{\beta} f(z=0)|_{\lambda=1} \, .
\end{equation}
The remaining tasks are $(i)$ make sense of the analytic continuation to $n=0$ and $(ii)$ extremize the functional $\CF[q,m]$.

\subsubsection{$n \to 0$ with RSB}

\label{sec:rsbn0}

\begin{figure}
\begin{center}
\includegraphics[height=6cm]{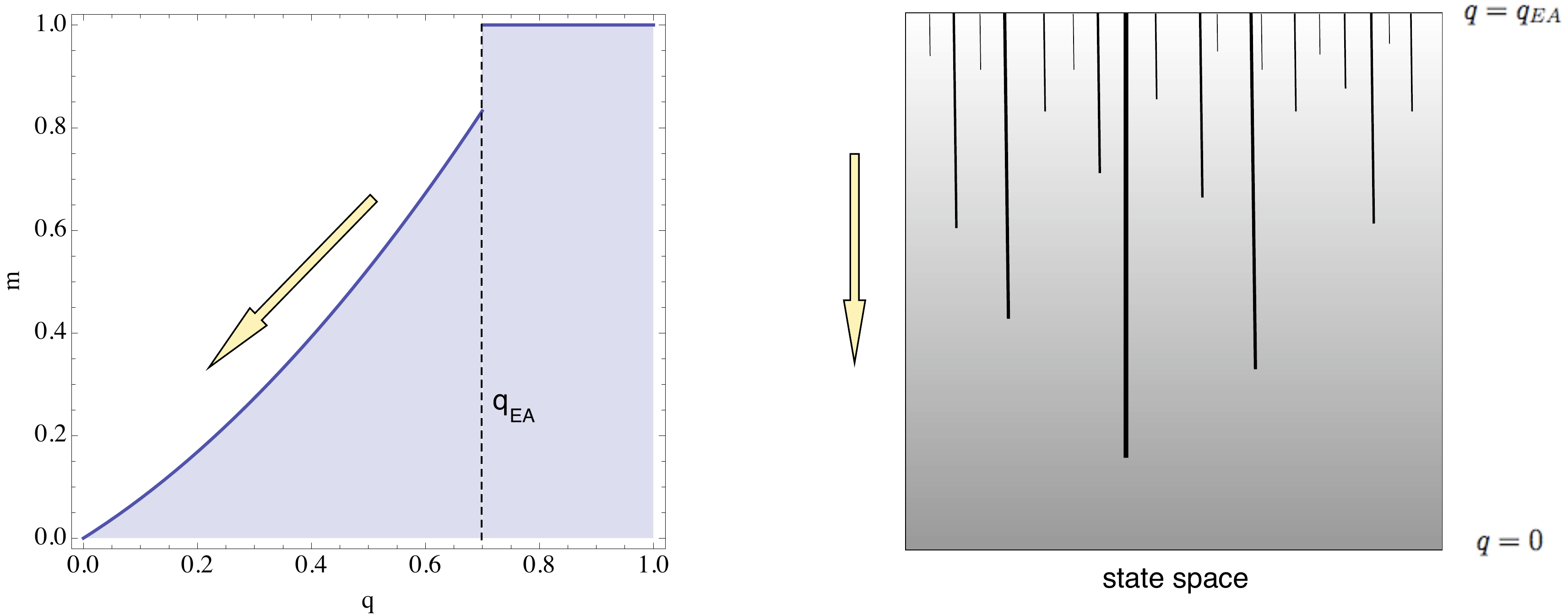}
\end{center}
\vskip-5mm
\caption{\footnotesize (Fictitious) example of $(m,q)$-line after continuation to $n=0$. The arrow again indicates the direction of the RG flow. The diagram on the right is a representation of the ultrametric state space tree for some choice of the disorder $J_{ij}$. The vertical lines are separating the branches of the tree. Different compartments at the top (the leaves) correspond to different pure states $\alpha$, and their width represents their probability weight $w_\alpha$. The value of $q$ where two branches separate sets their mutual overlap (so the taller the wall, the smaller the overlap). Thus, going up along the $q$-axis, $m(q)$ increases in discrete jumps. But different choices of $J_{ij}$ lead to different barrier structures, and averaging produces the smooth overlap distribution on the left. (The residual step up is due to the disorder independence of the self-overlap $q_{EA}$.)
\label{mq_n0}}
\end{figure}

Given that $1=m_0 < m_1 < m_2 < \cdots < m_{K} < m_{K+1} = n$ in the Parisi matrix construction, it is not obvious how to properly continue to $n=0<1$, to say the least. To get an idea of what could constitute a sensible continuation, consider the overlap distribution (\ref{Pqf2}), which in $K$-RSB becomes, taking into account the crucial overall sign of (\ref{Pqf2}),\footnote{To avoid having to duplicate everything, we restrict again to positive overlaps here, i.e.\ we condition the probabilities on $q \geq 0$.}
\begin{eqnarray} \label{Pqf2bis}
 \overline{P(q)} &=& \lim_{n \to 0} \sum_{i=0}^{K} (m_{i} - m_{i+1}) \, \delta(q-q_i) = \lim_{n \to 0} \partial_q m(q) \, .
\end{eqnarray}
For this to make sense as a probability density we need
$1=m_0>m_1 > m_2 > \cdots > m_{K+1} = n = 0$, i.e.\ the above cluster size inequalities must be inverted, and the $m_i$ obviously can no longer be integers; we will allow them to be arbitrary real numbers. Equivalently, $m(q)$ must become an \emph{increasing} function from the unit interval to the unit interval, to make $\partial_q m$ a proper probability density. This also inverts the interpretation of $m(q)$ from (\ref{mqdef}) to 
\begin{equation}
 \lim_{n \to 0} m(q) = \mbox{probability of finding an overlap $\leq q$} \, .
\end{equation}
In general the function $m(q)$ may have continuously increasing parts, which can be thought of as the $K \to \infty$ limit of the discretized construction. This is illustrated in fig.\ \ref{mq_n0}. Recalling the discussion of section \ref{sec:overlap}, getting a smooth function should indeed not surprise us: even if individual realizations of the disorder $J_{ij}$ produce a discrete overlap structure like the one depicted on the right of the figure, if this structure itself is sensitive to the disorder, averaging over the $J_{ij}$ will smooth out the discreteness.

This suggests what we should aim for. Looking back at the equations (\ref{bcond})-(\ref{CFlambda}), we see that in fact, the required structure is very naturally obtained by simply taking $n\to 0$ in (\ref{bcond}) while keeping (\ref{RGlambda}) and (\ref{CFlambda}) the same. Indeed, $m(\lambda)$ becomes now a decreasing function while $q(\lambda)$ remains decreasing. Everything else stays the same. In particular the intrinsic ultrametric structure is preserved. Notice that now the quadratic terms $\int_{\lambda=0}^1 dm \, q^2$ in (\ref{CFlambda}) become negative definite, with $q=0$ being the maximum. This is not an indication of thermodynamic instability, but rather the converse, as mentioned already in footnote \ref{invertedrule}. In the limit $n \to 0$, thermodynamically stable states correspond to maxima of the free energy $\CF(Q)$ instead of minima. 

Admittedly, the above is not a particularly solid justification for the proposed continuation to $n=0$. 
One can probably do better, but I will not try to do this here.\footnote{The continuation procedure is somewhat reminiscent of continuation in the $p$-adic numbers: A $p$-adic integral over a region with $p$-adic norm between 1 and $n$ is a finite sum with $p^n$ terms, and it naturally continues to a $p$-adic integral over a region with $p$-adic norm between $n=0$ and $1$, which is an infinite series. Similarities with $p$-adic and adelic cumbers were explored in chapter 9 of \cite{Cartier1992} and \cite{VAAvetisovetal1999,Parisi2000a}. However to reproduce the above one still needs to do a formal continuation $p \to 1$.} 

\subsubsection{The Parisi solution}

To summarize, the Parisi solution\footnote{To compare to \cite{Parisi1980}: $x$, $h$ there equals $m$ resp.\ $z$ here, and the apparent extra terms there are generated by the initial flow from $q=1$ to $q=q_{EA}$ during which $m(q)=1$, which is included here but not there. This part of the flow is easy to solve because it is basically exactly the heat equation: $f(q,z)=\frac{\beta^2}{2}(1-q) + \log(2 \cosh(\beta z))$, as is easily checked.} of the SK model is (going back to the gauge $\lambda=1-q$), in all its glory:
\begin{equation}
 F(\beta) = \max_{m(q)} \left[ \frac{\beta}{4} \biggl( 1 - \int_{q=0}^{1} dm \, q^2 \biggr) - \frac{1}{\beta} \, f_m(z=0)|_{q=0}  \right]  \, ,
\end{equation}
where $m(q)$ is an non-decreasing function on the unit interval satisfying the boundary conditions 
\begin{equation} \label{bcond2}
  m(0)=0, \qquad m(1)=1 \, ,
\end{equation} 
and $f_m(q,z)$ is the solution to the following flow equation running from $q=1$ to $q=0$: 
\begin{equation} \label{RGlambda2}
 \frac{\partial f}{\partial q} = -\frac{1}{2}  \biggl[ m(q) \biggl( \frac{\partial f}{\partial z} \biggr)^2 + \frac{\partial^2 f}{\partial z^2} \biggr] \, ,\qquad f(z)|_{q=1} = \log\bigl(2 \cosh(\beta z) \bigr) \, .
\end{equation}
Although this may still look somewhat unwieldy, in practice a simple trial family of $m(q)$ functions (for example stepwise constant functions with just a few steps $K$) can be treated easily and already lead to excellent approximations of the exact result. For example \cite{Parisi1980}, for $K=1$ one gets for the ground state free energy $F=-0.765$ and entropy $S=-0.01$, already significantly better than the RS result (\ref{RSBFS}). For $K=2$ this improves further to $F=-0.764$, $S=-0.004$. In particular we see that the entropy converges rapidly to the exact $S=0$. Detailed computations at various temperatures can be found in \cite{Crisanti2002a}. A recent stability analysis of the low $T$ solution is given in \cite{Crisanti2011}.

\subsection{Beyond spin glasses}

This is as far as we will go in describing the explicit solution of the Sherrington-Kirkpatrick model. We have left many topics untouched, in particular the various other ways in which the model can be solved (for which we refer to \cite{Mezard1987new,DeDominicis2006new} as a starting point). We hope we have made it clear that the hierarchical organization of the state space was the central structure in the story. This appears to be typical for disordered mean field (i.e.\ nonlocal) models. Other popular examples of such models include the Ising and spherical $p$-spin models \cite{Derrida1980,Gross1984a,Crisanti1992a}. The latter has continuous variables $s_i$ constrained to lie on an $N$-dimensional sphere $\sum_i s_i^2 = N$, with a Hamiltonian $H = \sum_{i_1 \cdots i_p} J_{i_1 \cdots i_p} s_{i_1} \cdots s_{i_p}$, where the $J$ are random couplings. It is known \cite{Crisanti1992a} that when $p=2$, the replica symmetric solution is exact, while for $p \geq 3$, the 1RSB solution is exact. In the limit $p \to \infty$, the model becomes equivalent to the ``random energy model'' \cite{Derrida1980} and can be solved exactly and explicitly  \cite{Gross1984a}. One can interpolate between spherical spin models and Ising spin models such as SK by switching on an additional non-random potential $V = \sum_i \lambda (s_i^2 - 1)^2$, where $p=2$, $\lambda \to \infty$ reproduces the SK model. Various generalizations to higher spins, $SU(N)$ spins, rotors, quantum spins etc have been studied too, for example in \cite{Binder1986,Bray1980,Ye1993,Cugliandolo2001}.

As mentioned in section \ref{sec:spinglassmodels}, such mean field models are not expected to be a good approximations for physical, short range spin glasses in spaces with a moderate number of dimensions (including three), and it is not clear to what extent the hierarchical organization of the mean field models is a useful zeroth order expansion point to understand the physics of such systems. But there are other systems of interest, in physics and beyond, where nonlocal interactions occur naturally, and where the conceptual framework developed for the mean field theory of spin glasses has turned out to be very useful. This includes the theory of neural networks, either as models for the brain or for various forms of machine learning, like the way Google figures out all the time what you really mean. One of the oldest and most famous such models is the Hopfield model \cite{Hopfield1982}, which is effectively described by a Hamiltonian $H = \sum_{ij} J_{ij} s_i s_j$, where the couplings $J_{ij}$ encode synaptic strengths and $s_i = \pm 1$ encodes the on/off firing state of the $i$-th neuron. Stable firing states of the neurons correspond to minima of $H$. $N$-component patterns $\xi_i$ with $\xi_i = \pm 1$ are stored according to a learning rule $J_{ij} \to J_{ij} - \xi_i \xi_j$ (or variants thereof), which energetically favors the stability of neuron states $s_i = \xi_i$. After $M$ patterns are learned, the couplings are thus $J_{ij} = J_{ij}^{(0)} - \sum_{\mu=1}^M \xi^\mu_i \xi^\mu_j$. However, as is well known, learning does not equal remembering. Memory retrieval dynamics in this context is modeled by a standard Monte Carlo type relaxation process, which tries to minimize the energy. Remembering a learned pattern $\xi^\mu$ then means that the change in coupling effectively has given rise to a stable minimum $s_\star$ at or very near the pattern $\xi^\mu$. Clearly this model shares many characteristics with mean field spin glass models, and mean field spin glass theory has indeed been very useful in the analysis of these models. For example in \cite{Amit1985} it was shown that with the above learning rule, the memory undergoes a phase transition when it tries to learn too much: for $M < M_c$ with $M_c \approx 1.4 \, N$ it remembers practically everything, while for $M > M_c$ it gets completely confused and remembers practically nothing.\footnote{It turns out that the key to retain the ability to remember new things is forgetting old things {\cite{Parisi1986}}.} In this framework, the phase in which learned patterns are remembered is similar to the spin glass phase (many highly metastable states), hierarchical clustering in state space gets interpreted as categorization and organization of memories in concepts, and so on. A collection of some of the original papers along these lines can be found in \cite{Mezard1987new}. 

Another important\footnote{As measured by the $2 \times 10^4$ \href{http://scholar.google.com/scholar?cites=4771121466026992245&as_sdt=5,33&sciodt=0,33&hl=en}{citations} of e.g.\ {\cite{Kirkpatrick1984}}.} branch of human activity where the ideas of spin glass theory have found fruitful applications is optimization algorithms for NP-hard problems. Canonical examples of such problems are the satisfiability problem and the traveling salesman problem, which find incarnations in practical problems going from optimal chip design to the problem of finding string vacua satisfying certain conditions \cite{Denef2007new}. For an early review we refer again to \cite{Mezard1987new}. Recent developments include for example the relation of clustering phase transitions to effective hardness of optimization problem instances and ways to exploit these insights in devising new efficient algorithms \cite{Mezard2005,Mezard2002}.

\subsection{Beyond spins}   \label{sec:beyondspins}

The statistical mechanics community is fond of Ising spins, and for a good reason: there is a lot in the real world that can be naturally modeled by large collections of interacting binary degrees of freedom. String theorists on the other hand face the sad reality that the complex systems that are natural in their world are rarely as simple. 
On the other hand, Parisi's idea of using the overlap distribution to characterize the emergence of different pure states / ergodic components / superselection sectors in an intrinsic, quantitative and computable way is an attractive one. Thus, when pondering for example the state space structure of complex D-brane systems \cite{AABDG} or de Sitter space \cite{ADnew}, one is led to the question what the analog is of the pure state spin overlap $q_{\alpha \beta} \equiv \frac{1}{N} \sum_i \langle s_i \rangle_\alpha \langle s_i \rangle_\beta$ defined in (\ref{defoverlap}) for general quantum systems. For the more general systems of interest the degrees of freedom could be fluxes, scalars, fermions, KK modes, or whatever. Since typically there is no canonical scalar product between such degrees of freedom, a naive direct generalization of the overlap will not work. 

A simple construction that is fully quantum and makes sense in general, and that is equivalent to the standard spin overlap in the Ising spin case, is the following. Since we want to consider systems for which we can at least formally take a thermodynamic limit, we assume we can randomly sample\footnote{Sampling is not necessary for the construction, but it spares us the trouble of having to know and work with the complete set of degrees of freedom, which may lead to UV divergences or may not even be known. Moreover the overlap $q_{\alpha\beta}$ and corresponding spin glass order parameter $P(q)$ could equally well have been defined by sampling say every other spin instead of all spins; ultrametricity implies that the results would have been identical. In general this need not be true of course, so the notion of overlap may depend on the choice of degrees of freedom that are sampled.} $N$ independent ``representative'' degrees of freedom $x_i$, $i = 1,\ldots,N$ of the system, with the thermodynamic limit of infinite system size corresponding to $N \to \infty$. For example consider a scalar field theory living in a large box of volume $V$. We can subdivide the large box into smaller identical boxes of some fixed volume $v$, and then uniformly sample at some fixed sampling density a subset of $N$ of the smaller boxes, which we label by $i=1,\ldots,N$. Finally, in each of the sampled boxes $i$, we compute the average value of the scalar field, and we call this $x_i$. Another example would be a $U(M)$ matrix quantum mechanics describing a wrapped D-brane system. Here we may sample a uniform subset of $N$ of the $M^2$ matrix degrees of freedom. A final example is again an Ising spin glass, for which we sample $N$ spins. 

The state of the complete system is assumed to be described by some density matrix $\rho$. Expectation values are computed as usual as $\langle A \rangle_\rho \equiv {\rm Tr} \, (\rho A)$. For each degree of freedom $i$ there is a corresponding reduced density matrix $\rho^{i}$ obtained by tracing out all other degrees of freedom, so
\begin{equation}
 {\rm Tr}_{\CH_i} \left( \rho^{i} A_i \right) = {\rm Tr} \left( \rho A_i \right) \, .
\end{equation}
Here $\CH_i$ is the Hilbert space associated to the degree of freedom $x_i$ (for the scalar field example this would be $\CH_i = \CL^2(\IR)$, whereas for (quantum) Ising spins it is $\CH_i = \IC^2$), and $A_i$ is an observable referring to $\CH_i$ alone (so for the scalar this could for example be $A_i = x_i^2$, or $A_i = -i \partial_{x^i}$). 

A natural definition of overlap between two states $\rho_\alpha$, $\rho_\beta$ is then
\begin{equation}
 Q_{\alpha \beta} \equiv \frac{1}{N} \sum_{i=1}^N {\rm Tr}_{\CH_i} \, \rho_\alpha^{i} \rho_\beta^{i} \, .
\end{equation}
When the system consists of Ising spins, this reduces to the standard magnetization (or spin) overlap, up to a rescaling and a shift. For indeed, the general density matrix for a spin 1/2 degree of freedom can be written as 
$\rho = \frac{1}{2}\left({\bf 1} + \vec{m} \cdot \vec{s}\right)$, 
where $\vec{m} = \langle \vec{s} \rangle_\rho$ is the magnetization, whence $Q_{\alpha \beta}=\frac{1}{N} \sum_i \frac{1}{2}(1+\vec{m}_\alpha^i \cdot \vec{m}_\beta^i)$.\footnote{This is valid for quantum Ising spins. Classical Ising spins are trivially obtained from this by restricting to the up/down eigenstates of $s_z$, yielding $Q_{\alpha \beta} = \frac{1}{2} \left(1 + q_{\alpha\beta} \right)$ with $q_{\alpha \beta}$ as defined in (\ref{defoverlap}).} For continuous degrees of freedom this can also be written in terms of the associated Wigner densities on phase space: 
\begin{equation}
 Q_{\alpha \beta} = \frac{1}{N} \sum_{i=1}^N 2 \pi \hbar \int dr dp \, W_\alpha^i(r,p) \, W_\beta^i(r,p) \, ,
\end{equation}
where $W_\alpha^i(r,p) \equiv \frac{1}{2 \pi \hbar} \int ds \, e^{i p s/\hbar} \langle r - s/2 |\rho_\alpha^i| r + s/2 \rangle $. In the classical limit this becomes (after suitable coarse graining to kill off highly oscillatory modes) the overlap of ordinary phase space probability densities, and we can think of $Q_{\alpha\beta}$ as the fraction of degrees of freedom $(x^i,p_i)$ found in the same elementary phase space cell (of size $2 \pi \hbar$) when randomly sampling from two states $\alpha$ and $\beta$. 

We can mimic (\ref{defoverlap}) even more closely, by writing the reduced density matrix $\rho^i$ as the expectation value $\langle E^i \rangle$ of the operator $E^i$ defined by 
\begin{equation}
 E^i_{rs} \, \equiv \, \CP_{x_i=r}^\dagger \CP_{x_i=s} \, = \, | x_i=r \rangle \langle x_i=s| \, ,
\end{equation}
where $\CP_{x_i=r}$ is the projection operator on the eigenspace $x_i=r$. Then we get
\begin{equation}
 Q_{\alpha \beta} = \frac{1}{N} \sum_{i=1}^N \sum_{rs} \langle E_{rs}^i \rangle_\alpha \langle E_{sr}^i \rangle_\beta  \, .
\end{equation} 

Finally, for some states $\rho$, an analog of the clustering property (\ref{clusterproperty}) will hold, for others not. We define this analog to be the property that correlation functions of ``local'' observables factorize in the thermodynamic limit for almost all evaluation points. That is, for any finite $r$
\begin{equation}
 \lim_{N \to \infty} \frac{1}{N^r} \sum_{i_1,\cdots,i_r} \biggl| \langle A_{i_1} B_{i_2} \cdots C_{i_r} \rangle_\rho -  \langle A_{i_1} \rangle_\rho  \langle B_{i_2} \rangle_\rho \, \cdots \langle C_{i_r} \rangle_\rho \biggr| = 0 \, .
  \label{clusterproperty2}
\end{equation} 
By ``local'' we mean here observables $A_i$ referring to $\CH_i$ alone (we could relax this definition by allowing dependence on other, ``nearby'' degrees of freedom, but this will be good enough for our purposes). 

With these definitions set up, we can proceed in complete parallel with the Ising spin glass case, and define ``pure states'' $\rho_\alpha$, decompositions $\rho=w_\alpha \rho_\alpha$ of non-pure states (such as $\rho = e^{-\beta H}$ for glassy systems at low temperatures), and their overlap distributions $P(q)=\sum_{\alpha\beta} w_\alpha w_\beta \delta(q-Q_{\alpha\beta})$. Whenever $\rho$ fails to satisfy cluster decomposition (signaling a nontrivial phase structure), $P(q)$ becomes nontrivial. For example for an isotropic quantum ferromagnet, the low temperature pure states will be labeled by a vector $\vec{m}$ such that for all $i$, $\rho^i_{\vec{m}}=\frac{1}{2}(1+\vec{m} \cdot \vec{s})$, and the overlap distribution $P(q)$ is uniform on the interval $[\frac{1-m^2}{2},\frac{1+m^2}{2}]$.

Moreover $P(q)$ will again be computable in principle for any known state $\rho$, even if we do not know how to describe its decomposition into pure states. Explicitly, the analog of (\ref{Pqrep}) that allows this is
\begin{equation} 
 P(q) = \bigl\langle \delta \left( q - \mbox{$\frac{1}{N} \sum_{irs} E_{1rs}^i E_{2sr}^i$} \right) \bigr\rangle_{n=2} \, ,
\end{equation}
where again the subscript `$n=2$' means we consider two replicas of the system, with density matrix $\rho \otimes \rho$, and $E_1$ depends on the first replica's degrees of freedom while $E_2$ depends on the second replica's degrees of freedom. In the classical limit, $P(q)$ becomes the probability distribution for finding a fraction $q$ of the degrees of freedom $(x^i,p_i)$ of the two replicas in the same elementary phase space cell.

The operator inside the delta-function has a more concrete interpretation. For a given $i$, let
\begin{equation}
 X_{12}^i \equiv \sum_{rs} E^i_{1rs} E^i_{2sr} = \sum_{rs} | x^i_1 = r, x^i_2 = s \rangle \langle x^i_1=s, x^i_2 =r | \, .
\end{equation}
Notice the $r \leftrightarrow s$ flip on the right hand side, which we can equivalently think of as a $x^i_1 \leftrightarrow x^i_2$ flip. Thus, what this operator $X_{12}^i$ does for a given $i$ (when acting on a wave function say) is exchange the replica variables $x^i_1$, $x^i_2$. 
The first moment of $P(q)$ is then
\begin{equation}
 \langle q \rangle = \frac{1}{N} \sum_i {\rm Tr}(\rho_1  X^i_{12} \, \rho_2) = 
 \frac{1}{N} \sum_i {\rm Tr}_{\CH_i} (\rho^i)^2 = \frac{1}{N} \sum_i e^{-S^i(2)} \, ,
\end{equation} 
where $S^i(2)$ is the second Renyi entropy \cite{Renyi1961} of $\rho^i$. If $\rho=|\Psi\rangle \langle \Psi|$, this gives information on how entangled the $x_i$ are in the wave function $\Psi$; if $\Psi$ factorizes (no entanglement), then $\langle q \rangle = 1$. For the isotropic ferromagnet (strong entanglement), we have $\langle q \rangle = \frac{1}{2}$. The second moment is
\begin{equation} 
 \langle q^2 \rangle = \frac{1}{N^2} \sum_{ij} {\rm Tr}(\rho_1  X^i_{12} X^j_{12} \, \rho_2) = 
 \frac{1}{N^2} \sum_{i \neq j} {\rm Tr}_{\CH_{ij}} (\rho^{ij})^2 + \frac{1}{N} = \frac{1}{N^2} \sum_{i \neq j} e^{-S^{ij}(2)} \, ,
\end{equation}
where $\rho^{ij}$ is the density matrix obtained by tracing out all degrees of freedom except $x_i$, $x_j$, and in the last expression we dropped the $1/N$ term since it is understood that the thermodynamic limit $N \to \infty$ is taken. The ratio $\langle q^2 \rangle/\langle q \rangle^2$ gives information about the degree of long range correlation (and failure of cluster decomposition) in the system. For a thermodynamic pure state it equals 1, for the isotropic ferromagnet it equals $1+\frac{m^4}{3}$. Moments of order $k$ are obtained analogously by fixing $k$ degrees of freedom.
 


Thus, for rather general systems, we have arrived at an infinite set of in principle computable order parameters that detect the emergence of a nontrivial state space landscape. Applications are discussed in \cite{ADnew,AABDG}.

\section{Supersymmetric quantum mechanics} \label{sec:susyQM}

\subsection{Introduction}

Complex systems are complex. As a result, it is hard in general to quantitatively determine even their basic features. When it can be done, such as for the Sherrington-Kirkpatrick model, the results often have a striking richness. Complex systems arising in string theory, such as the geometrically highly complex wrapped D-branes reviewed in section \ref{sec:Dbranes}, are often significantly more intricate than simple spin models. This makes straightforward analysis more daunting. Although in principle the probability overlaps defined in section \ref{sec:beyondspins} define an intrinsic order parameter that could be used to analyze those stringy systems along the lines of the SK model, in practice this may still be difficult. This is even true for much simpler quantities like the entropy. 

On the other hand, the existence of dualities in string theory and the presence of supersymmetry often tremendously simplify things. In this section we will focus on supersymmetry. The main simplification that arises due to this is that many quantities of interest can be computed exactly in a semiclassical limit. The primary example of this is the Witten index \cite{Witten1982}, which counts ground state degeneracies in supersymmetric quantum mechanics, and is generically invariant under continuous deformations of the theory. Famously this allowed Strominger and Vafa \cite{STROMINGER1996} to microscopically compute the entropy of certain classes of extremal black holes arising in string theory, by tuning the string coupling constant from the strongly self-gravitating black hole regime to a weakly coupled, non-self-gravitating regime. In this regime the system has a well-understood D-brane description, and the Witten index can be computed in a relatively straightforward and precise way.

The ground state degeneracy is not the only thing supersymmetry allows us to control. Through its relation with Morse theory, the structure of supersymmetric quantum mechanics gives useful insights into the saddle point structure of energy landscapes and the way they are dynamically connected, and makes it possible to systematically compute nonperturbative level splittings and their associated exponentially slow relaxation rates. Clearly this is of potential interest in the study of complex systems. In fact, through a formal map between nonsupersymmetric classical statistical mechanics and supersymmetric quantum mechanics, this has ramifications beyond the realm of supersymmetric systems (see section \ref{sec:susystochastic}).

We will review these fundamental aspects of supersymmetric quantum mechanics in what follows. I will not assume any familiarity with the topic, and start from scratch, building up the topic along the lines of 
chapter 10 of \cite{Hori2003}, an excellent introduction to supersymmetric quantum mechanics and its interplay with geometry. A review focusing more on relations to exactly solvable systems is 
\cite{Cooper1995}, and finally Witten's original \cite{Witten1982} is a classic. 

Specific applications to string theory are left to section \ref{sec:Dbranes}.

\subsection{Definition}

By definition, the Hamiltonian of a supersymmetric quantum mechanical system can be written as
\begin{equation}
 H = \frac{1}{2} \{ Q, Q^\dagger \} \, ,
\end{equation}
where $Q$, the supercharge, is an operator satisfying
\begin{equation}
 Q^2 = \left( Q^\dagger \right)^2 = 0 \, .
\end{equation}
The Hilbert space is $\IZ_2$ graded, splitting up in a even or ``bosonic'' part, and an odd or ``fermionic'' part:\footnote{At this level ``bosonic'' and ``fermionic'' are just conventional names. They do not necessarily mean that the quantum mechanical states represent bosonic or fermionic particles in the usual sense of the word, although in specific situations the notion may be correlated.}
\begin{equation}
 \CH = \CH_B \oplus \CH_F \, .
\end{equation}
The operator $(-1)^F$ is defined as $+1$ on $\CH_B$ and $-1$ on $\CH_F$. The operator $Q$ is odd/fermionic, i.e.\ it maps from $\CH_B$ into $\CH_F$ and from $\CH_F$ into $\CH_B$:
\begin{equation}
 [(-1)^F,Q]=-Q \, .
\end{equation}
Some immediate consequences are
\begin{itemize}
 \item  $[(-1)^F,H]=0$, $[Q,H] = 0$, i.e.\ $(-1)^F$ and $Q$ are symmetries.
 \item $H \geq 0$, all energies are positive or zero.
 \item $H|\alpha\rangle = 0 \Leftrightarrow Q|\alpha\rangle = 0 = Q^\dagger |\alpha\rangle$, zero energy states are supersymmetric.
 \item For each \emph{positive} energy level $\CH_E = \CH_{E,B} \oplus \CH_{E,F}$, $E>0$, the fermionic and bosonic subspaces are \emph{isomorphic}. The isomorphism is given by the Hermitian supercharge
 \begin{equation}
   Q_1 \equiv Q+Q^\dagger \, ,
 \end{equation}
 which satisfies $Q_1^2 = 2 E$ on $\CH_E$ and is therefore invertible when $E \neq 0$.
\end{itemize}
The latter property is particularly important. It means that all positive energy eigenstates come in boson-fermion pairs. This is not necessarily true for zero energy ground states. The Witten index $\Omega$ quantifies the degree to which this is not true:
\begin{equation} \label{Omdef}
 \Omega \equiv \dim \CH_{0,B} - \dim \CH_{0,F} \, .
\end{equation}
Since all positive energy states come in even-odd pairs, this can equally well be written in the following alternative ways:
\begin{eqnarray}
 \Omega &=& {\rm Tr} \, (-1)^F \\
 &=& {\rm Tr} \, (-1)^F e^{-\beta H} \\
 &=& \int \CD(\cdots) \, e^{-S(\cdots)} \, .
\end{eqnarray}
The last line represents the usual Euclidean path integral representation of the partition function,  except that now \emph{both} fermionic and bosonic variables satisfy periodic boundary conditions, instead of the usual antiperiodic boundary conditions on the fermions. This is due to the insertion of $(-1)^F$, which flips the sign of the fermionic variables. (The example below may make this more clear.)

The most important property of the Witten index is that it is invariant under generic continuous deformations. This is clear from its definition: although the \emph{total} number of supersymmetric ground states may vary when we vary the parameters of the model, as states move in and out of the $E=0$ level, all such arrivals and departures must come in bose-fermi pairs, and therefore the index will not be affected.\footnote{Although generically robust, there are important exceptions to this argument. In particular, when the gap we have implicitly assumed between the zero and first excited energy levels vanishes in the course of the deformation, the Witten index may jump. This can happen for example when the deformation passes through a degeneration allowing a ground state wave function to spread out all the way to infinite distance, becoming non-normalizable and therefore no longer a proper state in the Hilbert space, but rather part of the continuum of scattering states.}

\subsection{One dimensional example}

\begin{figure}
\begin{center}
\includegraphics[height=6.5cm]{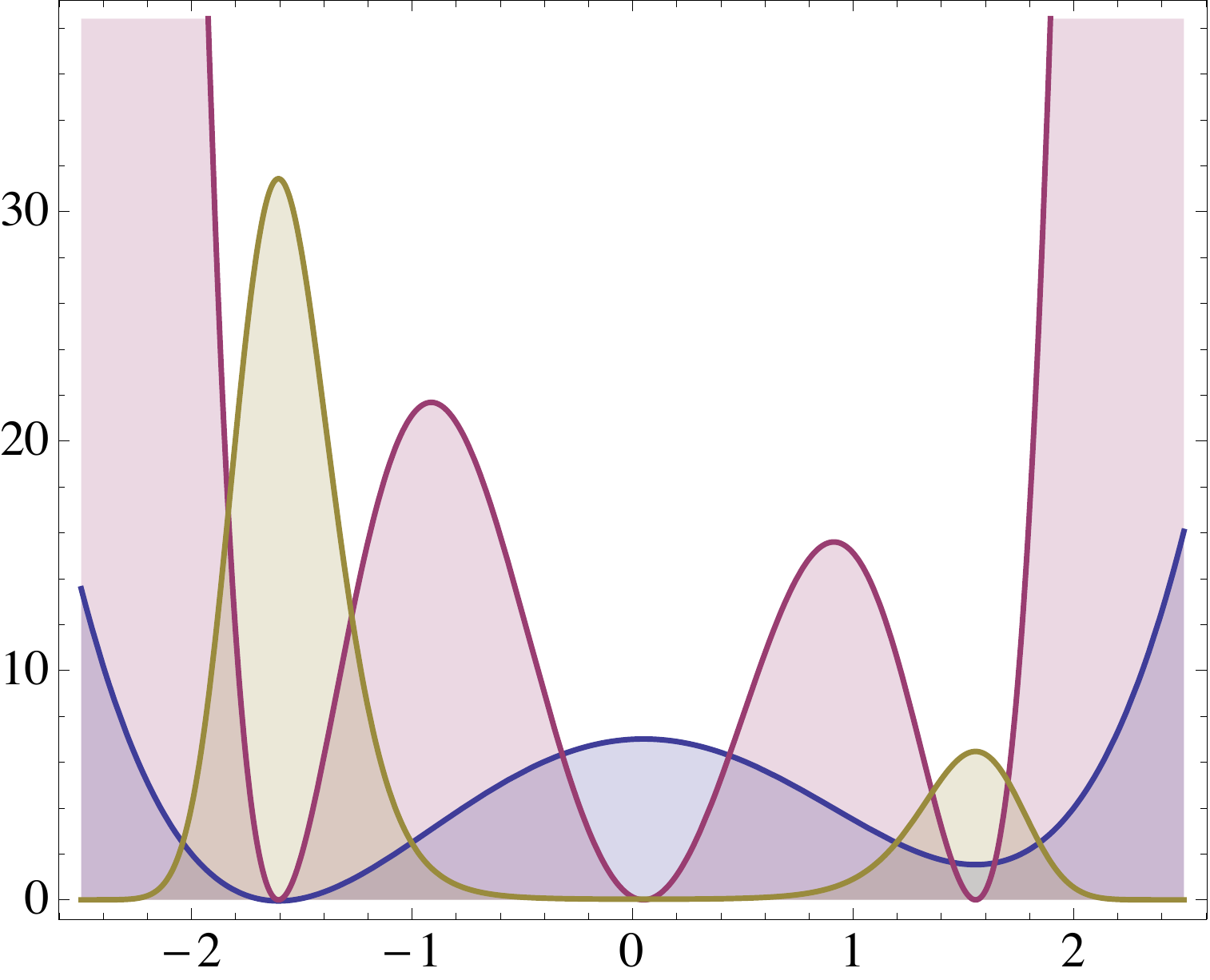}
\end{center}
\caption{\footnotesize 1d susy QM with $h(x)=x^4 - 5 \, x^2 + 7$. Blue: $h(x)$, red: $V(x)=h'(x)^2/2$, green: $\Phi_0(x) \propto e^{-h(x)}$.
\label{susyQM}}
\end{figure}

A simple example is obtained by considering a bosonic (i.e.\ commuting) variable $x$ and a fermionic (i.e.\ anticommuting) variable $\bar{\psi}$, wave functions 
\begin{equation} \label{wavefunction}
 \Phi(x,\bar{\psi}) = \Phi_B(x) + \Phi_F(x) \bar{\psi} \, , 
\end{equation}
and canonical commutation relations $[x,p]=i$, $\{\bar{\psi},\psi\}=1$, where $\psi \equiv \bar{\psi}^\dagger$. So acting on wave functions we have
\begin{equation}
 p = - i \partial_x \, , \qquad \psi = \partial_{\bar{\psi}} \, .
\end{equation}
Then we consider the supercharge
\begin{equation}
 Q \equiv \bar{\psi} \left(ip + h'(x) \right) \, , \qquad 
 Q^\dagger = \psi \left(-ip + h'(x) \right) \, .
\end{equation}
Here $h(x)$ is an arbitrary function, sometimes called the superpotential.
Expressed in components $\Phi={ \Phi_B \choose \Phi_F }$ this becomes
\begin{equation}
 Q = \begin{pmatrix}
  0 & 0 \\ \partial_x + h'(x) & 0 
 \end{pmatrix} \, ,
\end{equation}
and the Hamiltonian
\begin{eqnarray}
 H &=& \frac{1}{2} \{ Q,Q^\dagger \} \nonumber \\
 &=& \frac{1}{2}\left(-\partial_x^2 + h'(x)^2 + h''(x) [\bar{\psi},\psi] \right) \\
 &=& \frac{1}{2}
 \begin{pmatrix}
  -\partial_x^2 + h'(x)^2 - h''(x) & 0 \\ 0 & \partial_x^2 + h'(x)^2 + h''(x) 
 \end{pmatrix} \, . \nonumber
\end{eqnarray}
We can think of this as the Hamiltonian of a spin 1/2 particle of unit mass in a potential $V(x)=\frac{1}{2} h'(x)^2$ and a magnetic field $\frac{1}{2} h''(x)$. In the case $h(x)=\frac{\omega}{2} x^2$, this becomes the harmonic oscillator potential $V(x)=\frac{1}{2} \omega^2 x^2$ plus a constant magnetic field $\omega$. Assuming $\omega>0$, the bosonic (i.e.\ $\Phi_F=0$) energy levels are then $E_B=0,\omega,2\omega,3\omega,\ldots$ and the fermionic energy levels $E_F=\omega,2\omega,3\omega,\ldots$. We explicitly see the pairing of $E>0$ states and the mismatch at $E=0$, captured by the Witten index $\Omega=+1$. When on the other hand $\omega<0$, we get $E_B=|\omega|,2|\omega|,\ldots$ and $E_F=0,|\omega|,2|\omega|,\ldots$, so $\Omega=-1$.

More generally, it is easy to explicitly construct solutions to $Q \Phi =Q^\dagger \Phi = 0$:
\begin{equation}
 \Phi = A_B \, e^{-h(x)} + A_F \, e^{+h(x)} \, \bar{\psi} \, . \footnote{Here $\bar{\psi}$ is, as in (\ref{wavefunction}), a Grassman variable. Equivalently we could write $\Phi = A_B \, e^{-h(x)} |0\rangle + A_F \, e^{+h(x)} \, \bar{\psi} |0\rangle $, with $|0\rangle$ the ``fermionic vacuum state'', i.e.\ $\psi|0\rangle = 0$.}
\end{equation} 
For these to be true supersymmetric ground states, $\Phi$ has to be normalizable. When $\lim_{|x| \to \infty} h(x) = + \infty$, this requires $A_F=0$, and thus $\Omega=+1$. When $\lim_{|x| \to \infty} h(x)= - \infty$, we get instead $A_B=0$ and $\Omega=-1$. In all other cases, none of the solutions are normalizable, and $\Omega=0$.  An example is shown in fig.\ \ref{susyQM}.

\subsection{More degrees of freedom}

We can easily generalize this to multiple degrees of freedom $x^I$ and $\bar{\psi}^I$, $I=1,\ldots,N$, leading to $2^N$-component wave functions
\begin{equation}
 \Phi(\bar{\psi},x) = \sum_{b_1,\cdots,b_N = 0,1} \Phi_{b_1 \cdots b_N}(x) \, (\bar{\psi}^1)^{b_1} \cdots ({\bar{\psi}^N})^{b_N} \, ,
\end{equation}
and a fermion number quantum number $F$ counting the number of $\bar{\psi}^I$. We take the supercharge to be
\begin{equation}
 Q = \sum_I \bar{\psi}^I \left(i p_I + \partial_I h(x) \right) \, ,
\end{equation}
which satisfies $Q^2=0$. The Hamiltonian is then
\begin{equation}
 H = \sum_I \frac{1}{2} p_I^2 + \frac{1}{2}  (\partial_I h)^2 + \frac{1}{2} \sum_{IJ} [\bar{\psi}^I,\psi^J] \partial_I \partial_J h \, .
\end{equation}
It is now no longer possible in general to find explicit solutions to $Q \Phi=Q^\dagger \Phi = 0$, but when $h(x)$ is a nondegenerate quadratic function, say 
\begin{equation}
 h(x) = \frac{1}{2} \sum_{I=1}^N \omega_I (x^I)^2 \, ,
\end{equation}
then the problem just factorizes into $N$ copies of the 1d harmonic oscillator situation we solved above, and thus the unique normalizable supersymmetric ground state is
\begin{equation} \label{pertGS}
 \Phi_0(x) = e^{-\frac{1}{2} \sum_I |\omega_I| (x^I)^2} \, \prod_{I : \,\omega_I < 0} \bar{\psi}^I \, .
\end{equation}
This has fermion number $F$ equal to the number of negative eigenvalues of the Hessian of $h$ at the critical point $x=0$. This number $\mu$ is called the Morse index of the critical point. The Witten index is $\Omega=(-1)^\mu$.

Using this, we can compute the Witten index for any $h(x)$ that has only isolated, nondegenerate critical points.\footnote{Such a function is called a Morse function, and it corresponds to the generic case. We will always assume $h$ is a Morse functions in this section.} Indeed, since the Witten index is invariant under continuous deformations that do not change the asymptotics of $h(x)$, we can consider the deformation $h(x) \to \lambda h(x)$ and compute $\Omega$ in the limit $\lambda \to \infty$. This limit corresponds to a semiclassical limit. The classical supersymmetric ground states are given by the critical points of $h(x)$. The corresponding perturbative\footnote{More precisely, the perturbation theory for each critical point $x_\star$ of $h$ is a series in inverse powers of $\lambda$ and is obtained by changing variables from $x$ to $y$ defined by $x^I=x_\star^I + y^I/\sqrt{\lambda}$ so $H=\lambda (H_0 + \lambda^{-1/2} H_1 + \cdots)$ where $H_0$ is the harmonic oscillator part of the Hamiltonian, obtained by truncating $h$ at quadratic order in $y$, and $H_1$ the part corresponding to the higher order terms.} semiclassical wave functions are to lowest order Gaussians centered at the critical points, of the form (\ref{pertGS}). All of these remain supersymmetric to all orders in perturbation theory. (Roughly this is because perturbation theory around one critical point does not know about the presence of other critical points, and if there is only one critical point, the Witten index is 1.) However, nonperturbative tunneling effects may lift this degeneracy, just like tunneling splits up the doubly degenerate perturbative ground state energy level of a double well potential in ordinary quantum mechanics. But because all lifting must occur in bose-fermi pairs, it will not affect the total Witten index, so we can compute the index simply by adding the contributions of all critical points $x_\star$ of $h$:
\begin{equation}
 \Omega = \sum_{x_\star: \, h'(x_\star)=0} (-1)^{\mu(x_\star)} \, .
\end{equation}
Checking this for the 1d example shown in fig.\ \ref{susyQM}, we see that indeed there $h$ has two minima and one maximum, so $\Omega=+1+1-1=+1$, in agreement with our earlier result.

When $h$ is a harmonic function, as is the case when it is for example equal to the real part of a holomorphic function $W(z)$ of variables $z^a=x^{2a-1}+i x^{2a}$, $a=1,\cdots,n=N/2$, all critical points have Morse index $n$, and the Witten index is equal to $(-1)^n$ times the absolute number of critical points of $W$. This situation occurs naturally in the description of wrapped D-branes in Calabi-Yau compactifications of type II string theory.

\subsection{Curved spaces} \label{sec:curvedspaces}

\begin{figure}
\begin{center}
\includegraphics[height=5.5cm]{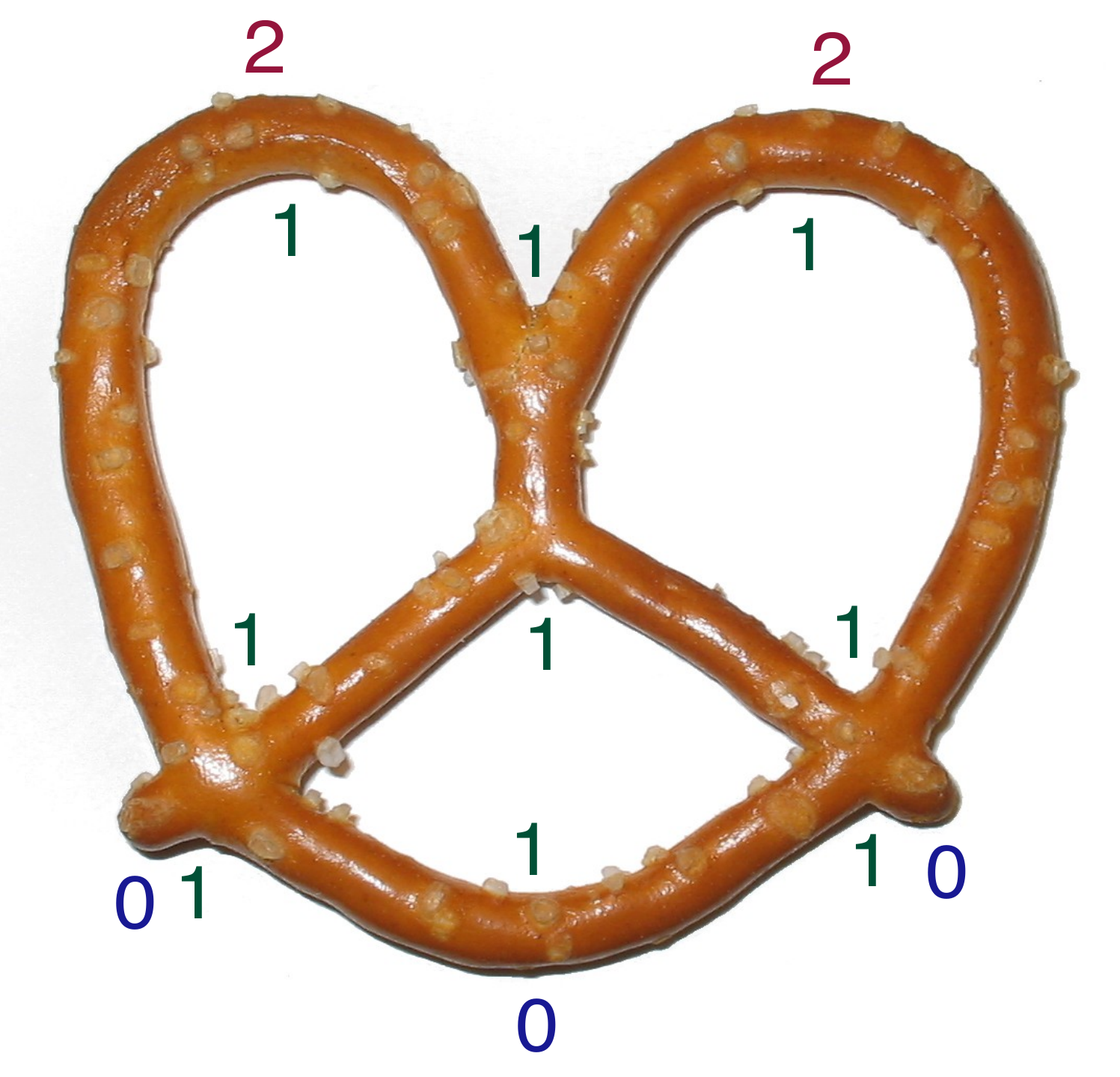}
\end{center}
\caption{\footnotesize Critical points of the vertical height function on a pretzel, with fermion number (Morse index) indicated. The Witten index is $\Omega = 3 - 9 + 2 = -4$. 
\label{pretzel}}
\end{figure}

On a compact $N$-dimensional Riemannian manifold $M$, we can naturally define a supersymmetric quantum mechanics with Hilbert space given by the space of complex valued differential forms $\Phi=\Phi_0 + \Phi_I dx^I + \Phi_{IJ} dx^I \wedge dx^J + \cdots$ on $M$. The elementary 1-forms $dx^I$ get identified with the Grassmann coordinates $\bar{\psi}^I$. The inner product of two differential forms $\Phi^{(a)}=\sum_{F=0}^N \Phi_{I_1 \cdots I_F} \, dx^{I_1} \wedge \cdots \wedge dx^{I_F}$, $a=1,2$ is 
\begin{equation}
 \langle \Phi^{(1)} | \Phi^{(2)} \rangle = \int \overline{\Phi^{(1)}} \wedge \star \Phi^{(2)} = \sum_{F=0}^N \int d^N x \, \sqrt{g} \, g^{I_1 J_1} \cdots
 g^{I_F J_F} \, \overline{\Phi^{(1)}_{I_1 \cdots I_F}} \, \Phi^{(2)}_{J_1 \cdots J_F} \, ,
\end{equation}
where $g_{IJ}$ is the metric on $M$. The fermion number $F$ is the form degree and the supercharge in the absence of a potential is the exterior derivative
\begin{equation}
 Q = d = dx^I \nabla_I \, , 
\end{equation}
where $\nabla_I$ is the covariant derivative. The Hamiltonian is the Laplace-Beltrami operator
\begin{equation}
 \Delta = \frac{1}{2} \left( d d^\dagger + d^\dagger d \right) \, .
\end{equation}
Comparing with our previous notation, we have, at the level of operators, the identifications
\begin{equation}
 \bar{\psi}^I = dx^I \wedge \, , \qquad \psi^I = g^{IJ} \frac{\partial}{\partial \bar{\psi}^J} = g^{IJ} \iota_{\partial_I} \, ,
\end{equation}
implying in particular the anticommutation relations $\{\psi^I,\bar{\psi}^J\}=g^{IJ}$. More explicitly, using the relation between curvature and covariant derivative commutators, the Hamiltonian can be written as
\begin{equation} \label{curvham}
 H = -\frac{1}{2} g^{IJ} \nabla_I \nabla_J + \frac{1}{2} R_{IJKL} \psi^I \bar{\psi}^J \psi^K \bar{\psi}^L  \, .  
\end{equation}
The supersymmetric ground states are the harmonic differential forms, and the Witten index is the alternating sum of the number of harmonic forms at each form degree, which is nothing but the Euler characteristic of $M$:
\begin{equation}
 \Omega = \sum_{F=0}^N (-1)^p \, b^p(M) = \chi(M) \, .
\end{equation}
The Betti number $b^p$ is also equal to the number of homologically independent $(N-p)-cycles$. On a compact manifold we have moreover $b^p=b^{N-p}$. An immediate consequence of this is that $\Omega=0$ when $N$ is odd. For a Riemann surface of genus $g$, we have $b^0=b^2=1$ and $b^1=2g$, so $\chi=2-2g$.

Adding a potential can be implemented by putting
\begin{equation}
 Q_h = e^{-h} d e^h = d + dh \, .
\end{equation}
This gives a potential term $V=g^{IJ}\partial_I h \partial_J h$ in the Hamiltonian. Switching on a potential will not change the Witten index, since the spectrum is guaranteed to be discrete on a compact space. In fact, it will not even change the absolute number of supersymmetric ground states in each fermion number. To prove this one uses the very useful general fact\footnote{This is not hard to show. First, for positive energy states, we have that if $Q|\alpha\rangle =0$, then $|\alpha\rangle = \frac{1}{2 E}(Q^\dagger Q + Q Q^\dagger) |\alpha \rangle = Q \left( \frac{Q^\dagger}{2 E}|\alpha \rangle \right)$, so all positive energy states are trivial in cohomology. For zero energy states on the other hand we have $Q|\alpha\rangle=0=Q^\dagger|\alpha\rangle$, so if $|\alpha\rangle=Q|\beta\rangle$, then $\langle \alpha|\alpha\rangle=\langle \beta|Q^\dagger| \alpha \rangle = 0$, i.e.\ $|\alpha\rangle=0$. So each zero energy state is nontrivial in cohomology.} that the space of supersymmetric ground states is isomorphic to the $Q$-cohomology:\begin{equation}
 \CH_{\rm susy} \simeq \frac{{\rm ker}\, Q}{{\rm im}\, Q} \, .
\end{equation}
This is combined with the observation that there is an isomorphism between the cohomologies of $Q_0=d$ and $Q_h=e^{-h}d e^h$ provided by the map $|\alpha\rangle \to e^{-h} |\alpha \rangle$.

An example of a Riemann surface, embedded in $\IR^3$ is shown in fig.\ \ref{pretzel}. This Riemann surface is known as a \emph{pretzel} and acquires a natural metric from its embedding in $\IR^3$. For the function $h$ we take the height along the vertical axis. There are 3 minima, 9 saddle points and 2 maxima, each of which gives rise in the large pretzel limit to a \emph{perturbative} susy ground state, here of fermion number $0$, $1$ and $2$ respectively. We know we can compute the Witten index in perturbation theory, yielding $\Omega=3-9+2=-4$, so the Riemann surface is of genus 3, which is confirmed by visual inspection. We know also that $b_0=1=b_2$, and thus $b_1=6$. Therefore, since the Betti numbers count the number of supersymmetric ground states at given fermion number, several of the perturbative ground states must be lifted by tunneling effects: only one linear combination of the $F=0$ and the $F=2$ states can remain supersymmetric, and six linear combinations of the $F=1$. The remaining three $F=1$ states pair up with two $F=0$ and one $F=2$ state and get lifted. The lifting will be exponentially small in the pretzel size (which sets the scale of $h$). 

\subsection{Lagrangian formalism} 

So far we have described everything in a Hamiltonian framework. Semiclassical tunneling amplitudes responsible for lifting perturbative ground states are however most easily computed in a Lagrangian framework. The Lagrangian framework and its associated path integral are also most convenient to efficiently obtain a differential geometric expression for the Witten index, as the integral over $M$ of a differential form made out of the curvature, namely the Euler density. 

The Lagrangian is obtained from the Hamiltonian in the usual way. We are primarily interested in the Euclidean Lagrangian. Starting from (\ref{curvham}) with the addition of the potential term, this is
\begin{eqnarray} \label{curvlag}
 L &=& \frac{1}{2} g_{IJ} \dot{x}^I \dot{x}^J + \frac{1}{2} g^{IJ} \partial_I h \partial_J h  \nonumber \\
 && + g_{IJ} \bar{\psi}^I D_\tau \psi^J 
 + \frac{1}{2} R_{IJKL} \psi^I \bar{\psi}^J \psi^K \bar{\psi}^L 
 + \nabla_I \partial_J h \, \bar{\psi}^I \psi^J
 \, .  
\end{eqnarray}
where $\tau$ is the Euclidean time conjugate to $H$ and $\dot{x}=\frac{dx}{d\tau}$,  $D_\tau \psi^I = \partial_t \psi^I + \Gamma^I_{JK} \partial_t x^J \psi^K$, $\nabla_I V_J = \partial_I V_J - \Gamma^K_{IJ} V_K$. The supersymmetries act on the fields as $\delta (
\cdots) = [\epsilon Q + \bar{\epsilon} Q^\dagger,\cdots]$:
\begin{eqnarray}
 \delta x^I &=& \epsilon \bar{\psi}^I - \bar{\epsilon} \psi^I \\
 \delta \psi^I &=& \epsilon(-\dot{x}^I-\Gamma^I_{JK} \bar{\psi}^J \psi^K + g^{IJ} \partial_J h) \label{susyvar1} \\
 \delta \bar{\psi}^I &=& \bar{\epsilon} (\dot{x}^I-\Gamma^I_{JK} \bar{\psi}^J \psi^K + g^{IJ} \partial_J h) \, , \label{susyvar2}
\end{eqnarray}
leaving the action invariant. 

As an application we derive a differential geometric expression for the Witten index, which is essentially the generalized Gauss-Bonnet formula \cite{Alvarez-Gaum1983}. The path integral representation of the Witten index is
\begin{equation}
 \Omega = {\rm Tr}\, (-1)^F e^{-\beta H} = \int \CD x \, \CD \bar{\psi} \, \CD \psi \, e^{-\int_0^\beta L \, dt} \, ,
\end{equation}
with periodic boundary conditions on all fields: $\phi(\beta)=\phi(0)$ for $\phi=x,\psi,\bar{\psi}$. In the limit $\beta \to 0$, only the constant trajectories contribute to the path integral. Thus, taking $h=0$ and working in an orthonormal frame, the path integral reduces to the finite dimensional integral
\begin{equation} \label{indfinPI}
 \Omega = \frac{1}{(2 \pi \beta)^{N/2}} \int d^N x \, d^N \bar{\psi} \, d^N \psi \, e^{-\frac{\beta}{2} R_{IJKL} \psi^I \bar{\psi}^J \psi^K \bar{\psi}^L } \, ,
\end{equation}
where the normalization can be fixed for example by comparing path integral and canonical expressions for the free propagator in flat space $\langle x|e^{-\beta H_0}|x\rangle$. Using the symmetries of the Riemann tensor and the Grassmann integral representation of the Pfaffian of a matrix $M$, ${\rm Pf \,} M = \int d\psi^1 \cdots d\psi^N \, e^{-\frac{1}{2} \psi^i M_{ij} \psi^j}$ (with $({\rm Pf} \, M)^2=\det M$), this becomes
\begin{equation} \label{eulerdensityformula}
 \Omega = \int_M {\rm Pf} \, \CR  \, , \qquad 
 {\CR^I}_J \equiv \frac{1}{4 \pi} {R^I}_{J K L} \, dx^K \wedge dx^L \, .
\end{equation}
This expression, which gives the Euler characteristic of $M$ in terms of the integrated Euler density, is known as the generalized Gauss-Bonnet theorem.

\subsection{Instantons and lifting} \label{sec:instantons}

\begin{figure}
\begin{center}
\includegraphics[height=7cm]{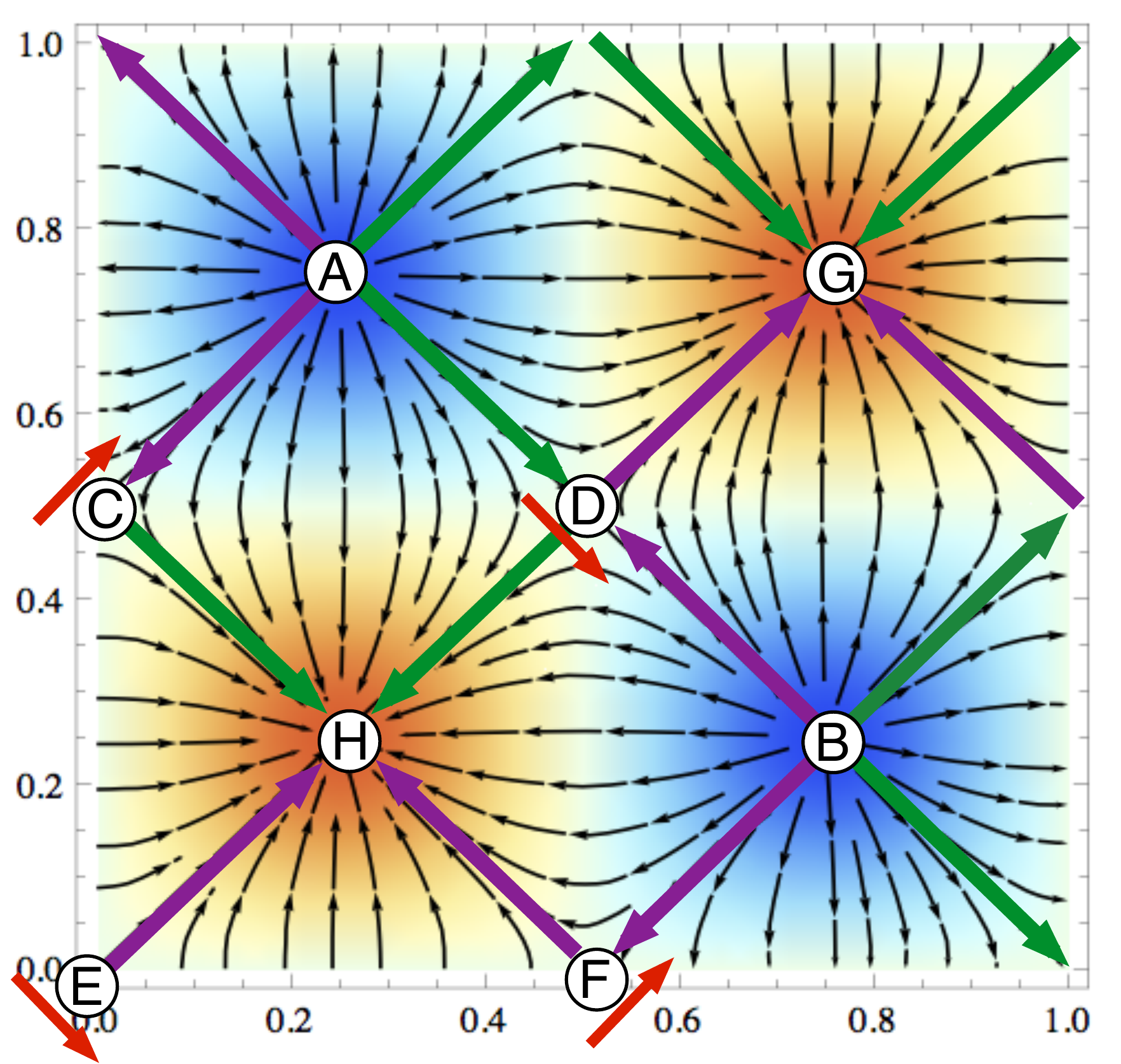}
\end{center}
\caption{\footnotesize Steepest ascent gradient flow lines for the function $h(x,y)=\lambda \, \sin(2 \pi x) \, \sin(2 \pi y)$ on the 2-torus defined by the identifications $x \simeq x+1$, $y \simeq y+1$. The critical points $A,B$ are minima, $G,H$ are maxima, $C,D,E,F$ are saddle points. Big green arrows are paths which give a positive contribution to $Q$, big purple arrows give negative contributions. Little red arrows indicate wave function orientations.
\label{gradflows}}
\end{figure}

By completing the squares, the bosonic part of the action (\ref{curvlag}) can be brought into the form
\begin{equation}
 S_B = s \bigl( h(x_f) - h(x_i) \bigr) + \int_{-T/2}^{T/2} d\tau \, \frac{1}{2} g_{IJ} \left( \dot{x}^I - s g^{IK} \partial_K h \right)\left( \dot{x}^J - s g^{JL} \partial_L h \right) \, ,
\end{equation}
where $s=\pm1$. The first term comes from integrating a total derivative, and $x(-T/2)=x_i$ and $x(T/2)=x_f$ are the start and end points of the trajectory. This implies the lower bound $S_B \geq s \bigl( h(x_f) - h(x_i) \bigr)$ for $s = \pm 1$, hence $S_B \geq |h(x_f) - h(x_i)|$, which is saturated for steepest descent or ascent trajectories: 
\begin{equation}
 \dot{x}^I = s g^{IJ} \partial_J h(x) \, , \qquad s = {\rm sign} \bigl( h(x_f) - h(x_i) \bigr) \, .
\end{equation}
These gradient flow trajectories (with the fermions put to zero) are automatically solutions to the Euclidean equations of motion, and moreover from the supersymmetry variations (\ref{susyvar1})-(\ref{susyvar2}) it can be seen that ascending ($s=+1$) trajectories are annihilated by the $Q$-supersymmetry, while descending trajectories are annihilated by the $Q^\dagger$-supersymmetry. Such minimal action trajectories, in the limit $T \to \infty$, are called instantons.

Instantons play a key role in the computation of nonperturbative lifting of ground states. For concreteness consider the example shown in fig.\ \ref{gradflows}, a two dimensional flat unit torus with superpotential $h(x,y)=\lambda \, \sin(2 \pi x) \, \sin(2 \pi y)$. There are 8 critical points: 2 minima, 4 saddle points and 2 maxima, giving $\Omega=2-4+2=0=\chi(T^2)$. The Betti numbers for $T^2$ are $b_0=1$, $b_1=2$, $b_2=1$, so again some of the perturbative ground states must be lifted by tunneling effects. The relevant tunneling trajectories responsible for this are precisely the instantons introduced above.

The general story goes as follows. Denoting the perturbative ground states by $|i\rangle$, lifting means that the matrix elements of the supercharge reduced to the perturbative zero energy level,
\begin{equation}
 \hat{Q}_{ij} \equiv \langle i | Q | j \rangle \, ,
\end{equation}
although identically zero in perturbation theory, is not identically zero when nonperturbative effects are taken into account. Notice that since $Q$ has fermion number 1, $Q_{ij}$ can only be nonzero if the fermion numbers of $|i\rangle$ and $|j\rangle$ statisfy the selection rule $F_i=F_{j}+1$. The matrix elements $Q_{ij}$ can be given a path integral representation. It can be shown that thanks to supersymmetry, this path integral (generically) \emph{only} gets contributions from instanton trajectories annihilated by $Q$, i.e.\ paths of steepest ascent, interpolating between critical points $j \to i$ with $F_i=F_{j}+1$. Moreover the Gaussian approximation to this contribution is exact, resulting in the very simple expression
\begin{equation}
 \hat{Q}_{ij} = \sum_{\gamma:j \to i} n_\gamma \, e^{-h_{ij}} \, ,
\end{equation}
where the sum is over steepest ascent flows from $j$ to $i$ satisfying $F_i-F_j = 1$. The prefactor $n_\gamma = \pm 1$ and $h_{ij}$ is the instanton action, $h_{ij}=|h(i)-h(j)|$. This simplification in the path integral is referred to as localization. We will not derive this here; see \cite{Hori2003} chapter 10 for a detailed pedagogical account. 

The sign of $n_\gamma$ is determined by orientation considerations. This is best explained by going back to the torus example. First we need to choose an orientation for the perturbative ground states $|i\rangle$. In the figure the critical points are denoted by the letters $A$ to $G$, and we will denote the perturbative ground states by these letters too. Then $|A\rangle$ and $|B\rangle$ are 0-forms, $|C\rangle, |D\rangle, |E\rangle, |F\rangle$ are 1-forms, and $|G\rangle, |H\rangle$ are 2-forms. For example, to lowest order in perturbation theory, 
\begin{eqnarray}
 |A\rangle &=& \alpha \, e^{-\frac{\lambda}{2} (2\pi)^2(u^2+v^2)} \, , \qquad \alpha = \sqrt{2 \pi \lambda} \, , \quad u=x-\mbox{$\frac{1}{4}$} , \quad v=y-\mbox{$\frac{3}{4}$}  \, ,
 \\
 |C\rangle &=& \alpha \, e^{-\frac{\lambda}{2}(2 \pi)^2 (u^2+v^2)} \, du \, , \qquad u=x+y-\mbox{$\frac{1}{2}$} , \quad v= x-y+\mbox{$\frac{1}{2}$} \, , \\
 |G\rangle &=& \alpha \, e^{-\frac{\lambda}{2} (2 \pi)^2(u^2+v^2)} \, du \wedge dv \, , \qquad u= x-\mbox{$\frac{3}{4}$} , \quad v= y-\mbox{$\frac{3}{4}$} \, .
  \end{eqnarray}
Note that whereas the top and bottom forms can be given a natural orientation inherited from the orientation of the torus itself, this is not so for the 1-forms. We choose them such that at the critical point, the form is positive in the positive $x$-direction, i.e.\ the local coordinate $u$ that is such that $h \sim -u^2 + v^2$ near the critical point is taken to increase when going to the right. In the figure, the orientation of the local $u$-coordinate is indicated by the red arrows. The arrow is a local frame for the unstable direction. More generally, the choice of ordered standard local coordinates near each critical point determines a local frame for the unstable directions of $h$ at each critical point. On the other hand when moving along an instanton trajectory, we can transport this frame from the initial point to the final point, and then add to it a new vector pointing in the direction of the path, to obtain a new frame for the unstable directions around the final point (which indeed generically has one more unstable direction than the initial point, in the direction of the arriving path). If this new frame has the same orientation as the frame we already had there, then $n_\gamma=+1$, and otherwise $n_\gamma=-1$.

Using these rules, and noticing that in this example $\Delta h=\lambda$ for successive critical points, we get for example $\hat{Q}|A\rangle=e^{-\lambda} (|D\rangle + |F\rangle - |C\rangle - |E\rangle)$, $\hat{Q}|C\rangle = e^{-\lambda} (|H\rangle-|G\rangle)$, $\hat{Q}|G\rangle = 0$. In matrix form, we thus compute $\hat{Q}$, $\hat{Q}^\dagger$ and $\hat{H}=\frac{1}{2}\{\hat{Q},\hat{Q}^\dagger \}$ (writing zeros as blanks):
\begin{equation*}
\scriptsize
\hat{Q} = e^{-\lambda} 
\begin{pmatrix}
  &  & -1 & 1 & -1 & 1 &  &  \\
 ~ & ~ & 1 & -1 & 1 & -1 & ~ & ~ \\
 ~ & ~ & ~ & ~ & ~ & ~ & -1 & 1 \\
 ~ & ~ & ~ & ~ & ~ & ~ & -1 & 1 \\
 ~ & ~ & ~ & ~ & ~ & ~ & 1 & -1 \\
 ~ & ~ & ~ & ~ & ~ & ~ & 1 & -1 \\
 ~ & ~ & ~ & ~ & ~ & ~ & ~ & ~ \\
 ~ & ~ & ~ & ~ & ~ & ~ & ~ & ~
\end{pmatrix} \, , \quad
\hat{H} = 2 \, e^{-2 \lambda} \begin{pmatrix}
  1 & -1 & ~ & ~ & ~ & ~ & ~ & ~ \\
 -1 & 1 & ~ & ~ & ~ & ~ & ~ & ~ \\
 ~ & ~ & 1 & ~ & ~ & -1 & ~ & ~ \\
 ~ & ~ & ~ & 1 & -1 & ~ & ~ & ~ \\
 ~ & ~ & ~ & -1 & 1 & ~ & ~ & ~ \\
 ~ & ~ & -1 & ~ & ~ & 1 & ~ & ~ \\
 ~ & ~ & ~ & ~ & ~ & ~ & 1 & -1 \\
 ~ & ~ & ~ & ~ & ~ & ~ & -1 & 1
\end{pmatrix}
\end{equation*}
The space of true supersymmetric ground states is isomorphic to the cohomology ${\rm ker} \, \hat{Q}/{\rm im} \, \hat{Q}$ of the matrix $\hat{Q}$, which is relatively easily computed. This is called the Morse-Witten cohomology. Alternatively, we can find the actual supersymmetric ground states (within the approximation of working within the finite dimensional space of perturbative supersymmetric ground states, i.e.\ to leading order in $e^{-\lambda}$), as the joint null space of $\hat{Q}$ and $\hat{Q}^\dagger$, or equivalently the null space of $\hat{H}$. The reduced Hamiltonian $\hat{H}$ furthermore gives the dynamical transition rates between different perturbative ground states, and diagonalizing it gives the energy level splittings.
  
The conclusion in our example is that the supersymmetric ground states in this approximation are given by the symmetric combinations $|A\rangle + |B\rangle$, $| C \rangle + |F \rangle$, $|D \rangle + |E \rangle$, $|G\rangle + |H\rangle$, consistent with the Betti numbers $(1,2,1)$. The four antisymmetric combinations orthogonal to those all have lifted energies equal to $4 e^{-2 \lambda}$.

Of course in more general examples, the splitting pattern and linear combinations will be less simple. But in principle at least there is a general algorithm to compute it in terms of simple, generically isolated, ``atomic'' ($\Delta F = 1$) gradient flows, which is quite remarkable; in nonsupersymmetric quantum mechanics in dimensions greater than 1, doing something similar is typically very hard if not impossible. 

For complex energy landscapes $h(x)$ in high dimensional spaces, it will typically be more appropriate to compute statistical, thermodynamic quantities, but again the structure of supersymmetry will lead to significant simplifications.

As a check on the above results, and also to get further insight into the nature of the approximation, observe that it is actually easy to write down the exact bottom and top fermion number ground state wave functions; they are:
\begin{equation}
 |F=0\rangle = e^{-h} \, , \qquad | F={\rm dim} \rangle = e^{+h} \, \mbox{$\prod_I \bar{\psi}^I$}\, .
\end{equation} 
The approximation we made replaces this to lowest order in perturbation theory by an even superposition of Gaussians centered at the critical points. This is indeed an excellent approximation at large $\lambda$. 

Note that in this limit perturbative ground states are exponentially close to being real ground states, and physically the number of perturbative ground states would therefore be a better measure for the ground state degeneracy than the Witten index. However, at generic, order 1 couplings, this is no longer the case, and one may reasonably expect all states that are not protected to be lifted to finite energies, making the Witten index an accurate, physically meaningful measure for the number of true ground states. On the other hand, as many examples in string theory have taught us, one must of course be careful that in the apparent strong coupling regime there isn't a dual weak coupling description hiding, which would again potentially produce many new near-exact ground states, not captured by the Witten index. 

This is the case in particular in thermodynamic, ``large $N$'' limits. For example the SK model and other glassy systems develop many pseudo ground states in the thermodynamic limit. Although not supersymmetric, there is an interesting map from classical statistical mechanics systems to supersymmetric quantum mechanics systems, showing this is not just an analogy, but really the same thing. To this we turn next.

\subsection{Supersymmetry, stochastic systems and glasses} \label{sec:susystochastic}

Supersymmetric quantum mechanics can be interpreted as a BRST description of a classical stochastic system \cite{Parisi1979b,Cecotti1983new,Damgaard1987,Kurchan1992,Kurchan2002a}. This goes as follows. Consider a stochastic dynamical Langevin equation for a particle with position $x(t)$ in a potential $h(x)$:
\begin{equation} \label{langevin}
 \frac{dx}{dt} = -\partial_x h(x) + \eta(t) \, ,
\end{equation}
where $\eta$ is a Gaussian random noise function with variance $\alpha$, $P(\eta) \propto e^{-\eta^2/2\alpha}$ (the noise can be thought of as thermal fluctuations with temperature proportional to $\alpha$).  Starting from some initial point $x_i$ at time $t=0$, the probability density of finding the configuration at a point $x_f$ at time $t$ is given by the path integral
\begin{equation}
 P(x_f,t|x_i,0) = \int \CD x \, \CD \eta  \, e^{-\int_0^t \frac{1}{2 \alpha} \eta^2} \, \delta\bigl(\frac{dx}{dt} + h'(x) - \eta\bigr) \, |\det \bigl( \frac{d}{dt} + h''(x) \bigr)| \, ,
\end{equation}
appropriately normalized and with boundary conditions $x(0)=x_i$, $x(t)=x_f$. The combination of delta function and determinant ensures that the path integration over $x$ simply picks out with unit weight the unique solution to the first order linear differential equation (\ref{langevin}) for given $\eta$. It is the functional analog of the finite dimensional formula $\int dx \, \delta(A \cdot x + B) \, |\det A| = 1$ where $A$ is a constant matrix and $B$ is a constant vector. Ignoring for the time being the absolute value signs, we can write the determinant as a fermionic path integral, so that amusingly, after also integrating out $\eta$, we obtain exactly the Euclidean propagator of supersymmetric quantum mechanics with superpotential $h(x)$:
\begin{eqnarray}
 P(x_f,t|x_i,0) &=& \int \CD x \, \CD \bar{\psi} \, \CD \psi  \, e^{-\int_0^t \frac{1}{2 \alpha} ( \dot{x} + h'(x) )^2 + \bar{\psi} \dot{\psi} + \bar{\psi} h''(x) \psi }  \, \\
 &=& e^{(h(x_i) - h(x_f))/\alpha} \, 
 \int \CD x \, \CD \bar{\psi} \, \CD \psi  \, e^{-\int_0^t \frac{1}{2 \alpha} (\dot{x}^2 + h'(x)^2) + \bar{\psi} \dot{\psi} + \bar{\psi} h''(x) \psi } \, \\
 &=& e^{(h(x_i) - h(x_f))/\alpha} \, \langle x_f| e^{-t H_{\rm susy}/\alpha} |x_i\rangle \, . 
\end{eqnarray}
The last line is the propagator between states in the zero fermion number sector, with elapsed Euclidean time $t$ and Planck's constant $\hbar=\alpha$ (before we were setting $\hbar \equiv 1$). Inserting a complete set of ($F=0$) energy eigenstates $|n\rangle$, we have
\begin{equation} \label{timeevol}
 \langle x_f| e^{-t H_{\rm susy}/\alpha} |x_i\rangle = \sum_n \langle x_f | n\rangle \langle n|x_i\rangle e^{-t E_n/\alpha} \, .
\end{equation}
In the long time limit $t \to \infty$, only the ground state survives. As we have seen, when $Z \equiv \int dx \, e^{-2h(x)/\alpha} < \infty$, the Witten index is 1 and the unique supersymmetric ground state is  $\langle x|0\rangle = \frac{1}{\sqrt{Z}} e^{-h(x)/\alpha}$. We conclude
\begin{equation}
 P(x_f,\infty|x_i,0) = \frac{1}{Z} e^{-\beta \, h(x_f)} \, , \qquad\beta \equiv \frac{2}{\alpha} \, , 
\end{equation} 
with $Z = \int dx \, e^{-\beta h(x)}$. This is indeed the expected Boltzmann equilibrium distribution with inverse temperature $\beta$, which alternatively can be derived from the Fokker-Planck equation associated to the Langevin equation. In fact the Fokker-Planck equation can easily be seen to be equivalent to the Schr\"odinger equation in the zero fermion number sector (identifying $P(x) \propto e^{h(x)} \psi(x)$), providing an alternative derivation of the above identifications. 

Note that when the Witten index is $-1$, i.e.\ $e^{+ h(x)}$ is normalizable, there is no future equilibrium state, but there is a past equilibrium state. When it is zero, there is no future and no past equilibrium. If there are multiple \emph{perturbative} supersymmetric ground states with fermion number zero (i.e.\ local minima of $h(x)$), then in the small $\alpha$ (i.e.\ large $\beta$) limit, there will be highly metastable states. According to (\ref{timeevol}), the time scale for decay of these metastable states is set by the nonperturbatively lifted energy, which as we have seen in the previous section can be computed from simple instantons and the Morse-Witten complex.

All this generalizes in a straightforward way to higher dimensional and curved spaces. If $h(x)$ is the Hamiltonian of some classical statistical mechanical system, then in the thermodynamic limit $N \to \infty$ there may be degenerations that change the number of exact supersymmetric ground states at a given fermion number and even the Witten index. This is because degeneracy lifting instanton effects may vanish altogether, thus turning perturbative ground states into exact ground states. When this happens at fermion number zero, ergodicity is broken and we get multiple exact equilibrium states. Whether or not this happens will depend on the distribution of energy barriers, i.e.\ the distribution of supersymmetric ground states with low fermion number, and how they are connected by instantons, i.e.\ the Morse-Witten complex. This should capture, if present, the ultrametric structure of the state space.

Some discussion of spin glasses in the language of supersymmetric quantum mechanics can be found in e.g.\ \cite{Kurchan1992,Franz1992,Kurchan2002a,DeDominicis2006new}. Recent considerations in string theory involving the above connection between stochastic equations and supersymmetry include \cite{Cecotti2010,Dijkgraaf2010}. A nice example of the relation between transition rate distributions and universal properties of relaxation dynamics can be found in \cite{Amir2008}, where the typical $\log(1+t_w/t)$ behavior of return to equilibrium after a perturbation of duration $t_w$ of a glass
was derived from general arguments based on the form of a transition matrix with exponentially suppressed off-diagonal matrix elements. Such a transition matrix and Hamiltonian are typical for systems with a wide distribution of relevant instanton actions.

\section{D-brane landscapes} \label{sec:Dbranes}

\subsection{Introduction}  

We now move on to the description of complex systems that appear in the context of string theory, and review some of the tools that have been used to describe their basic features, such as ground state degeneracies and distributions. One of the interesting things string theory adds to conventional analysis is the power of holographic duality \cite{Aharony2000}, which in favorable circumstances can reduce a quantitative understanding of the strong coupling physics of the model under consideration to simple computations in gravity. This works particularly well for thermodynamics, as black holes are the holographic duals to thermodynamic states, capturing the strong coupling thermodynamics of the model with often rather astonishing simplicity.

More specifically, we will have a look at the complex configuration spaces that arise when wrapping D-branes around various cycles of compactification manifolds, with various worldvolume fields turned on. Such brane configurations arise in type II constructions of string vacua, where they are among other things responsible for the particle physics content of the compactification. They also arise in the description of charged black hole microstates, extrapolated to weak coupling. In the weak coupling limit (large volumes, small string coupling), the only difference between the two as far as their perturbative description is concerned is the fact that in the former case, branes are filling the observable, noncompact space, while in the latter, they are localized at a point. But in particular the geometric internal space description of the configuration is mathematically identical in the two cases. Of course the actual dynamics and other physics of the two systems will be very different, one describing the universe we may live in, the other an object we may look at, but the fact that they are related at the level of classical configuration spaces is one of the beautiful incarnations of the unifying nature of string theory. Thus, in particular, the largeness of the landscape and the largeness of the entropy of a black hole are intimately connected. 

Despite this map, the fact that branes fill all of space has one dramatic consequence, and that is that the charges (brane wrapping numbers etc) are tightly constrained by so-called tadpole cancellation conditions. What this means is simply that whenever we have charged objects (such as a branes) with compact transversal dimensions (as is the case for space-filling branes but not space-localized ones), the sum of all charges must necessarily be zero. This is a consequence of Gauss' law, or more colloquially, the fact that flux lines otherwise have no place to go to; when there's a source, there must be a sink. Combined with the requirement of stability (typically realized by staying close to a supersymmetric configuration), this imposes strict bounds on the allowed charges.\footnote{These bounds are set by the background curvature or orientifold contributions to the charge, which in turn are set by topological invariants such as the Euler number of the compactification manifold. Although these numbers can get fairly high, for example there are known \cite{Kreuzer} elliptic CY 4-folds suitable for F-theory compactifications allowing up to $\CO(10^5)$ transversal D3-branes, constructions get increasingly sparse at the higher end, leading many to suspect the set is finite. For reasons we do not understand any deeper than through case by case examination of various conspiracies, the string theory landscape is uncannily ``boxed in'', making it for example very hard, and plausibly impossible \cite{Acharya2006}, to find infinite families of parametrically controlled solutions with positive cosmological constant.} On the other hand, for space localized branes, no such charge restrictions arise, and in particular we can take the charges to infinity and consider thermodynamic limits, which can be expected to have universal properties. This motivates us to consider localized brane systems. 

At finite coupling another difference between the two creeps in: the presence of gravity. This has a very different effect in the two settings. In the space-filling case, ``vacuum'' solutions with slightly different internal brane configurations can give rise to very different asymptotic geometries. As emphasized by Tom Banks at this school \cite{Banks2010}, for various reasons including the fact that our most successful descriptions of quantum gravity use fixed spacetime asymptotics in an essential way, this makes it unclear if we can even think of these different solutions as being part of the same theory, or, as he would put it, it makes it clear that these solutions are not part of the same theory. This does not mean that the presence of a space of brane configurations is of no physical relevance here of course; they do all \emph{exist} after all, and it is certainly of interest to classify the set of all possibilities. Furthermore local fluctuations as well as larger fluctuations in the form of bubble nucleations can certainly probe --- albeit in a limited sense --- the configuration space. But it does mean that the situation is conceptually much more subtle than, say, a collection of manganese atoms in copper. In contrast, for space-localized branes, these  issues do not arise. Gravitational backreaction just turns the weakly coupled brane system into a black hole, leaving the asymptotics unaffected. Even better, thanks to the effects of gravity, we get a host of thermodynamic parameters for free, courtesy of the Bekenstein-Hawking entropy formula. This gives another motivation to study brane systems associated to black holes.

Besides their intrinsic interest as models of complex, glassy systems with intricate landscapes and known holographic dual descriptions, the study of such localized brane systems can also teach us things about the space-filling systems. Given their close physical and mathematical relation, it is clear that they contain information about each other. For example, their Witten indices, suitably defined, are the same. Indeed, Witten indices of supersymmetric quantum field theories are traditionally computed by reduction to a quantum mechanical model. In the present context one can go further, and use the computational power of gravity (manifested through for example the Bekenstein-Hawking formula or its refinements \cite{Ooguri}) to extract information about ensembles of possible string vacua. This idea was made precise in \cite{Denef2010a}; I will not have much to say about it in these lectures, except that it serves as yet another motivation to study space-localized branes.
  
The analysis of D-brane systems as mean-field glassy systems, or in other words connecting this section and section \ref{sec:spinglass}, will be addressed in \cite{AABDG}.    
  
\subsection{Low energy physics of the weakly coupled wrapped D4} \label{sec:D4}

In this section we derive in detail the quantum mechanics describing the low energy dynamics of a D4-brane wrapped on a 4-cycle, in the limit of small string coupling and large volume. Besides being an interesting and accessible example of a complex system in string theory with applications to black hole physics, mathematics and string model building, it also provides a clean example illustrating various useful Kaluza-Klein reduction techniques, in particular the relation between fluxes and superpotentials. (The idea to consider such an open string landscape was proposed in \cite{Gomis2005}.)

\subsubsection{Actions}

Consider type IIA string theory on $\IR^4 \times X$, where $X$ is some compact 6-dimensional manifold. The part of the IIA bulk action relevant for our purposes is, in units with the string scale $\ell_s = 2 \pi \sqrt{\alpha'} \equiv 1$:
\begin{eqnarray}
 S_{\rm IIA} &=& S_{\rm IIA, NSNS} + S_{\rm IIA, RR} \\
 S_{\rm IIA, NSNS} &=&  2 \pi \int_{\IR^4 \times X} d^{10} x \, \sqrt{-g} \left( \frac{1}{g_s^2} R - \frac{1}{g_s^2} H_{MNP} H^{MNP} \right) \\
 S_{\rm IIA, RR} &=& - 2 \pi \int_{\IR^4 \times X} d^{10} x \, \sqrt{-g} \, \left(
  \frac{1}{2} F^{(2)}_{MN} F^{(2) MN} \, + \, \frac{1}{2} F^{(4)}_{MNPQ} F^{(4) MNPQ} \, \right) \, ,
\end{eqnarray}
where $H=d B$ is the NSNS 3-form field strength, $F^{(p+2)} = d C^{(p+1)}$ are the RR field strengths and $g_s$ is the string coupling constant. The theory contains D$p$-branes with $p$ even. In particular a single D4 wrapped around a 4-cycle $\Sigma$ in the compact manifold $X$ has, in the limit of large volume and weak string coupling, an effective worldvolume action
\begin{eqnarray}
 S_{D4} &=& S_{\rm D4,NSNS} + S_{\rm D4,RR} \\
 S_{\rm D4, NSNS} &=& -\frac{2 \pi}{g_s} \int_{\IR \times \Sigma} d^5 x \, \sqrt{-h}  \left( 1  + \frac{1}{2}  F_{IJ} F^{IJ} \right) \label{SD4NSNS} \\
 S_{\rm D4, RR} &=& 2\pi \int_{\IR \times \Sigma} F \wedge C^{(3)} + \left(\frac{1}{2} F \wedge F + \frac{1}{24} e(R) \right)\wedge C^{(1)} \, , \label{SD4RR}
\end{eqnarray}
where $h_{IJ}$ is the induced brane metric, $F=dA+B$, $A$ is the $U(1)$ gauge connection living on the brane and $e(R)$ is a quadratic polynomial in the tangent and normal bundle curvature forms. When $X$ is a Calabi-Yau manifold, this is the Euler density $e(R)$:
\begin{equation} 
 e(R) = {\rm Pf} \, \CR  \, , \qquad 
 {\CR^I}_J \equiv \frac{1}{4 \pi} {R^I}_{J K L} \, dx^K \wedge dx^L \, ,
\end{equation}
which we encountered earlier in (\ref{eulerdensityformula}).

The couplings to the RR fields given by $S_{\rm D4,RR}$ are topological in the sense that they are metric-independent. They specify the gauge flux and curvature induced D2- and D0-charges on the brane. 

We have ignored curvature corrections to $S_{\rm D4,NSNS}$, but in the end we will fix this by matching the energy required by supersymmetry for a given charge. A slightly more precise version of this action and references to derivations in the literature can be found in section 2 of \cite{Denef2007a}. 

\subsubsection{Bulk Kaluza-Klein reduction}

The low energy dynamics of the wrapped D4 system is obtained by Kaluza-Klein reduction. This is done most easily, and leads to the richest structure, when there is residual supersymmetry. An $\CN=1$ theory in four dimensions does not have pointlike objects that preserve part of the supersymmetry, so one has to consider at least $\CN=2$, i.e.\ compactifications with eight preserved supercharges, with the wrapped D-brane breaking half of those. When no bulk magnetic fluxes are turned on, this requires the compactification manifold $X$ to be Calabi-Yau, i.e.\ Ricci-flat and K\"ahler. Concretely this means that there exist complex coordinates $y^m$, $m=1,2,3$ and a holomorphic complex orthonormal frame $\theta^m={\theta^m}_n dy^n$, such that the metric is of the form
\begin{equation}
 ds^2 = |\theta^1|^2 + |\theta^2|^2 + |\theta^3|^2 \, ,
\end{equation}
with associated K\"ahler form $J$ and covariantly constant holomorphic 3-form $\Omega$ given by
\begin{equation}  \label{JOm}
 J = \frac{i}{2} \left( \theta^1 \wedge \bar{\theta}^1+\theta^2 \wedge \bar{\theta}^2+\theta^3 \wedge \bar{\theta}^3 \right) \, , \qquad
 \Omega = \theta^1 \wedge \theta^2 \wedge \theta^3 \, .
\end{equation}
The Calabi-Yau condition means that $J$ and $\Omega$ are globally well-defined. The K\"ahler form has one holomorphic and one anti-holomorphic index and is therefore called a $(1,1)$-form. Similarly $\Omega$ is a $(3,0)$-form. With the above definitions we have 
\begin{equation} \label{OmJnormalization}
 i \int_{X} \Omega \wedge \overline{\Omega} = 8 \int_{X} \frac{J^3}{6} = 8 V_X \, ,
\end{equation}
where $V_X$ is the volume of $X$.

As a concrete example we can take $X$ to be the quintic Calabi-Yau, described by a homogeneous degree 5 polynomial equation in $\IC\IP^4$. The four complex dimensional projective space $\ICP^4$ is the set of all nonzero $(x_1,x_2,x_3,x_4,x_5) \in \IC^5$, modulo the equivalence relation $x \simeq \lambda x$, $\lambda \in \IC^*$. The simplest choice of polynomial is of Fermat form:
\begin{equation} \label{Q5}
 X:Q_5(x) \equiv x_1^5 + x_2^5 + x_3^5 + x_4^5 + x_5^5 = 0 \, .
\end{equation}
Deformations of the metric preserving Calabi-Yauness are called (geometric) moduli. The four dimensional low energy effective field theory will contain these moduli as massless scalars. In general it is not possible to write down explicit expressions for the Ricci flat Calabi-Yau metric, but Yau's theorem states that for a given K\"ahler class (i.e.\ the cohomology class of the K\"ahler form) and complex structure (i.e.\ a choice of complex coordinates), there is a unique Ricci flat K\"ahler metric. Thus, there are two kinds of moduli, those specifying the cohomology class of the K\"ahler form and those specifying the complex structure. In the case at hand there is just one K\"ahler modulus, which parametrizes the overall scale of the metric, i.e.\ the volume of $X$ --- the second cohomology of $X$ is one-dimensional. The different complex structures are simply parametrized by the choice of defining polynomial $Q_5(x)$, modulo linear coordinate transformations of the $x_i$. This space is 101-dimensional. 

More generally, the dimension of the second cohomology of $X$ will be $b^2(X)$-dimensional, i.e.\ there are $b^2(X)$ independent harmonic 2-forms on $X$. Equivalently, because of Poincar\'e duality,\footnote{Poincar\'e duality is a natural isomorphism between $p$-cycle homology classes and $(n-p)$-form cohomology classes in an $n$-dimensional space. For a $p$-cycle $C$ locally given by $n-p$ equations $f_i(x)=0$, the associated dual cohomology class can be represented by the closed delta-function $(n-p)$-form $\hat{C} = \delta(f_1) \, df_1 \wedge \cdots \wedge \delta(f_{n-p}) \, df_{n-p}$. Note that for two cycles $C$ and $C'$ with dimension adding up to $n$, we can thus write $\int \hat{C} \wedge \hat{C}' = \int_C \hat{C}' = \#(C \cap C')$, where the last expression counts intersection points for generic representatives, with signs according to orientations. Usually we will drop the hat in the dual to avoid cluttering. \label{Poincdual}} there are $b^2(X)$ independent homology 4-cycles in $X$. Let $D_A$, $A=1,\ldots,B_2 \equiv b^2(X)$ be an integral basis of harmonic 2-forms. Then 
\begin{equation} \label{JDA}
 J=J^A D_A \, ,
\end{equation} 
and the $J^A$ are the K\"ahler moduli. We choose the signs of the 2-forms $D_A$ such that $J^A > 0$.

The reduction of the RR potentials produces, among other fields, electrostatic potentials in four dimensions coupling to D2 and D0 charges:
\begin{equation} \label{RRreduction}
 C^{(3)} = \phi^{A} \, dt \wedge D_A  + \cdots\, , \qquad C^{(1)} = \phi^{0} \, dt  + \cdots\, .
\end{equation}


\subsubsection{Brane Kaluza-Klein reduction}

In the weak coupling limit $g_s \to 0$, the backreaction of the wrapped D4 on the bulk geometry is negligible, due to the lower power in $1/g_s$ appearing in the D4-action. We can therefore consider the D-brane to be a probe in a background geometry specified by arbitrary, fixed bulk moduli. For the wrapped brane to be supersymmetric, it must wrap a minimal volume 4-cycle, or somewhat stronger even, it must be holomorphic \cite{Marino2000}. In the case of the quintic this  
is equivalent to saying it can be described by a degree $N$ homogeneous polynomial equation
\begin{equation}
 \Sigma \, : \,P_N(x_1,x_2,x_3,x_4,x_5) = 0 \, .
\end{equation}
The degree $N$ can be identified with the multiplicity of the D4-charge. For example for $N=1$, the most general polynomial is $P_1(x) = a_1 x_1 + \cdots + a_5 x_5$. The complex coefficients $c$ are deformation moduli of $\Sigma$. Since overall rescaling of the coefficients does not change the zero set, we have a four complex dimensional deformation moduli space, topologically $\IC\IP^4$. An example of a degree $N$ polynomial is $\left(P_1(x)\right)^N$, which corresponds to $N$ coincident branes wrapping $P_1(x)=0$, with gauge group enhanced to $U(N)$. The most general degree $N$ homogeneous polynomial on the other hand gives a smooth singly wrapped brane with gauge group $U(1)$. In the case at hand it has $d(N) = \frac{5 N^3}{6} + \frac{50 N}{12} - 1$ independent deformation moduli $z^a$, as can be verified by direct monomial counting, taking into account that when $N \geq 5$, adding a term $q_{N-5}(x) Q_5(x)$ to $P_N(x)$, with $Q_5$ as in (\ref{Q5}), does not change the zero set. More generally one can deduce the number of deformations using standard algebraic geometry techniques \cite{Maldacena1997}. For a D4-brane wrapping a homology class\footnote{Here the $D_A$ are the Poincar\'e duals to the $D_A$ introduced in (\ref{JDA}), for which as mentioned in footnote \ref{Poincdual} we use the same notation.} $[\Sigma] = N^A D_A$ with $N^A > 0$
this gives
\begin{equation} \label{deDNNN}
 d = \frac{1}{6} D_{ABC} N^A N^B N^C + \frac{1}{12} c_{2,A} N^A - 1 \, . 
\end{equation}
Here the ``triple intersection numbers'' $D_{ABC}$ are defined as
\begin{equation} \label{DABC}
 D_{ABC} \equiv \int_{X} D_A \wedge D_B \wedge D_C = \#(D_A \cap D_B \cap D_C) 
\end{equation}
and the $c_{2,A}$ are topological numbers (the second Chern class of $X$) depending on $X$ only. For the quintic we see by comparing that $D_{111}=5$ and $c_{2,1} = 50$. The same techniques allow straightforward computation of the generic Euler characteristic of $\Sigma$, namely
\begin{equation}
 \chi = D_{ABC} N^A N^B N^C + c_{2,A} N^A \, ,
\end{equation}
which for the quintic leads to $\chi=5 \, N^3+50 \, N$. The second Betti number, i.e.\ the number of independent 2-cycles or 2-forms on $\Sigma$ is obtained from this as $b_2(\Sigma) = \chi-2-b_1-b_3$. When $N^A > 0$ the Betti numbers $b_1=b_3$ of $\Sigma$ are inherited from $X$, so they vanish except in the higher susy cases when $X$ is $T^6$ ($b_1=6$) or $T^2 \times K3$ $(b_1=2)$. In any case, $b_2$ grows as $N^3$, and therefore, crucially, we will get a huge magnetic flux degeneracy on these branes in the large $N$ limit. Indeed we can turn on harmonic gauge field strengths $F=dA$ on $\Sigma$ characterized by $b_2$ integers $S_i$, the flux quanta: 
\begin{equation} \label{sigmai}
 F = \sum_{i=1}^{b_2(\Sigma)} S_i \, \sigma^i \, ,
\end{equation}
where the $\sigma^i$ form a basis of integrally quantized harmonic 2-forms on $\Sigma$. In terms of Poincar\'e dual 2-cycles, we can equivalently write
\begin{equation}
 S_i = \eta_{ij} \int F \wedge \sigma^j = \eta_{ij} \int_{\sigma^j} F \, , \qquad
 \eta^{ij} \equiv \int_{\Sigma} \sigma^i \wedge \sigma^j = \#(\sigma^i \cap \sigma^j) \, ,
\end{equation}
with $\eta_{ij}$ the inverse of $\eta^{ij}$. The matrix $\eta^{ij}$ is called the intersection form of $\Sigma$. It is integral and unimodular, but in general not positive definite --- in fact as we will see later at large $N$ it has signature $(b_2^+,b_2^-) \propto (\frac{1}{3},\frac{2}{3}) \times N^3$.

The flux and curvature induced D2 and D0 brane charge can be read off from (\ref{SD4RR}) with the reduction (\ref{RRreduction}):
\begin{eqnarray} 
 S_{\rm D4,RR} &=&  2\pi \int dt \, \left( -q_0 \, \phi^0  + q_A \, \phi^A \right) \\
 q_0 &=& -\frac{1}{24} \, \chi(\Sigma) -\frac{1}{2} \int_{\Sigma} F \wedge F  = -\frac{\chi}{24} - \frac{1}{2} \eta^{ij} S_i S_j \label{QD0} \\
 q_A &=& \int_{\Sigma} F \wedge D_A =  D_A^i S_i \,  
 \, ,
 \label{QD2}
\end{eqnarray}
where the integers $D_A^i$ are given by
\begin{equation}
 D_{A,i} = \int_{\Sigma} \sigma^i \wedge D_A = \# (\sigma^i \cap D_A) \, .
\end{equation}

\begin{figure}
\begin{center}
\includegraphics[height=5cm]{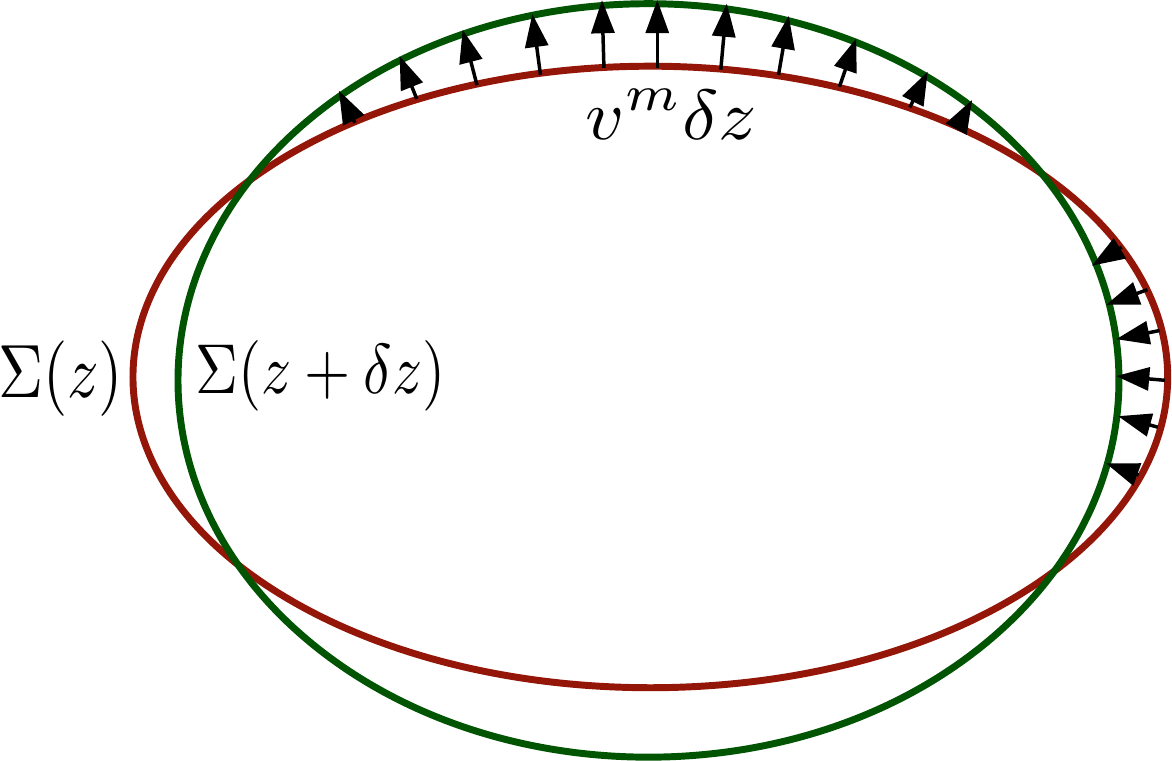}
\end{center}
\caption{\footnotesize Infinitesimal variation of $\Sigma$ along its moduli space.
\label{deltasigma}}
\end{figure}

Kaluza-Klein reduction of the D4 produces a supersymmetric quantum mechanics with $d(\Sigma)$ complex continuous degrees of freedom $z^a$ and $b_2(\Sigma)$ discrete flux degrees of freedom $S_i$. The moduli $z^a$ parametrize a moduli space $\CM$. The discrete fluxes can be thought of as quantized momenta dual to periodic coordinates that do not explicitly appear in the Hamiltonian. The wrapped D-brane with $F=0$ preserves 4 supercharges; it is the dimensional reduction from 4 to 1 of a four dimensional $\CN=1$ theory. In a sector with fixed flux we therefore expect the low energy effective action for the $z^a$ to be of the general chiral multiplet form \cite{Wess1992}\footnote{Note that this is of the form (\ref{curvlag}) (in its Lorentzian version) with $h=\frac{1}{2}(W + \bar{W})$. The special form of $h$ is due to the presence of four supersymmetries $Q_\alpha$, $Q_\alpha^\dagger$, $\alpha=1,2$, rather than  the generic minimal amount of 2 supercharges.}
\begin{equation} \label{D4QMgenform}
 S =  \frac{2 \pi}{g_s} \int dt \, \frac{1}{2} \left( g_{a\bar{b}} \dot{z}^a \dot{\bar{z}}^b 
 - g^{a \bar{b}} \partial_a W(z) \bar{\partial}_{\bar{b}} \bar{W}(\bar{z})
  \right) \, + \cdots \, ,
\end{equation}
where the $+ \cdots$ consists of terms independent of $z$, $W(z)$ is a holomorphic superpotential (absent when $F=0$ but generically nonzero when $F \neq 0$), and $g_{a \bar{b}}$ a K\"ahler metric:
\begin{equation}
 g_{a \bar{b}} = \partial_a \bar{\partial}_{\bar{b}} \, K(z,\bar{z}) \, .
\end{equation}
We have explicitly retained the overall prefactor $\frac{2 \pi}{g_s}$ instead of absorbing it in the metric and superpotential. In this way $g_s$ can be thought of as Planck's constant in the supersymmetric quantum mechanics.

To identify the metric, we should take the brane to be slowly moving along its moduli space and expand the first term in (\ref{SD4NSNS}) to second order in the velocities. An infinitesimal displacement $\delta z^a$ along the moduli space causes an infinitesimal normal displacement $\delta y^m = v_a^m \delta z^a$ of $\Sigma$ inside $X$, where the $v_a^m$ depend holomorphically on the coordinates of $\Sigma$. This is illustrated in fig.\ \ref{deltasigma}. Thus, when slowly moving along the moduli space with velocity $\dot{z}^a$, the first term in (\ref{SD4NSNS}) becomes
\begin{equation}
 -\int_{\IR \times \Sigma} d^5 x \, \sqrt{-h} =  \int dt \, 
 \int_{\Sigma} dV \left( -1 + \frac{1}{2} g_{m \bar{n}}  v^m_a \bar{v}^{\bar{n}}_{\bar{b}} \, \dot{z}^a \dot{\bar{z}}^{\bar{b}} \right) \, ,
\end{equation}
where $dV$ is the volume element on $\Sigma$. Choosing the orthonormal frame  $\theta$ appearing in (\ref{JOm}) such that $\theta^1$ and $\theta^2$ lie along $\Sigma$ (i.e.\ they span $T^* \Sigma$) while $\theta^3$ is normal to it, we can write $dV = \left(\frac{i}{2} \theta^1 \wedge \bar{\theta}^1 \right) \wedge \left(\frac{i}{2} \theta^2 \wedge \bar{\theta}^2 \right)$ so the kinetic term becomes
\begin{equation}
 \left( \frac{1}{8}  \int_{\Sigma}  \theta^1 \wedge \theta^2 v^3_a 
 \wedge \bar{\theta}^1 \wedge \bar{\theta}^2 \bar{v}^3_{\bar{b}}
 \right) \dot{z}^a \dot{\bar{z}}^{\bar{b}} =
 \left( \frac{1}{8}  \int_{\Sigma}  \omega_a \wedge \bar{\omega}_{\bar{b}} 
 \right) \dot{z}^a \dot{\bar{z}}^{\bar{b}} \, ,
\end{equation}
where
\begin{equation} \label{omdef}
 \omega_a = \Omega \cdot v_a = \theta^1 \wedge \theta^2 \, v^3_a
\end{equation} 
is the contraction of $\Omega$ with the vector field $v_a$. It is a well-defined holomorphic $(2,0)$ form on $\Sigma$. This map is an isomorphism between deformations of $\Sigma$ and harmonic $(2,0)$-forms on $\Sigma$. By comparing to (\ref{D4QMgenform}) we thus identify 
\begin{equation} \label{gabomab}
 g_{a\bar{b}} =  \frac{1}{4} \, \int_{\Sigma}  \omega_a \wedge \bar{\omega}_{\bar{b}} \, .
\end{equation}
\begin{figure}
\begin{center}
\includegraphics[height=7cm]{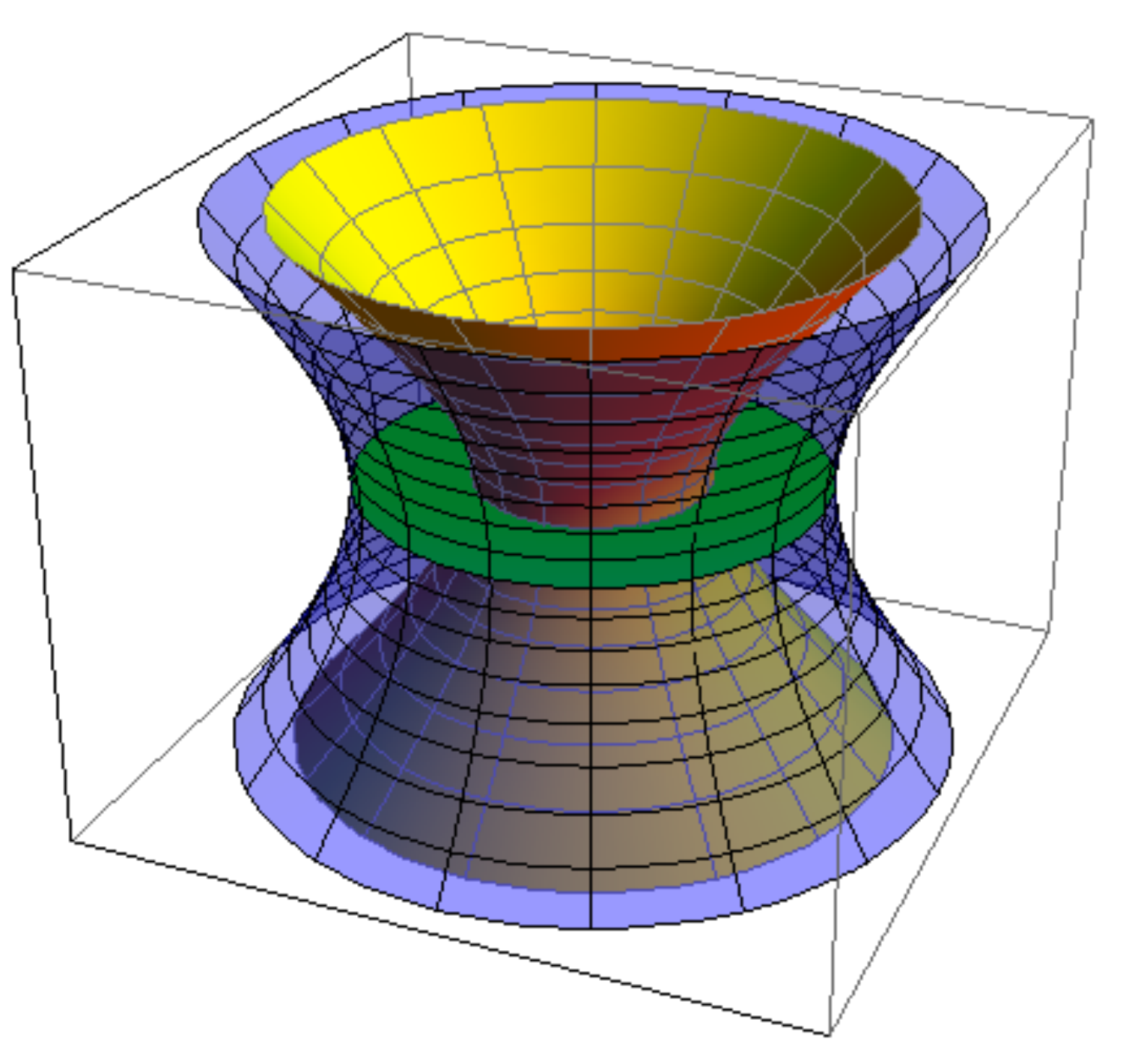}
\end{center}
\caption{\footnotesize The inner (yellow) hyperboloid represents $\Sigma(0)$, the outer (blue) one is $\Sigma(z)$, the green horizontal disk stretched between the two represents the 3-chain $\Gamma(z)$, and its boundary in $\Sigma(z)$ is $\sigma(z)$.
\label{nested}}
\end{figure}
To show that this is K\"ahler, we first expand the $\omega_a$ in the integral basis introduced in (\ref{sigmai}): 
\begin{equation} \label{omexp}
 \omega_a = \omega_{ai} \sigma^i \, , \qquad \omega_{ai} = \eta_{ij} \int_{\sigma^j} \omega_a \, .
\end{equation}
Now fix an arbitrary reference point $z^a \equiv 0$ in the moduli space $\CM$ parametrizing the holomorphic cycles $\Sigma(z)$. If we move away from this point, the 2-cycles $\sigma^i$ will sweep out 3-chains $\Gamma^i(z)$. This is illustrated in figure \ref{nested}. We can integrate the holomorphic 3-form over these 3-chains, producing the holomorphic 3-chain periods
\begin{equation}
 \Pi^i(z) = \int_{\Gamma^i(z)} \Omega \, .
\end{equation}
Because $\Sigma$ is holomorphic this is independent of the choice of representative of $\sigma^i$ inside $\Sigma(z)$. By the definition of the $\omega_a$, we then have
\begin{equation}  \label{dPiom}
 \partial_a \Pi^i(z) = \int_{\sigma^i(z)} \Omega \cdot v_a(z) = \int_{\sigma^i} \omega_a \, ,
\end{equation}
and combining this with (\ref{omexp}) and (\ref{gabomab}) gives us
\begin{equation}
 g_{a\bar{b}} =  \frac{1}{4} \, \partial_a \Pi^i \, \eta_{ij} \, \bar{\partial}_{\bar{b}} \bar{\Pi}^j = \partial_a \bar{\partial}_{\bar b} K(z,\bar{z}) \, , \qquad K = \frac{1}{4} \, \Pi^i \, \eta_{ij} \,  \bar{\Pi}^j \, .
\end{equation}
This makes it manifest that the metric is K\"ahler.

To identify the superpotential, we consider the second term in (\ref{SD4NSNS}), which for a given magnetic flux $F$ gives a potential energy 
\begin{equation} \label{potenergy}
 \int_{\Sigma} \frac{1}{2} F \wedge \star F =  \int_{\Sigma} \frac{1}{4}(F+\star F) \wedge (F+ \star F) -  \int_{\Sigma} \frac{1}{2} F \wedge F \, ,
\end{equation}
where $\star$ is the Hodge star operator on $\Sigma$ and we used $\star^2=1$. The second term is topological --- it does not depend on the $z^a$, and (\ref{QD0}) shows it is nothing but the D0-charge induced by the fluxes. Thus, the magnetic potential energy is bounded below by the flux induced D0-charge, with the bound saturated by anti-self-dual flux $F=-\star F$. In the following we will show that the first term is (almost) the $|\partial W|^2$ term of (\ref{D4QMgenform}). 

The $\star$ operator commutes with the decomposition of 2-forms according to their number of (holomorphic, antiholomorphic) indices.\footnote{This is true because $\Sigma$ is a K\"ahler manifold, being a complex submanifold of the K\"ahler manifold $X$.} It acts as $+1$ on $(2,0)$ forms: $\star \omega^{(2,0)} = \omega^{(2,0)}$. This is easily checked by considering a holomorphic orthonormal frame of $\Sigma$: $\star (\theta^1 \wedge \theta^2) = + \theta^1 \wedge \theta^2$. The complex conjugate $(0,2)$-forms are self-dual as well. For $(1,1)$-forms there are two cases to consider: $(1,1)$-forms proportional to the K\"ahler form are again self-dual, while all $(1,1)$ forms orthogonal to that are anti-selfdual, i.e.\ $\star=-1$. This is checked by considering $\theta^1 \wedge \bar{\theta}^1 + \theta^2 \wedge \bar{\theta}^2$ resp.\ $\theta^1 \wedge \bar{\theta}^1 - \theta^2 \wedge \bar{\theta}^2$.

Accordingly the first term in (\ref{potenergy}) is
\begin{eqnarray} \label{FFexp}
  \int_{\Sigma} \frac{1}{4} (F + \star F) \wedge (F + \star F) &=& \int_{\Sigma} 2 \, F^{(2,0)} \wedge F^{(0,2)} + F^{(1,1)}_+ \wedge F^{(1,1)}_+ \, .
\end{eqnarray}
The component $F^{(1,1)}_+$ is obtained by projection onto the K\"ahler form $J=J^A D_A$ (pulled back to $\Sigma$):
\begin{equation} \label{F11plusexp}
 F^{(1,1)}_+ = f_+ J \, , \qquad f_+ = \frac{1}{\int_{\Sigma} J^2} \int_{\Sigma} F \wedge J = \frac{1}{\int_{\Sigma} J^2} \, q_A J^A \, , \qquad \int_{\Sigma} J^2 = D_{ABC} N^A J^B J^C \, ,
\end{equation}
where we used (\ref{QD2}). This expression is again independent of the D4 moduli $z^a$. The second term in (\ref{FFexp}) thus equals $\frac{1}{\int_{\Sigma} J^2} (q_A J^A)^2$. Note this is independent of the overall scale of $J$.

The remaining term does depend on the moduli $z^a$ and we will see it can be identified with $|\partial W|^2$. Expand $F^{(2,0)}$ in the basis of $(2,0)$-forms defined in (\ref{omdef}):
\begin{equation} \label{F20decomp}
 F^{(2,0)} = f^a \omega_a \, , \qquad f^a = \frac{1}{4} \, g^{a \bar{b}} \int_\Sigma F \wedge \bar{\omega}_{\bar{b}} \, ,
\end{equation}
with $g_{a\bar{b}}$ defined in (\ref{gabomab}) and $g^{a \bar{b}}$ its inverse. Using the decomposition (\ref{sigmai}) and the relation to chain period derivatives (\ref{dPiom}), we write
\begin{equation}
 \int_\Sigma F \wedge \bar{\omega}_{\bar{b}} =S_i \int_{\sigma^i} \bar{\omega}_{\bar{b}} =  \bar{\partial}_{\bar{b}} \left( S_i \bar{\Pi}^i(\bar{z}) \right) \, ,
\end{equation}
and the first term in (\ref{FFexp}) becomes $\frac{1}{2} g^{a \bar{b}} \partial_a ( S_i \Pi^i ) \bar{\partial}_{\bar{b}} ( S_j \bar{\Pi}^j )$. Comparing to (\ref{D4QMgenform}), we identify the flux induced superpotential 
\begin{equation}
 W(z) = S_i \Pi^i(z) \, .
\end{equation}
We put everything together below.

\subsection{D4 supersymmetric quantum mechanics} \label{sec:D4susyQM}

\subsubsection{Bosonic action}

To summarize, the low energy degrees of freedom at weak coupling and large volume of a generic\footnote{Generic means here that the brane worldvolume is smooth and the gauge group is $U(1)$. Coincident branes will give rise to an enhanced nonabelian gauge symmetry and intersecting branes will give rise to additional light bifundamentals. We also do not include pointlike D0-branes yet. We will return to this in section \ref{sec:constrsusy}.} D4-brane wrapped on a 4-cycle $\Sigma=N^A D_A$ in a generic Calabi-Yau 3-fold with triple intersection numbers $D_{ABC}$ and K\"ahler form $J=J^A D_A$ ($J^A, N^A > 0$) are $d = b_{2+} - 1$ complex deformation moduli $z^a$ and $b_2 = \chi-2$ integers $S_i$ parametrizing the $U(1)$ flux $F=S_i \sigma^i$, with $b_{2+} = \frac{1}{6} D_{ABC} N^A N^B N^C + \frac{1}{12} c_{2,A} N^A $ and $\chi = D_{ABC} N^A N^B N^C + c_{2,A} N^A$. Up to a constant energy term the bosonic part of the action is
\begin{eqnarray}
 S &=& S_{\rm dyn} +  S_{\rm top} \\
 S_{\rm dyn} &=& \frac{2 \pi}{g_s} \int dt \left[ \frac{1}{2} g_{a\bar{b}} \dot{z}^a \dot{\bar{z}}^b 
 - \frac{1}{2} g^{a \bar{b}} \partial_a W \bar{\partial}_{\bar{b}} \bar{W} \right] \label{Sdynn} \\
    S_{\rm top} &=&  - \frac{2 \pi}{g_s} \int dt \left[ \left( q_{0} + (q_A \tilde{J}^A)^2  \right) + g_s \left( q_0 \, \phi^0  - q_A \, \phi^A \right) \right] \, ,
\end{eqnarray}
with superpotential and K\"ahler metric given by
\begin{eqnarray}
 W(z) &=& S_i \Pi^i(z) , \qquad \Pi^i(z) = \int_{\Gamma^i(z)} \Omega \\
 g_{a \bar{b}} &=& \partial_a \bar{\partial}_{\bar{b}} \bigl( \frac{1}{4}  \Pi^i \, \eta_{ij} \, \overline{\Pi^j} \bigr) \, , \label{gabrev}
\end{eqnarray}
flux-dependent D0- and D2-charges
\begin{eqnarray}
q_0 &=& -\frac{\chi}{24} -\frac{1}{2} \eta^{ij} S_i S_j = -\frac{\chi}{24} - \frac{1}{2} \int_{\Sigma} F \wedge F \, \\
q_A &=& D_A^i S_i = \int_{\Sigma} D_A \wedge F \, ,
\end{eqnarray}
and unit K\"ahler form
\begin{eqnarray} \label{tildeJdef}
 \tilde{J} &=& \frac{1}{(D_{ABC}N^A J^B J^C)^{1/2}} \, J \, =
 \frac{1}{(\int_\Sigma J^2)^{1/2}} \, J \, .
\end{eqnarray}
The matrix $\eta^{ij}=\#(\sigma^i \cap \sigma^j)$ is the intersection form of $\Sigma$ and $D_A^i = \#(D_A \cap \sigma^i)$. The 3-chains $\Gamma^i(z)$ are 3-chains swept out by the 2-cycles $\sigma^i$ on $\Sigma$ when moving from a fixed reference point to $z$ in the moduli space. We recall that for the quintic $D_{111}=5$, $\tilde{J}^1 = \frac{1}{\sqrt{5 N}}$, $c_{2,1}=50$. 

This is a reliable description at weak coupling and low energies. What exactly does weak coupling mean? 
We recall from our general discussion of supersymmetric quantum mechanics in section \ref{sec:susyQM} that weak coupling corresponds to a limit in which the superpotential $h$ is scaled up as $h \to \lambda h$ with $\lambda \to \infty$, while keeping the metric constant. To see how this relates to other scalings, consider the simplest case of a single variable with bosonic Lagrangian $L = A \dot{x}^2 + B (dh/dx)^2$. Then we can redefine $x=y/\sqrt{A}$ such that $L=\dot{y}^2+AB (dh/dy)^2$. So weak coupling means $AB \to \infty$. Applying this to the case at hand and remembering that because of (\ref{OmJnormalization}) the metric and periods scale as $g_{a\bar{b}} \propto V_X$, $\Pi^i \propto \sqrt{V_X}$ with the volume of $X$, we have effectively $A \propto V_X/g_s$ and $B \propto 1/g_s$. Thus weak coupling means
\begin{equation} \label{weakcoupling}
 \frac{V_X}{g_s^2} \to \infty \, .
\end{equation}
The left hand side is essentially the ratio of the string length squared over the 4d Planck length squared, $\ell_s^2/\ell_4^2$. Hence the weak coupling regime is the regime in which the string length is much larger than the 4d Planck length, and the low energy supergravity description breaks down.

We see from (\ref{Sdynn}) that switching on the flux quanta $S_i$ generates a highly complex potential energy function for the $z^a$, leading at fixed $q_0$ and $q_A$ (and large $q_0$) to a vast energy landscape with exponentially many minima, parametrized by the values of $S_i$ and $z^a$ for which $\partial W_S(z) = 0$. This is kinematically very similar to the landscape of flux vacua in string theory, except of course that here we have a quantum mechanical system rather than a universe.

The potential is in general not a single valued function on the moduli space $\CM$, due to possible monodromies $\sigma^i \to {M^i}_j \sigma^j$ of the 2-cycle basis when going around noncontractible loops in $\CM$. It does become single valued on the covering space $\widetilde{\CM}$, also known as Teichm\"uller space, but then one has to quotient the theory by the covering group, which acts as a discrete gauge symmetry on $z^a$ and $S_i$. The metric $g_{a \bar{b}}$ and the charges $q_0$ and $q_A$ on the other hand are single valued, since they can be expressed without reference to the basis $\sigma^i$.

The different zero energy classical minima of the potential can mix quantum mechanically through tunneling instantons. By completing the squares in the Euclidean version of the action one finds a bound $S_{\rm dyn}^E \geq \frac{2 \pi}{g_s} \, {\rm Re}(e^{-i\alpha} \Delta W)$ for any real $\alpha$, saturated when
\begin{equation} \label{thetainst}
 \dot{z}^a = - e^{-i \alpha} g^{a \bar{b}} \bar{\partial}_{\bar b} \bar{W}(\bar{z}) \, .
\end{equation}
The strongest bound is obtained with $\alpha = \arg (\Delta W)$, so this is what we should take $\alpha$ to be in order to be able to find instanton solutions. The instanton action is then
\begin{equation} \label{SDW}
 S_{\rm dyn}^E = \frac{2 \pi}{g_s} |\Delta W| \, ,
\end{equation}
and the trajectories are straight lines when projected to the $W$-plane.

\subsubsection{Geometry of moduli space} \label{sec:geomM}

Before turning to the supersymmetric completion of the model, it may be useful to quickly review some basic formulae in K\"ahler geometry as well as more specialized expressions applicable to the model under consideration. When a metric is K\"ahler, i.e.\ $g_{a \bar{b}} = \partial_a \bar{\partial}_{\bar b} K$, all Christoffel symbols $\Gamma^K_{IJ} = \frac{1}{2} g^{KL}\left( \partial_I g_{LJ} + \partial_J g_{IL} - \partial_L g_{IJ} \right)$ vanish except when all indices are holomorphic or all indices are anti-holomorphic. When all indices are holomorphic, we have
\begin{equation}
 \Gamma^a_{bc} = g^{a \bar{d}} \partial_b g_{c \bar d} = \tfrac{1}{4}  \bar{\partial}^{a} \bar{\Pi}^i \, \partial_b \partial_c \Pi_i = \tfrac{1}{4} \int_X \bar{\omega}^a \wedge \partial_b \omega_c \, .
\end{equation}
We are lowering and raising indices with $g_{a \bar b}$ and $\eta_{ij}$. The first expression is true for general K\"ahler manifolds, the remainder specializes to (\ref{gabrev}). Equivalently, $\int \bar{\omega}_{\bar a} \wedge \nabla_b \omega_c = 0$. The Riemann curvature, defined by $[\nabla_I,\nabla_J] X^K={R_{IJL}}^K X^L$, also simplifies: 
\begin{equation} \label{RabcdPiPi}
 R_{a\bar{b} c \bar{d}} = g_{e \bar{d}} \bar{\partial}_{\bar b} \Gamma^e_{ac} = \tfrac{1}{4} \bar{\nabla}_{\bar b} \bar{\partial}_{\bar d} \bar{\Pi}^i \,  \nabla_a \partial_c \Pi_i =
 \tfrac{1}{4} \int_X \bar{\nabla}_{\bar b} \bar{\omega}_{\bar d} \wedge \nabla_a \omega_c \, .
\end{equation}
Again, the first equation is valid for general K\"ahler manifolds. Components of the curvature tensor not related by symmetries to the above ones (such as $R_{ab c \bar{d}}$) all vanish. Finally, in addition to the usual curvature symmetries $R_{IJKL}=-R_{JIKL}=-R_{IJLK}=R_{KLIJ}$ and $R_{I[JKL]}=0$, a K\"ahler manifold also satisfies $R_{a\bar{b}c\bar{d}}=R_{c\bar{b}a\bar{d}}$.

Geometrically, the fact that $\int \bar{\omega}_{\bar a} \wedge \nabla_b \omega_c = 0$ means that $\nabla_b \omega_c$ is of type $(1,1)$. Indeed, taking a derivative of a $(2,0)$-form turns it into a $(2,0)$-form plus a $(1,1)$-form (because the pullback of a $(2,0)$ form from $\Sigma(z)$ to $\Sigma(z+\delta z)$ will at most produce one extra antiholomorphic leg to first order in $\delta z$). The same is then true for $\nabla_b \omega_c$, and $\int \bar{\omega}_{\bar a} \wedge \nabla_b \omega_c = 0$ further implies that the $(2,0)$ part is actually zero, leaving only a $(1,1)$ part. 

Keeping in mind the $(p,q)$-type of various forms, the following orthogonality properties hold:
\begin{equation}
 \partial_a \Pi^i \,  \partial_b \Pi_i = 0 \, , \qquad \partial_a \Pi^i \,  \nabla_b \partial_c \Pi_i = 0 \, , \qquad \partial_a \Pi^i \, D_{Ai} = 0 \, .
\end{equation}
Further orthogonality properties can be derived from these by differentiation, for example $\nabla_b \partial_a \Pi^i \, D_{Ai} = 0$.

\subsubsection{Hamiltonian and supersymmetric completion}

To find the supersymmetric completion of this model, we could either reduce the fermionic part of the full D4-action, or we can infer it from the structure of the bosonic part. We will go for the latter route. To do so, we first write down the bosonic Hamiltonian derived from the action given earlier. This is $H^{(0)}= H^{(0)}_{\rm dyn} + H^{(0)}_{\rm top}$ with
\begin{eqnarray}
 H^{(0)}_{\rm dyn} &=&  \tfrac{g_s}{\pi} \, g^{a \bar{b}} p_a \bar{p}_{\bar{b}}  
 + \tfrac{\pi}{g_s} \, g^{a \bar{b}} \, \partial_a \Pi^i S_i \, \bar{\partial}_{\bar{b}} \bar{\Pi}^j S_j , \label{Hdynn} \\
 H^{(0)}_{\rm top} &=& \tfrac{2 \pi}{g_s} \bigl( q_0 + (\tilde{J}^A q_A)^2  \bigr) \, .
\end{eqnarray}
Here $p_a = \frac{\pi}{g_s} g_{a\bar{b}} \dot{\bar{z}}^a$ and we dropped the ``chemical potential terms'' proportional to $\phi^0$ and $\phi^A$ here. If we consider the $S_i$ as constants, this is of the form of a 4d $\CN=1$ theory of chiral superfields reduced to 1d, and the supersymmetric completion would be immediate. However the $S_i$ are \emph{not} dynamically conserved in this system, since motion from a point $z$ in the moduli space $\CM$ to itself along a noncontractible loop will in general change $S_i \to {M_i}^j S_j$ due to monodromy. On the other hand $q_0$, $q_A$ and  $H^{(0)}_{\rm top}$ are monodromy invariant, and hence they are conserved. In particular for an instanton, the change $\Delta W$ that appears in (\ref{SDW}) is due to a change $\Delta z$ with constant $S_i$ on \emph{Teichmuller space}, but it will in general lead to a change $\Delta S$ when reduced back to the moduli space.\footnote{Incidentally, an open D2-instanton wrapping a special Lagrangian 3-chain $\Gamma$ in $X$ with boundary $\partial \Gamma = \sigma_+ - \sigma_-$ on $\Sigma$ has an action $S_{\rm D2}= \frac{2 \pi}{g_s} \, |\int_\Gamma \Omega| = \frac{2 \pi}{g_s} |\Delta W|$, where $\Delta W$ is the change in superpotential when $F \to F + \sigma_+ - \sigma_-$. This is exactly the same as for monodromy instantons, suggesting a possible identification of the two.} Therefore we should consider the $S_i$ to be dynamical. We can think of the $S_i$ as quantized momenta conjugate to angular coordinates $\vartheta^i$ that do not appear explicitly in the Hamiltonian, i.e.\ $S_i = - i \partial_{\vartheta^i}$. The $\vartheta^i$ can then be viewed as the potentials obtained by KK reduction of the 2-form potential on the D4 that is dual to the $U(1)$ gauge field. The appropriate metric for these coordinates can be read off from the above expression for the Hamiltonian, which can be cast in the form 
\begin{equation} \label{HSS}
 H^{(0)} = \tfrac{g_s}{\pi} \, g^{a \bar{b}} p_a \bar{p}_{\bar b}  + \tfrac{\pi}{g_s} g^{ij} S_i S_j  \, ,
\end{equation}
where we have defined a metric on flux space
\begin{equation} \label{gij}
 g^{ij} \equiv 2 \left( \Pi^i_{a} \bar{\Pi}^{ja} + \bar{\Pi}^i_{\bar{a}} \Pi^{j\bar{a}} + \tilde{J}^i \tilde{J}^j \right) - \eta^{ij} \, , \qquad \Pi^i_a \equiv \frac{1}{2} \partial_a \Pi^i \, ,
 \qquad \tilde{J}^i \equiv \tilde{J}^A D_A^i \, .
\end{equation}
In this expression we use $\eta_{ij}$ and $g_{a\bar{b}}$ for index raising and lowering. The three terms inside the brackets are projectors onto $(2,0)$-forms, $(0,2)$-forms and the $(1,1)$ direction parallel to the K\"ahler form; together they project to the space of self-dual forms. That these terms are properly normalized projectors follows from the relation (\ref{gabomab}) and from the definition (\ref{tildeJdef}). 

In fact this metric is nothing but the the Hodge $\star$-product, and the decomposition into projections is just (\ref{potenergy}) with (\ref{FFexp}) again. Consistent with this, we have that ${g^i}_j {g^j}_k = \delta^i_k$, which is the property $\star^2=1$. This also shows that the inverse of $g_{ij}$ is $g^{ij}=\eta^{ik} \eta^{jl} g_{kl}$. 

The above Hamiltonian is just that of a free supersymmetric quantum mechanics in a curved space. Accordingly we could construct the usual supercharges $Q=d$ and $Q^\dagger=d^\dagger$ and take $H = \frac{1}{2} \{Q,Q^\dagger\}$, i.e.\ the Laplacian. However then the $Q=0$ supersymmetric ground states of the system would have $H_{\rm dyn} = H_{\rm top} = 0$, which corresponds to D-brane states without any flux. This is too restrictive. We only need $H_{\rm dyn}=0$ for a state to be supersymmetric, so we need to find supercharges that square to $H_{\rm dyn}$ only. These are basically the dimensionally reduced 4d $\CN=1$ supersymmetry generators, except that we interpret the $S_i$ now as dynamical momenta: $S_i = - i \partial_i$. Introducing the fermionic operators $\bar{\psi}^a = dz^a \wedge$, $\bar{\psi}^{\bar a} = d \bar{z}^{\bar a} \wedge$ and their conjugates $\bar{\psi}^{\bar{a}}=(\psi^a)^{\dagger}$, $\bar{\psi}^a=(\psi^{\bar{a}})^{\dagger}$, we define the supercharges
\begin{equation}
 Q_- = {\sqrt{\tfrac{g_s}{\pi}}} \, \bar{\psi}^a \nabla_a + \sqrt{\tfrac{\pi}{g_s}} \, \bar{\psi}^{\bar a} \, \bar{\partial}_{\bar a}\bar{\Pi}^i \, \partial_i  \, , \qquad \bar{Q}_+ = \sqrt{\tfrac{g_s}{\pi}} \, \bar{\psi}^{\bar a} \bar{\nabla}_{\bar a} + \sqrt{\tfrac{\pi}{g_s}} \, \bar{\psi}^a \, \partial_a \Pi^i \, \partial_i \, .
\end{equation}
and their conjugates $\bar{Q}_- = (Q_-)^\dagger$, $Q_+ = (\bar{Q}_+)^\dagger$. Equivalently 
\begin{eqnarray}
 Q_- = \sqrt{\tfrac{g_s}{\pi}} \, \partial + \sqrt{\tfrac{\pi}{g_s}} \, \bar{\partial}\bar{\Pi}^i \, \partial_i  \, , \qquad \bar{Q}_+ = \sqrt{\tfrac{g_s}{\pi}} \, \bar{\partial} + \sqrt{\tfrac{\pi}{g_s}} \, \partial \Pi^i \, \partial_i \, .
\end{eqnarray}
The supercharges satisfy the extended supersymmetry algebra\footnote{Because a 4d $\CN=1$ theory has four supersymmetries, the system has twice as much supersymmetry as the minimal susy QM case studied in section \ref{sec:susyQM}. As ``the'' $Q$-operator (satisfying $Q^2=0$, $\{Q,Q^\dagger\}=H_{\rm dyn}$), we can therefore choose from a family of linear combinations of the supercharges. The phase $\alpha$ appearing in the instanton flow equation (\ref{thetainst}) determines which supersymmetry is preserved by the instanton, and therefore which supersymmetry gets corrected. 
}
\begin{equation} \label{Hsusydyn}
 \{Q_\alpha,\bar{Q}_\beta\} = \delta_{\alpha \beta} H_{\rm dyn}, \qquad
 \{Q_\alpha,Q_\beta\} = 0 = \{\bar{Q}_\alpha,\bar{Q}_\beta \} \, ,
\end{equation}
where $H_{\rm dyn}$ reduces to (\ref{Hdynn}) in the zero fermion number sector. This defines the summetric completion of $H^{(0)}_{\rm dyn}$. Working out e.g.\ $\{Q_-,Q_-^\dagger \}$ 
we get explicitly, acting on wave functions $\Phi(\bar{\psi},x)$ (i.e.\ differential forms):
\begin{equation} 
 H_{\rm dyn} = -\tfrac{g_s}{\pi} \bar{\nabla}^a \nabla_a  + \tfrac{g_s}{\pi} R_{a \bar{b} c \bar{d}} \, \bar{\psi}^{a} \bar{\psi}^{\bar b} \psi^{c} \psi^{\bar d}  
 - \tfrac{\pi}{g_s} \, \partial_a \Pi^i \bar{\partial}^a \bar{\Pi}^j \partial_i \partial_j \nonumber \label{Hdynexpl}
 - \bar{\psi}^a \psi^b \nabla_a \partial_b \Pi^i \, \partial_i 
 + \bar{\psi}^{\bar b} \psi^{\bar a}  \bar{\nabla}_{\bar a} {\bar \partial}_{\bar b} \bar{\Pi}^i \partial_i  \, .
\end{equation}
Here, again acting on differential forms, $\nabla_a = \partial_a + \Gamma^b_{ac} \bar{\psi}^c \psi_b$, and $p_a = -i \partial_a$, $S_i=-i\partial_i$. Evidently neither the $\vartheta^i$ nor their fermionic superpartners (i.e.\ something like $d \vartheta^i$) appear explicitly in the Hamiltonian. Nevertheless they do not completely decouple from the $z$-degrees of freedom, again because of global monodromies --- the $\vartheta$-torus fibration over the moduli space $\CM$ has nontrivial identifications when going around nontrivial loops. 

To summarize, the full supersymmetric Hamiltonian including the RR potential terms is, in terms of the conserved charges:
\begin{equation}
 H = H_{\rm dyn} + \tfrac{2 \pi}{g_s} \bigl( q_0 + (\tilde{J}^A q_A)^2  \bigr) + 2\pi( \phi^0 q_0 + \phi^A q_A ) \, . \label{HHH1} 
\end{equation} 
Making the $\vartheta$ momenta $S_i$ explicit in the topological part of the Hamiltonian, this becomes
\begin{equation}
  H = H_{\rm dyn}  - 2 \pi \bigl(\tfrac{1}{g_s} + \phi^0 \bigr) \bigl(\tfrac{\chi}{24} + \tfrac{1}{2} \eta^{ij}S_i S_j  \bigr) + \tfrac{2 \pi}{g_s} (\tilde{J}^i S_i)^2 
 +  2 \pi \phi^i S_i \, . \label{HHH2} 
\end{equation} 
Unpacking the whole thing gives us 
\begin{eqnarray}
 H &=& -\tfrac{g_s}{\pi} \bar{\nabla}^a \nabla_a  + \tfrac{g_s}{\pi} R_{a \bar{b} c \bar{d}} \, \bar{\psi}^{a} \bar{\psi}^{\bar b} \psi^{c} \psi^{\bar d}   -
 2 \pi \mbox{$(\frac{1}{g_s} + \phi^0 ) \tfrac{\chi}{24}$} \nonumber \\
 && + \pi \bigl(\tfrac{1}{g_s} g^{ij} - \phi^0 \eta^{ij}\bigr) S_i S_j 
  + 2 \pi L^i S_i   
  \, . \label{HHH3}
\end{eqnarray}
Here $\phi^i = \phi^A D_A^i$, $g^{ij}$ is the metric (\ref{gij}), and 
\begin{equation}
 L^i = \phi^i
 - \frac{i}{2 \pi} \bar{\psi}^a \psi^b \nabla_a \partial_b \Pi^i 
 + \frac{i}{2 \pi} \bar{\psi}^{\bar b} \psi^{\bar a}  \bar{\nabla}_{\bar a} {\bar \partial}_{\bar b} \bar{\Pi}^i \, .
\end{equation}
It is often convenient to split quantities like $S_i$ and $L_i$ into self-dual and antiself-dual parts, as this leads to more transparent ``left-moving'' and ``right-moving'' expressions. Thus, for instance
\begin{equation}
\bigl(\tfrac{1}{g_s}  g^{ij} - \phi^0 \eta^{ij}\bigr) S_i S_j  = (\tfrac{1}{g_s}-\phi^0) S_+^2 - (\tfrac{1}{g_s}+\phi^0) S_-^2 \, ,
\end{equation}
where squares denote contraction with $\eta$, i.e.\ $X^2=X_i X^i = \eta^{ij} X_i X_j$ (so $X_+^2>0$, $X_-^2<0$). The projectors introduced in (\ref{gij}) can of course be used to give an explicit expression for $X_+$ and hence also for $X_-=X-X_+$.

\subsubsection{Supersymmetric Lagrangian}

Legendre transforming the Hamiltonian, with $\dot{z}^a=\frac{\partial H}{\partial p_a}$, $\dot{\vartheta}^i = \frac{\partial H}{\partial S_i}$, we get the corresponding Lagrangian 
\begin{eqnarray}
 \CL &=&  \frac{\pi}{g_s} g_{a \bar{b}} \dot{z}^a \dot{\bar{z}}^{\bar b} + i \bar{\psi}_{a} D_t{\psi}^{a} + i \bar{\psi}_{\bar b} D_t{\psi}^{\bar b} 
 -  \frac{g_s}{\pi} R_{a \bar{b} c \bar{d}} \, \bar{\psi}^{a} \bar{\psi}^{\bar b} \psi^{c} \psi^{\bar d} 
 + 2 \pi (\tfrac{1}{g_s} + \phi^0 ) \frac{\chi}{24}  \nonumber
 \\
 &&+ \frac{1}{4 \pi(\frac{1}{g_s}-\phi^0)} 
  \bigl( \dot{\vartheta}_+ - 2 \pi L_+ \bigr)^2
 - \frac{1}{4 \pi(\frac{1}{g_s}+\phi^0)}
 \bigl( \dot{\vartheta}_- - 2 \pi L_- \bigr)^2 \, , \label{Lagfinal}
\end{eqnarray}
where $D_t \psi^a = \dot{\psi}^a + \dot{z}^b \Gamma^a_{bc} \psi^c$. Note that $L_+=\phi_+$ as $(\nabla_a \partial_b \Pi)_+ = 0$.

\subsection{Constructing supersymmetric flux configurations} \label{sec:constrsusy}


\label{sec:explconstr}

We have constructed the supersymmetric quantum mechanics describing the low energy physics of smooth D4-branes with $U(1)$ gauge symmetry and abelian fluxes $F$ turned on. In the thermodynamic limit $N \to \infty$ we get $\CO(N^3)$ discrete flux degrees of freedom $S_i$ interacting with each other in a complicated, nonlocal way through coupling to $\CO(N^3)$ continuous degrees of freedom $z^a$. The effective Hamiltonian (\ref{HSS}) is somewhat reminiscent of models for spin glasses or the Hopfield model for neural networks, but more intricate, and a priori without quenched disorder, making it more like a structural glass than a spin glass.\footnote{A spin glass has quenched disorder built into the Hamiltonian. A structural glass has a simple Hamiltonian, and spontaneously generates its own disorder. This includes the material normal people refer to when they use the word glass.} In any case this suggests the presence of many effectively stable states. We will now make this quantitative, ignoring for the time being the possibility of brane degenerations and nonabelian gauge group enhancement. 

\begin{figure}
\begin{center}
\includegraphics[height=7cm]{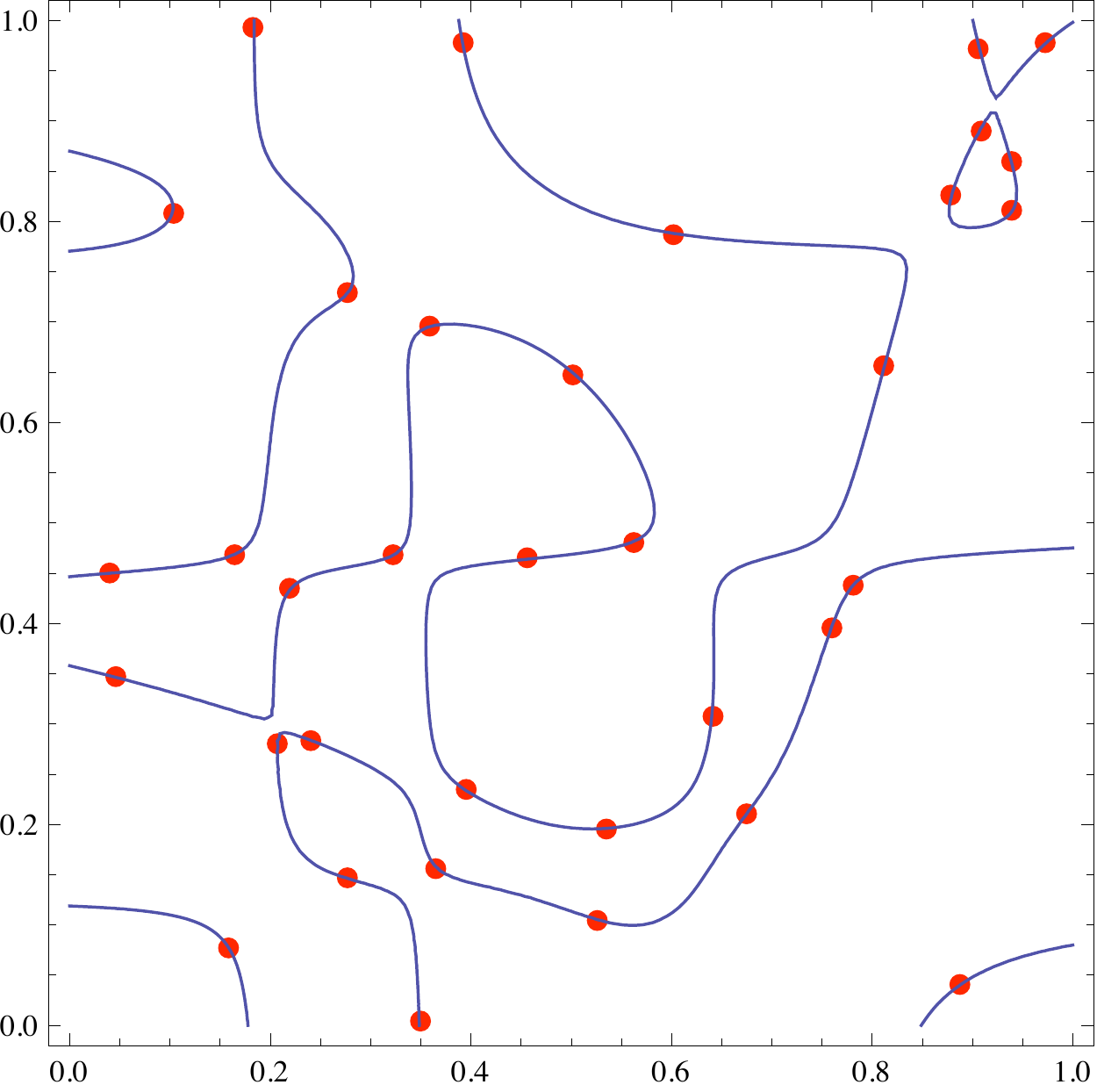}
\end{center}
\caption{\footnotesize Analog in 2d of the problem of fitting a holomorphic 4-cycle to contain a given collection of rigid holomorphic 2-spheres in a 6-dimensional space: here we are fitting a 1-cycle given by a degree 5 polynomial in $x$ and $y$ to contain 36 randomly generated points, thus fixing all moduli of the 1-cycle. 
\label{curvefit}}
\end{figure}

Classical supersymmetric brane configurations are those for which $\Pi_a^i S_i =0$, i.e.\ $\partial W_F(z) = 0$. As the 3-chain periods are very hard to compute explicitly, it might seem hopeless to try to construct any of those configurations explicitly. But a little thought shows otherwise. The condition $\partial W_F=0$ is equivalent to
\begin{equation}
 F^{(0,2)} = F^{(2,0)} = 0 \, .
\end{equation}
In other words if we switch on a flux $F=S_i \omega^i$, the brane moduli must adapt themselves to make the flux of type $(1,1)$. A geometric interpretation of this is that the flux becomes Poincar\'e dual to a linear combination of holomorphic 2-cycles. Indeed the Poincar\'e dual of a holomorphic 2-cycle $\sigma$ is clearly of type $(1,1)$ since for every $(2,0)$-form $\omega$ we have $\int \sigma \wedge \omega = \int_\sigma \omega = 0$. The last equation is true because $\sigma$ is the image of some map $y^m(u)$ with $u$ a single complex coordinate, and so the pullback of a $(2,0)$ form will be proportional to $du \wedge du = 0$. Thus, any linear combination (with arbitrary signs) of fluxes Poincar\'e dual to holomorphic 2-cycle will be of type $(1,1)$. The converse is also true as a special (proven) case of the Hodge conjecture. The subspace of moduli space where extra integral harmonic forms become of type $(1,1)$ is sometimes called the (generalized) Noether-Lefschetz locus in algebraic geometry. 

This interpretation gives a relatively simple way of explicitly constructing classical supersymmetric ground states. Let us illustrate this with an example. Consider again the quintic Calabi-Yau $Q_5(x)=0$ and let us assume that we have chosen the polynomial $Q_5$ such that it contains the following holomorphic 2-cycle, parametrized by $(u,v) \in \ICP^1$:
\begin{equation}
 \gamma: (x_1,x_2,x_3,x_4,x_5) = (u,-u,v,-v,0) \, .
\end{equation}
The Fermat quintic $\sum_i x_i^5=0$ for example contains $\gamma$. The topology of $\gamma$ is that of $\ICP^1$, i.e.\ it is a 2-sphere. Such ``degree 1 rational curves'' are generically isolated in a Calabi-Yau. This can be checked in the case at hand by simple counting: the most general linear map $(u,v) \to (x_1,\cdots,x_5)$, has 10 coefficients, 4 of those can be eliminated by linear coordinate transformations of $(u,v)$, and requiring $Q_5(x(u,v))=0$ for all $(u,v)$ produces 6 equations  on the remaining coefficients (namely the coefficients of $u^5$, $u^4v$, \ldots, $v^5$ after expanding out), so we generically expect a discrete solution set. Indeed in the generic quintic there are 2875 isolated degree 1 rational curves. 

Now consider our single wrapped D4 ($N=1$), described by a linear equation
\begin{equation}
 \Sigma:\sum_i a_i x_i = 0 \, .
\end{equation}
Requiring $\Sigma$ to contain $\gamma$ produces the equations $a_1=a_2$, $a_3=a_4$. Therefore, if we switch on a flux Poincar\'e dual to $\gamma$ on $\Sigma$, rigidity of the rational curve means that in order to preserve supersymmetry, these equations must remain satisfied when deforming $\Sigma$. Hence the D4 moduli space is reduced from $\ICP^4$ to $\ICP^2$. If we also require say the rational curve $\gamma':x=(0,u,-u,v,-v)$ to lie on $\Sigma$, enforced by switching on $F=\gamma'$, we end up with $a_1=a_2=a_3=a_4=a_5$, i.e.\ all moduli have been lifted. A simple 2d analog of this is illustrated in fig.\ \ref{curvefit}. By combining different pairs of the 2875 degree 1 rational curves we can thus explicitly construct $\frac{2875 \times 2874}{2} = 4131375$ different isolated supersymmetric configurations (assuming there are no duplicates or degenerates). In fact for each choice of pair $(\gamma,\gamma')$ there is an infinite number of different flux states, obtained by switching on a more general linear combination of the harmonic forms $\gamma$ and $\gamma'$:\footnote{Actually we should shift $F$ by the ``half-flux'' $\frac{1}{2} D_1$, because for $N=1$, $\Sigma$ is not spin \cite{Freed1999,Minasian1997}. This ensures proper charge quantization. We will ignore this here.} 
\begin{equation}
 F= k \gamma + k' \gamma' \, ,
\end{equation}
where $k,k' \in \IZ$. Of course these are not all degenerate in total energy. They all have $H_{\rm dyn}=0$ but they will have different $H_{\rm top}$. To compute this we should find the D2- and D0-charges. They are  
\begin{equation}
 q_1(F) = \int_\Sigma D_1 \wedge F = k + k'  \, , \qquad
 q_0(F) = -\frac{\chi}{24} -\frac{1}{2} \int_{\Sigma} F \wedge F = -\frac{\chi}{24} + \frac{3}{2} (k^2 + k'^2) \,,
\end{equation}
with $\chi=55$. Here we used the fact that the self-intersection number of $\gamma$ inside $\Sigma$ equals $-3$. This can be computed either by explicitly deforming $\gamma$ inside $\Sigma$ (necessarily nonholomorphically) and counting intersections, or by using some algebraic geometry (as in section 4.3 of \cite{Denef2007a}). The total energy $H=H_{\rm top}$ is given by
\begin{equation}
 \frac{g_s}{2 \pi} \, H= q_0 + \frac{q_1^2}{5} = -\frac{\chi}{24} + \frac{3}{2} (k^2 + {k'}^2) + \frac{1}{5}(k+k')^2 \, .
\end{equation}
Thus we get a nondegenerate lattice of vacua for each generic pair of degree 1 rational curves.

This can be generalized in various ways. Instead of degree 1 rational curves, we can consider degree $d$ rational curves. The same simple counting argument as before indicates again that they are generically isolated: there are $5(d+1)-4$ coefficients modulo reparametrizations, and there are $5d + 1$ equations, which happens to be the same number. The number of rational curves grows exponentially with the degree. There are for example $609250$ degree 2 curves and 704288164978454686113488249750 degree 10 curves. Furthermore, we can consider arbitrary D4-charge $N$. Requiring such a degree $N$ 4-cycle to contain a degree $d$ 2-cycle freezes $Nd+1$ moduli, out of $\frac{5 N^3+25 N}{6}$. So at large $N$ the 4-cycle can ``store'' up to $\frac{5N^2}{6d}$ degree $d$ rational curves by switching on the appropriate fluxes. If $d$ is sufficiently large so that the number of degree $d$ curves $N_d \sim e^{\kappa d}$ is much larger than this number, then this leads, naively at least, to up to $\sim e^{\frac{5 \kappa}{6} N^2}$ susy configurations.
Switching on one degree $d$ flux quantum increases $q_0$ by $\frac{d N}{2} + 1$, and $q_2$ by $\pm d$ depending on the sign of the flux (see section 4.3 of \cite{Denef2007a} for a derivation). So at large $N$ the energy in the maximal amount of $\frac{5N^2}{6d}$ stored curves is $\frac{g_s}{2 \pi} H \approx q_0 \approx  -\frac{\chi}{24} + \frac{5 N^2}{6d} \times \frac{d N}{2} \approx -\frac{5 N^3}{24} + \frac{5 N^3}{12} = \frac{5 N^3}{24}$.\footnote{Notice in particular that the energy becomes positive before all moduli are frozen. If what we were studying were not D4-branes but D7-branes in IIB orientifolds, this would basically mean that we cannot fix all moduli in this way while respecting tadpole cancelation, going against common genericity arguments. This is not conclusive though since this construction certainly does not exhaust all possibilities.}


Although the above discussion is for weak coupling, there is strong evidence that some of this structure carries over the strong coupling black hole regime, in the form of the structure of multicentered black hole bound states (cf section \ref{sec:BHduals}). We refer to \cite{Denef2007a} for more discussion on this.


\subsection{Counting supersymmetric states}

Although the above construction is explicit, and the number of ground state configurations that can be built in this way is huge, it is still only a small subset, especially at large $q_0$. In this regime the generic ground state configuration is isolated, and their total number and distribution over the moduli space can be computed using the statistical methods to count flux vacua developed in \cite{Douglas2003,Ashok2004,Denef2004a} and reviewed in section 6 of \cite{Denefb}. Those methods map the classical critical point counting problem, after making a continuum approximation, to the computation a supersymmetric finite dimensional integral, essentially a finite dimensional version of what we did is section \ref{sec:susystochastic}. In the present case we already have a supersymmetric quantum mechanics to start from. Not surprisingly, it is closely related to the effective supersymmetric quantum mechanics used in the statistical approach. 

In what follows we will consider the problem by applying the general machinery of supersymmetric quantum mechanics reviewed is section \ref{sec:susyQM}. This clarifies and complements the results obtained by the methods reviewed in \cite{Denefb} (and applied to the case at hand in appendix G of \cite{Denef2007a}). In particular it allows us to go beyond the continuum approximation. The computations will get a little technical, for which I apologize. I chose to include them here because it may be useful for some readers to see a nontrivial example worked out, and because as far as I know this has not been done in the literature (the treatment has some overlap with \cite{Denef2007a}, but we will not use S-duality as an input, but rather derive it directly from the susy quantum mechanics). 


\begin{figure}
\begin{center}
\includegraphics[height=4cm]{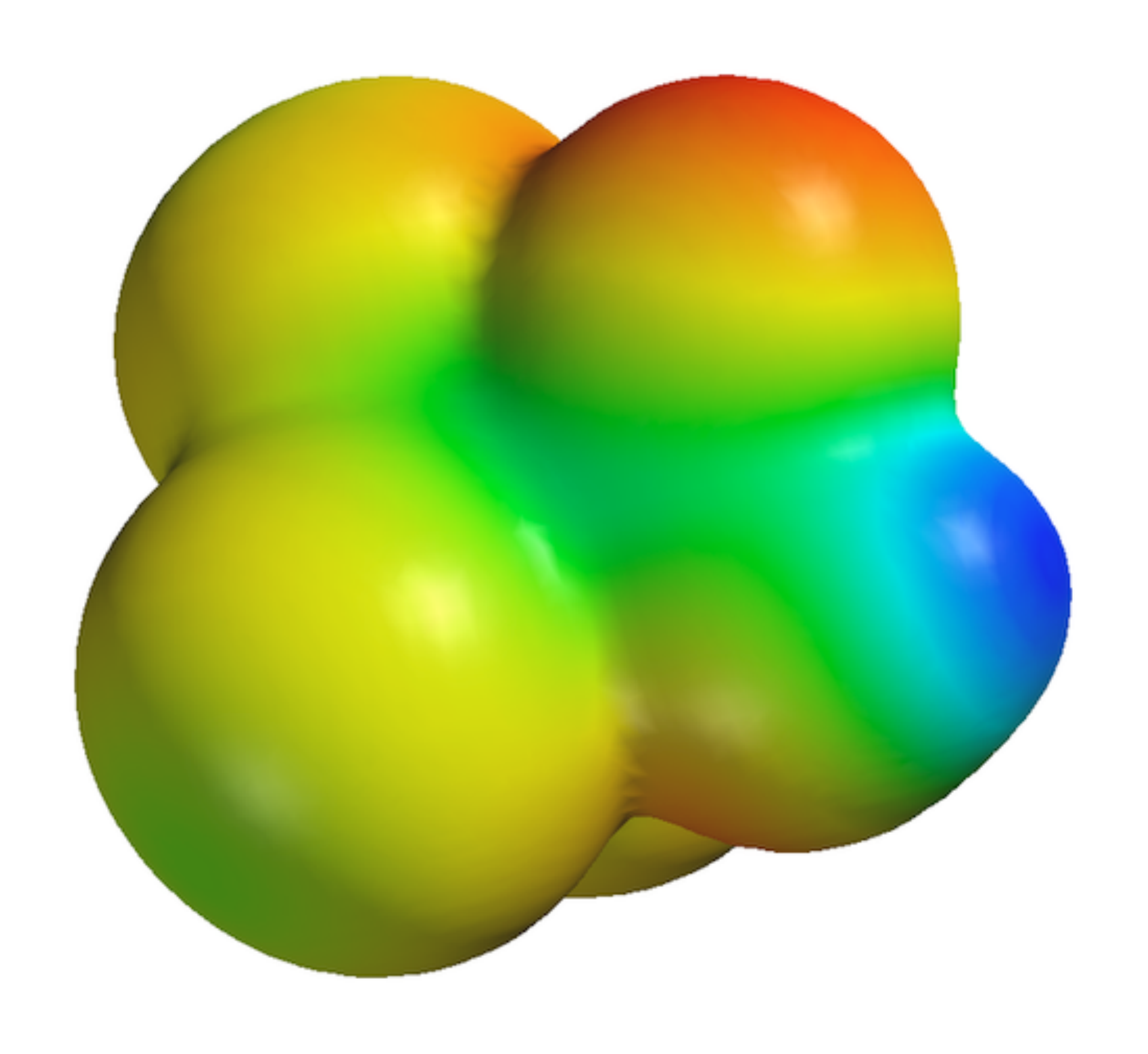}
\end{center}
\caption{\footnotesize Artist impression of a smooth abelian D4 configuration with $U(1)$ magnetic fluxes turned on (flux density represented by colors). 
\label{D4flux}}
\end{figure}

From our general consideration of supersymmetric quantum mechanics in the previous section, it is not hard to come up with a formula for the Witten index $\Omega(q)$ as a weighted sum of Euler characteristics of critical point loci $\CM_S$ of the superpotential $W_S$, summed over fluxes/momenta $S$ with the given total charge $q$:
\begin{equation} \label{OmS}
 \Omega(q) = \sum_{S} (-1)^{d-d_S} \chi(\CM_S) \, .
\end{equation}
The sign $(-1)^{d-d_S}$ appears because $m$ transversal dimensions contribute $m$ (the Morse index) to the fermion number. This is a useful formula for supersymmetric configurations with small D0-charge, but not for large D0-charge. We will try to extract the large charge asymptotics from the partition function. 

The supersymmetric partition function generating the Witten indices $\Omega(q)$ in each conserved charge sector is, with $H$ as in (\ref{HHH1})-(\ref{HHH3}):
\begin{equation} \label{Zdefin}
 Z(\beta,g_s,\phi) = {\rm Tr} \, {\scriptsize (-1)^F} \, e^{-\beta H(g_s,\phi)} \\
= \sum_{q} \Omega(q) \, e^{-\frac{2 \pi \beta}{g_s} \left( q_0 + (\tilde{J}^A q_A)^2 \right) - 2 \pi \beta (\phi^0 q_0 + \phi^A q_A)}  \, .
 \end{equation}
Introducing rescaled RR potentials 
\begin{equation}
  \varphi \equiv \beta \phi \, , 
\end{equation}  
we obtain the following expression for the Witten index $\Omega(q)$:
\begin{equation} \label{OmZ}
 \Omega(q) = \lim_{\beta \to 0} \int d \varphi \, e^{2 \pi \varphi \cdot q} \, Z(\beta,\varphi) \, ,
\end{equation} 
where the $\varphi$ integrals in principle are over a unit imaginary interval, e.g.\ from $-i/2$ to $i/2$. For computational purposes it is more convenient to work with integrals over the entire imaginary axis, because it allows easy implementation of contour integration techniques. Extending the integration ranges in the integrals above only produces trivial additional delta-function factors, which in the end can be stripped off manually. 



We will now try to compute the Witten indices from the Euclidean path integral representation of $Z$ in the limit $\beta \to 0$, analogous to the derivation leading to (\ref{eulerdensityformula}):  
\begin{equation}
 Z = \int \CD z \, \CD \psi \, \CD \vartheta \, e^{-S[z,\psi,\vartheta]} \, ,
\end{equation}
with periodic boundary conditions on all fields: $z^a(\tau+\beta)=z^a(\tau)$, $\psi^a(\tau+\beta)=\psi^a(\tau)$, $\vartheta^i(\tau+\beta)=\vartheta^i(\tau) + 2 \pi n^i$, with $n^i \in \IZ$.  
By Wick rotating (\ref{Lagfinal}) to rescaled Euclidean time
\begin{equation}
 t = -i \beta \tau \, ,
\end{equation}
we obtain the Euclidean action
\begin{equation}
 S = \int_0^1 d\tau \biggl( \frac{\pi}{g_s \beta} \, |\dot{z}|^2  + \bar{\psi} D_t{\psi}  + \frac{g_s \beta}{\pi} \, R \, \bar{\psi} \bar{\psi} \psi \psi - 2 \pi  u  \, \frac{\chi}{24}
 + \frac{1}{4 \pi v } 
  \bigl( \dot{\vartheta} + 2 \pi i \ell \bigr)_+^2
 - \frac{1}{4 \pi u }
 \bigl( \dot{\vartheta} + 2 \pi i \ell \bigr)_-^2
   \biggr) , \label{SeuclhighT}
\end{equation}
where we suppressed indices for clarity, denoted $\ell \equiv L/\beta$, i.e.\
\begin{equation} \label{tildeLdef}
 \ell^i \equiv  \varphi^i 
 - i \beta \, \tfrac{1}{2\pi} \bar{\psi} \psi \nabla \partial \Pi^i + i \beta \, \tfrac{1}{2\pi} \bar{\psi} \psi  \bar{\nabla} {\bar \partial} \bar{\Pi}^i \, ,
\end{equation}
and introduced ``light cone'' coordinates
\begin{equation}
 u \equiv  \frac{\beta}{g_s}+\varphi^0  \, , \qquad v \equiv  \frac{\beta}{g_s}-\varphi^0  \, . 
\end{equation}
As in (\ref{indfinPI}), in the limit $\beta \to 0$, the path integral localizes to constant paths $(z,\psi)$, with the fluctuation determinants for $z$ and $\psi$ canceling each other. The path integral over $\vartheta$ does not localize to constant paths, since the $\vartheta$ kinetic terms remain finite when $\beta \to 0$ (for $\varphi^0 \neq 0$). The $\vartheta$ path integral is essentially the partition function of a free particle on a $b_2$-dimensional torus. Making the sum over different winding sectors $\vartheta^i(\tau+1)=\vartheta^i(\tau)+2 \pi n^i$ explicit, this is
\begin{equation} \label{windingsum}
  \sum_n 
  e^{-\frac{\pi}{ v } 
  (n + i \ell)_+^2
 + \frac{\pi}{ u } (n + i \ell)_-^2}
  \int \CD \vartheta \, e^{-\frac{1}{4 \pi} \int_0^1 d\tau \, 
  ( \frac{1}{ v } 
  \dot{\vartheta}_+^2
 - \frac{1}{ u } \dot{\vartheta}_-^2 )}\, .
\end{equation}
The first factor corresponds to the action weights for the classical straight line trajectories in each sector, and the 
remaining path integral corresponds to the fluctuations from the classical paths and thus has strictly periodic boundary conditions $\vartheta^i(\tau+1)=\vartheta^i(\tau)$. This is just the standard Euclidean free particle propagator integrated over the $\vartheta$-torus:
\begin{eqnarray}
 \int \CD \vartheta \, e^{-\frac{1}{4 \pi} \int_0^1 d\tau \, 
  [ \frac{1}{ v } 
  \dot{\vartheta}_+^2
 - \frac{1}{ u } \dot{\vartheta}_-^2 ]} 
 &=& u ^{-b_{2}^-/2} \, v ^{-b_{2}^+/2} \, .
\end{eqnarray}
By making use of (\ref{RabcdPiPi}) and the various orthogonality and (anti-)self-duality properties discussed there, as well as the projectors introduced in (\ref{gij}) to write explicit expressions for $n_+$ and $n_-=n-n_+$, the sum in (\ref{windingsum}) together with the 4-fermion curvature term in (\ref{SeuclhighT}) reduces after some tender and care to 
\begin{equation}
  \sum_n e^{-\frac{\pi}{  v  } (n + i\varphi)_{J}^2+\frac{\pi}{ u } (n + i \varphi)_{\perp J}^2} \,
  e^{-\beta\left(
  \frac{\pi}{ g_s u v  } |n \Pi'|^2
  -\frac{1}{ u } \,  (\bar{\psi} \psi \, n \Pi'' + {\rm cc}) 
  -\frac{g_s v }{\pi u } \, R \, \bar{\psi} \bar{\psi} \psi \psi
   \right) } \,,
\end{equation}
where we separated out a factor in the summand that is independent of the dynamical variables $z,\psi$, in which we also introduced the notation $X^i_J \equiv \tilde{J}^i X_i$ for the projection of $X$ in the direction of $J$, and $X_{\perp J}$ for its orthogonal complement (w.r.t.\ $\eta_{ij}$). Now consider the $z,\psi$ integral of the other factor:
\begin{equation}
 \int d^{2d}z \,d^d\psi \, d^d\bar{\psi} \, 
 e^{-\beta\left(
  \frac{\pi}{ g_s u v  } |n \Pi'|^2
  -\frac{1}{ u } \,  (\bar{\psi} \psi \, n \Pi'' + {\rm cc}) 
  -\frac{g_s v }{\pi u } \, R \, \bar{\psi} \bar{\psi} \psi \psi
   \right) }\, .
\end{equation}
This integral has the typical zero dimensional supersymmetric form with superpotential $W_n(z) = n^i \Pi_i(z)$. It localizes on the critical point locus $\CM_n$ of $W_n$, meaning the  Gaussian approximation to the integral is exact.\footnote{Actually, because $n$ can transform by a monodromy when going around a loop in moduli space, this only holds up to boundary terms for a given $n$, but these cancel between different values of $n$. To see localization more directly, note that since the Witten index is invariant under a change of $g_s$, we can make the integrand as sharply peaked on $\CM_n$ as we want, with vanishing boundary terms when $g_s \to 0$.} The Gaussian integral for quadratic fluctuations normal to $\CM_n$ produces a factor $\frac{g_s  u v  }{\beta}$ for each complex normal direction, while the corresponding normal fermions produce a factor $\frac{\beta^2}{ u^2}$. Together this gives a factor $ \frac{\beta v }{ u }$ for each normal direction. The tangential directions to $\CM_n$ remain integrated over, and the corresponding tangent fermion integral gives for each complex tangent dimension a factor $-\frac{g_s \beta v }{ \pi u } R$. Finally, there will be an overall path integral normalization factor $(2 \pi g_s \beta)^{-d}$ as in (\ref{indfinPI}). Altogether the above integral thus reduces to $(-1)^{d_n} e(\CM_n) \, v^d \, u^{-d}$, where $d_n \equiv \dim \CM_n$ and $e(\CM_n)$ is the Euler density on $\CM_n$, as in (\ref{eulerdensityformula}). The sign is physically meaningful; we expand on this below. 

Putting everything together, we find the generating function: 
\begin{equation} \label{Z2uv}
 Z = (u/v)^{1/2} \, u^{-b_2/2} \, e^{2 \pi u \frac{\chi}{24}} \sum_n (-1)^{d_n} \chi_n \, e^{-\frac{\pi}{  v  } (n + i\varphi)_{J}^2 + \frac{\pi}{ u } (n + i \varphi)_{\perp J}^2}  \, ,
\end{equation}
where we defined the differential geometric Euler characteristics
\begin{equation}
 \chi_n \equiv \int_{\CM_n} e(R) \, ,
\end{equation}
and we used $d=\frac{b_{2+}-1}{2}$. Note that since by definition $\Omega_n$ is nonzero only if $(n_{\perp J})_+$ can be made to vanish, all terms in this series have $n_{\perp J}^2 \leq 0$. 

Defining the dual charges $\tilde{q}_A = D_{Ai} n^i$, $\tilde{q}_0=-\frac{\chi}{24}-\frac{1}{2} n^2$, we can also express this as 
\begin{equation} \label{Zmodtr}
 Z =  (\tfrac{u}{v})^{1/2} \, u^{-b_2/2} \,  e^{2 \pi \frac{\chi}{24} (u-\frac{1}{u})-\frac{\pi}{u} \varphi_-^2 + \frac{\pi}{v} \varphi_+^2 } (-1)^d  \sum_{\tilde{q}} \Omega(\tilde{q}) \, e^{-\frac{2\pi}{u} \tilde{q}_0 - \pi (\frac{1}{u}+\frac{1}{v}) \tilde{q}_+^2 + \frac{2\pi i}{u} \tilde{q}_- \tilde{\varphi}_- - \frac{2\pi i}{v} \tilde{q}_+ \tilde{\varphi}_+ 
 }  \, ,
\end{equation}
with as in (\ref{OmS}) $\Omega(\tilde{q}) = \sum_{n} (-1)^{d_n-d} \chi_n$, summing over the values of $n$ with the given charges $\tilde{q}$. Comparing this to the original (\ref{Zdefin}), which can be written as
\begin{equation}
 Z = \sum_q \Omega(q) \,  e^{-2\pi u q_0 - \pi (u+v) q_+^2 - 2\pi q \varphi}  \, ,
\end{equation}
we see that what we have shown is roughly that $Z$ is a modular (Jacobi) form under the modular transformation $u \to 1/u$, $v \to 1/v$, transforming similar to an ordinary theta function (although what we have here is much more nontrivial than a theta function, due to the nontrivial moduli dynamics). The duality exchanges winding and momentum modes, and there are many ways of understanding it:  electromagnetic duality of the D-brane theory, S-duality of the parent string theory, modularity of the parent CFT in M-theory, T-duality of the torus, a version of Poisson resummation, etc. The modular transformation relates the large and small flux / D0-charge regimes.

In any case, we can now extract the large $q_0$ asymptotics of $\Omega(q)$ from $Z$. For simplicity we will consider here the case $q_A = 0$, but this is easily extended. Performing the Gaussian integrals over the $\varphi^A$ in (\ref{OmZ}) is easy. We get
\begin{equation}
 \int_{-i\infty}^{i \infty} d\varphi^1 \cdots d\varphi^{B_2} \, Z(u,v,\varphi) = c \, u^{-k} e^{2 \pi u \frac{\chi}{24}} \sum_n (-1)^{d_n} \chi_n \, e^{\frac{\pi}{ u } n_{\perp}^2} \, .
\end{equation}
Here $k \equiv \frac{b_2-B_2}{2}$ with $b_2=\dim H^(\Sigma)$, $B_2=\dim H^2(X)$, and $n_\perp$ is the component of $n$ orthogonal to all $\varphi = \varphi^A D_A$, i.e.\ the component of $n$ orthogonal to the pullback of $H^2(X)$ in $H^2(\Sigma)$. Explicitly $n_\perp^i = n^i - (n^j D^A_j)D_A^i$, where $D^A = D^{AB} D_B$, $D^{AB}$ being the inverse of $D_{AB} \equiv D_{ABC} N^C$, with $D_{ABC}$ defined in (\ref{DABC}). The constant $c$ is the Gaussian determinant factor: $c=(\det D_{AB})^{-1/2}$. For the example of the quintic we have $k=(5N^3+50N-3)/2$, $D_{11}=5 N$, $c=\frac{1}{\sqrt{5N}}$.
The sum over $n$ is trivially divergent because adding an arbitrary $m=m^A D_A \in H^2(X)$ to $n$ does not alter $n_\perp$. From the point of view of the integral over $\varphi^A$ this is due to the redundancy $\phi^A \to \phi^A-i m^A$, $n \to n + m^A D_A$, which we introduced ourselves a little earlier when we extended the integration domain from an interval to the full imaginary axis. The upshot is that we can take this into account simply by restricting the sum over $n \in H^2(\Sigma,\IZ)$ to a sum over the quotient $n \in H^2(\Sigma,\IZ)/H^2(\Sigma,\IZ)$. 

The $q_0 \to \infty$ asymptotics of $\Omega$ will be captured by the $u \to 0$ behavior of $Z$. In this regime all $n \neq 0$ exponential corrections can be dropped. The $\varphi^0$ integral can then be evaluated exactly by closing the contour. There is a pole of order $k$ at $u=0$ i.e.\ $\varphi^0=-\beta<0$. If $q_0' \equiv q_0 + \frac{\chi}{24}<0$ we must close the contour to the right and the integral vanishes. If $q_0'>0$, the contour is closed to the left, and the residue theorem gives
\begin{equation}
 \Omega(q_0,0) \approx 2 \pi |c| \, (-1)^d \chi(\CM) \,  \frac{(2 \pi q_0')^{k-1}}{(k-1)!} \sim \biggl( \frac{2 \pi e q_0'}{k} \biggr)^k \, .
\end{equation}  
The last approximation is valid for $k \gg 1$. It is obtained as a Stirling approximation, or equivalently as the saddle point approximation of the integral. The saddle point lies at $\varphi^0 = 2 \pi q_0'/k$.  This implies we need at least $q_0' \gg k$ to justify dropping  the $n \neq 0$ terms,  or equivalently $q_0 \gg \chi \sim N^3$. 

This expresses the number of supersymmetric flux configurations as the volume of a $2k$-dimensional shell of radius squared $2 q_0'$, which loosely (but not literally because $\eta_{ij}$ is indefinite) can be thought of as the shell in flux space for which the D0-charge $-\frac{\chi}{24} - \frac{1}{2} \eta^{ij} S_i S_j$ equals $q_0$ and the D2-charge is constrained to be zero. This reproduces the large $q_0$ asymptotics for the number of flux vacua found in various contexts in \cite{Douglas2003,Ashok2004,Denef2004a,Denefb,Denef2007a}, based on an approximation in which fluxes were taken to be continuous.

Note that $\chi(\CM) = \int_\CM e(R)$ is the Witten index for the pure D4 without the flux degrees of freedom. Since this space is topologically $\ICP^d$, one may expect $\chi(\CM)=\chi(\ICP^d)=d+1\sim N^3/6$. This is not obviously correct because of possible singularities in $\CM$, but arguments were given in \cite{Denef2007a} that this is nevertheless the correct identification. The sign factor $(-1)^d$ has a physical meaning: In the large $q_0$ limit, almost all supersymmetric configurations will be isolated critical points of the superpotential $W_S(z)$. Isolated critical points lead to susy states with fermion number equal to their Morse index, which here always equals $d$ because the superpotential is holomorphic. This explains the $(-1)^d$. The sign $(-1)^{d_n}$ in the terms with $n \neq 0$ further suggests (but does not prove) that these correspond to the contributions from non-isolated configurations, with $d-d_n$ residual moduli.

It is of course easy to get exponentially large numbers of flux configurations out of these estimates, even for modest charges. For example for the quintic with, say, $N=5$, $q_0=365$, we get $\Omega \sim 10^{500}$, widely considered to be a very large number. (Similar estimates in the context of mathematically very similar constructions of type IIB string vacua with D7-branes form the basis for suspicions that there exists a staggeringly huge landscape of string vacua sweeping out a for all practical purposes dense set in parameter space, giving a simple solution to the cosmological constant problem but obliterating hopes of top-down predictivity  \cite{Strominger1986,Bousso2000,Kachru2003,Susskind2003,Douglas2003,Ashok2004,Denef2004a,Douglas2007,Denefb}.) 

On the other hand note that when $2 \pi e q_0' < k$, the above estimates give exponentially \emph{small} estimates. Clearly then, in this regime, it must be that the $n \neq 0$ terms dominate the degeneracy. If it is indeed true, as suggested above, that $n \neq 0$ contributions can be identified with non-isolated configurations, this means that in this regime, non-isolated configurations become entropically dominant.

For $q_0' \ll k$ we could try to build up the spectrum along the lines of section \ref{sec:explconstr}, and this could in turn be used to infer the corrections in the regime $q_0' \gg k$. However in the intermediate regime $q_0' \sim k$, the system is extremely complex.\footnote{This happens to be the relevant regime for applications to type IIB string theory vacua \cite{Denefb,Denef2010a}. In this context space-localized D4-branes are replaced with space-filling D7-branes, carrying $U(1)$ fluxes inducing D5- and D3-charges. The D3-tadpole cancelation condition dictates that the total D3-charge vanishes, hence $q_0' = \frac{\chi}{24} \approx \frac{k}{12}$ and $\frac{2 \pi e q_0'}{k} \approx \frac{\pi e}{6} \approx 1.42$, barely above the threshold.}
Trying to count degeneracies in this regime may be like trying to compute the boiling point of water from first principles. In the case of water we can just measure the boiling point and be done with it. We let nature do the computation for us. In the case of D-branes, this would not appear to be an option. However, there is something analogous, provided we enlarge our task to the problem of counting \emph{all} D4-D2-D0 bound states, not just those corresponding to abelian flux configurations of the D4. In that case, we can construct the corresponding BPS black hole solutions, and simply read off their Bekenstein-Hawking entropy $S = \log \Omega(q)$. We let gravity do the computation for us! To include all entropically relevant D4-D2-D0 bound states we need in particular consider bound states with localized D0-branes. We turn to this next.

\subsection{Bound states with D0-branes}

\begin{figure}
\begin{center}
\includegraphics[height=4cm]{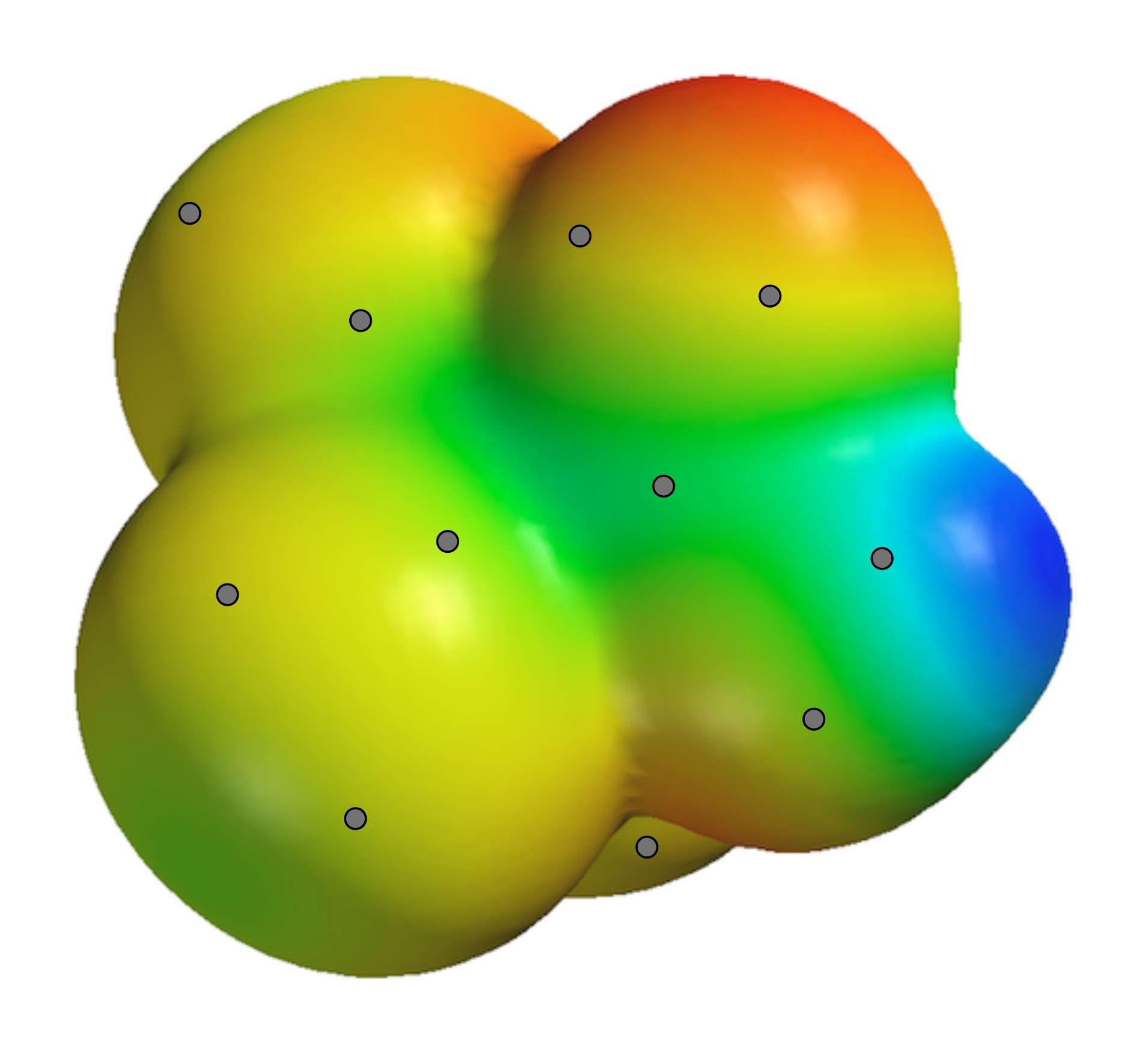}
\end{center}
\caption{\footnotesize Artist impression of a smooth abelian D4 configuration with $U(1)$ magnetic fluxes and mobile D0-branes bound to it. 
\label{D4fluxD0}}
\end{figure}

So far we have only considered D4-brane and $U(1)$ worldvolume flux degrees of freedom. A simple but entropically important extension is to include bound states with pointlike D0-branes. For a D4 wrapped on a fixed smooth $\Sigma$, the number of such bound states is easy to compute \cite{Vafa1997b}. We can think of the D0-branes as a gas of noninteracting particles. D0-branes form bound states among themselves of arbitrary D0-charge $k$ (the easiest way to see this is to consider their uplift to M-theory, where they are KK modes along the M-theory circle, $k$ being the KK momentum), so each particle in the D4-D0 gas is characterized by a charge quantum number $k>0$, as well as by the supersymmetric 1-particle state $|p\alpha\rangle$ it occupies. As we have seen in section \ref{sec:curvedspaces}, these 1-particle states are given by the harmonic differential forms on $\Sigma$. The label $p=0,1,2,3,4$ in $|p\alpha\rangle$ indicates the fermion number (form degree) and $\alpha=1,\ldots,b_p$. By the usual Fock space construction, an arbitrary multi-particle state can be represented by specifying occupation numbers $n_{k,p,\alpha}$, which take values in $\IZ^+$ if $p$ is even and in $\{0,1\}$ is $p$ is odd. In the absence of fluxes, the total D0-charge of the D4-D0 state is $Q=-\frac{\chi}{24} + \sum_{k,p,\alpha} k\, n_{k p \alpha}$ and the fermion number is $F=\sum_{k,p,\alpha} p \, n_{k p \alpha}$. Here we are still ignoring the moduli degrees of freedom of $\Sigma$, considering those frozen for the time being. The generating function for the degeneracies $d_{QF}$ of supersymmetric ground states of total D0-charge $Q$ and fermion number $F$ is then
\begin{eqnarray}
\sum_{QF} d_{QF} \, y^F t^Q &=& t^{-\frac{\chi}{24}} \sum_{\{n_{k\alpha}\}}
 y^{\sum_{kp\alpha} p \, n_{kp\alpha}} \, 
 t^{\sum_{kp\alpha} k \, n_{kp\alpha}}  \nonumber \\
 &=& t^{-\frac{\chi}{24}} \prod_{kp\alpha} \sum_n y^{n p} t^{n k} \nonumber \\
 &=& t^{-\frac{\chi}{24}} \prod_{k}  \frac{\prod_{p \, \rm odd} (1+y^p t^k)^{b_p}}{\prod_{p \, \rm even} (1-y^p t^k)^{b_p}} \, .
\end{eqnarray}
A generating function for the Witten indices $\Omega_{\rm D0}(Q) = \sum_F (-1)^F d_{QF}$ is obtained by setting $y=-1$ in the above:
\begin{equation}
 \sum_{Q} \Omega_{\rm D0}(Q) \, t^Q =  t^{-\frac{\chi}{24}} \prod_{k} (1-t^k)^{-\chi} = \eta(t)^{-\chi}  ,
\end{equation}
where $\chi = \sum_p (-1)^p b_p$ is the Euler characteristic of $\Sigma$, and $\eta(t)$ is the Dedekind eta-function. 

If we ignore the fact that there can be nontrivial interplay between the D4 and D0 moduli,\footnote{Ignoring this is actually not a good idea at small $q_0$, but at large $q_0$ it is a good approximation since most D4 flux configurations will be isolated anyway in this regime.} adding the D0-branes to the D4 simply has the effect of multiplying the partition function $Z$ in (\ref{OmZ}) by $e^{-2 \pi u \frac{\chi}{24}} \eta(e^{-2 \pi u})^{-\chi}$. Using the modularity property of the eta function, we furthermore have
\begin{equation} \label{etamodular}
 \eta(e^{-2 \pi  u })^{-\chi} = u^{\chi/2} \, \eta(e^{-\frac{2 \pi}{ u }})^{-\chi} \, .
\end{equation} 
Combining this with the D4 partition function (\ref{Z2uv}), using $\chi=b_2+2$, and for notational simplicity at the cost of slight formality setting $v = -u = -\varphi^0$ in (\ref{Z2uv}), we get, within our no D4-D0 interplay approximation:
\begin{eqnarray}
 Z_{\rm D4+D0} &=& \eta(e^{-2\pi \varphi^0})^{-\frac{\chi}{24}} \sum_n (-1)^{d-d_n} \chi_n \, e^{2\pi \varphi^0 \frac{n^2}{2} - 2 \pi \varphi^0 n } \\
 &=& \varphi^0 \, e^{\frac{2\pi}{\varphi^0} (\frac{\chi}{24} - \frac{\varphi^2}{2} )} \prod_k (1-e^{-\frac{2 \pi k}{ \varphi^0 }})^{-\chi}   \, \sum_n (-1)^{d_n} \chi_n \, e^{\frac{2\pi}{ \varphi^0 } (\frac{n^2}{2} + i n \varphi ) } \,  .
\end{eqnarray}
Again we see that $Z$ transforms as a modular form under $\varphi^0 \to 1/\varphi^0$. Notice in particular that the large positive and negative weights of the D4 resp.\ the D0-part have canceled.

To extract the large $q_0$ asymptotics, we retaining only the leading terms at small positive $\varphi^0$. After integrating out $\varphi$:
\begin{equation} \label{microS}
 \Omega(q) \approx \int d\varphi^0 \, \varphi_0^{\frac{B_2}{2}+1} \,  e^{2 \pi \hat{q}_0 \varphi^0  + \frac{2 \pi \chi}{24 \varphi^0}} \,
 (-1)^{d} \chi(\CM) \, .
\end{equation}
where $\hat{q}_0 \equiv q_0 + \frac{D^{AB} q_A q_B}{2}$.
The saddle point of this integral is at $\varphi^0 = \sqrt{\frac{\chi}{24 \, \hat{q}_0}}$, with value
\begin{equation} \label{OmD0}
 \Omega(q) \approx e^{2 \pi \sqrt{\frac{\hat{q}_0 \chi}{6}}} \, .
\end{equation}
We recall that $\chi=D_{ABC}N^AN^BN^C+c_{2,A}N^A$, so this is an explicit formula for the index as a function of the charges. The logarithm of this expression can therefore directly be compared to the corresponding black hole entropy. 

This is reliable when the saddle point value of $\varphi^0$ is small, i.e.\ $\hat{q}_0 \gg \chi$. 
When the latter is not the case, the saddle point value of $\varphi^0$ is not small, and so, as in the case without D0-branes, even within the model limitations we have made (such as ignoring singular or nonabelian D-branes), the above approximation for $\Omega(q)$ is not necessarily reliable.  We will see in the next section that the agreement with the black hole entropy is excellent when $\hat{q}_0 \gg \chi$. In fact for the leading order matching in this regime, we do not even need to include the fluxes and moduli degrees of freedom, and a much simpler derivation is possible \cite{STROMINGER1996,Maldacena1997,Vafa1997b}. The match does get better at subleading order when the D4 degrees of freedom are included. But when $\hat{q}_0 \lesssim \chi$, the correct entropy is not anywhere near (\ref{OmD0}). In this regime, the flux and D4-moduli degrees of freedom dominate the entropy. 

\subsection{Some extensions}

With the goal of introducing in detail a complex, glassy system that occurs naturally in string theory and has a precise geometric description and well-controlled holographic counterparts, we have given a fully explicit construction of the supersymmetric quantum mechanics describing the ground state sector of wrapped D4-branes bound to D0-branes, and discussed explicit constructions of supersymmetric ground states and their counting in some detail. Some of this was a review of parts of \cite{Denef2007a} and the ideas and results used there, with some simplifications and some points worked out more explicitly. Our discussion was necessarily incomplete in scope. Rather than try to give an overview of the huge related literature, let me mention a few immediate extensions that could have directly followed this part, with a small, non-representative, biased sample of possible starting points for further reading:
\begin{enumerate}
 \item The lift to M-theory as an M5 wrapped on $S^1 \times \Sigma$, reducing to a $(0,4)$ CFT. This is in some ways a more natural framework as it unifies D4 and D0 degrees of freedom and allows for a holographic dual description beyond the ground state sector \cite{Maldacena1997,Minasian1999,Boer2008a}. 
 \item Quiver quantum mechanics: these are 1d gauged linear sigma models, providing a simple but very rich class of models describing complex D-brane systems. The field content is represented by quiver diagrams, nodes being partonic branes and arrows open strings between them \cite{Douglas1996}. Quantization allows explicit interpolation between the weakly coupled geometric regime and the strongly coupled black hole bound state regimes \cite{Denef2002} (see also below).
  \item A more refined description of D4-D0 bound states that takes into account the interplay between D4 and D0 moduli, nonabelian degrees of freedom, and so on. One conceptually simple but efficient approach is the brane-anti-brane tachyon condensation picture \cite{Witten1998,Brunner2000,Oz2001,Gaiotto2005,Gaiotto2006,Denef2007a,Collinucci2009a}.
 \item Counting of ground states away from the regime $\hat{q}_0 \gg \chi$. Relation to Gromov-Witten, Gopakumar-Vafa and Donaldson-Thomas invariants and the OSV conjecture  \cite{Ooguri,GUICA2007, Gaiotto2007,Beasley2006,Gaiotto2007a,Boer2006,Denef2007a,Collinucci2008}. 
 \item Wall crossing phenomena: e.g.\ the possibility for a D4 to split into a D6 and an anti-D6 away from the strict large volume limit \cite{Kachru1999,Brunner2000,Denef2000,Denef2002,Denef2007a,Pioline2011,Manschot2010}.
 \item Applications to global aspects of IIB model building \cite{Collinucci2009a,Cecotti2010a}
\end{enumerate}
It would certainly be useful to have a fully explicit example in which in particular the period vectors can be computed exactly. This may be possible along the lines of \cite{Aganagic2009}.

We end with some references to key papers in the history of the subject. The idea that branes can be wrapped on nontrivial compact cycles to obtain charged particles in four dimensions appeared first in \cite{Strominger1990} (and shown to be necessary for the consistency of string theory in \cite{Strominger1995a}). The discovery of the perturbative string description of D-branes \cite{Polchinski1995} made it possible to quantize and count states at weak coupling, leading to the computation of the Bekenstein-Hawking entropy in \cite{STROMINGER1996}. Many of the ideas in the more systematic development of the physics and mathematics of wrapped D-branes in Calabi-Yau manifolds, including stability and wall crossing, originated in \cite{Brunner2000,Douglas2001}, with precursors in \cite{Berkooz1996,Kachru1999}.

\section{Black hole duals} \label{sec:BHduals}

\subsection{Introduction}

In the appropriate regime, D-branes wrapped on compact cycles manifest themselves as black holes in the low energy effective field theory. It turns out that an intricate zoo of stationary supersymmetric bound states of such black holes exists, like giant molecules, all of which collapse to a D-brane localized at a single point in space in the limit $g_s \to 0$.

In this final section we summarize the explicit solutions of such black hole bound states, and briefly discuss their entropy compared to the microscopic picture developed in the previous section. The complexity of the black hole solutions is correlated with the complexity of the D-brane landscapes that arise at weak coupling. We will focus again on type IIA Calabi-Yau compactifications.

\subsection{BPS solutions of $\CN=2$ supergravity}

The four dimensional effective theory of type IIA string theory compactified on a Calabi-Yau manifold $X$ is  ${\cal N}=2$ supergravity coupled to $B_2=b_2(X)$ abelian $\CN=2$ vector multiplets. There are $B_2 + 1$ gauge fields $\CA^\Lambda$, $\Lambda=0,1,\cdots, B_2$, obtained as in (\ref{RRreduction}) by KK reduction of RR potentials: $C^{(1)}=\CA^0$, $C^{(3)}=\CA^A \wedge D_A$. The $B_2$ vector multiplets  furthermore each contain a complex scalar, obtained by KK reduction of the complexified K\"ahler form: $B+iJ = t^A D_A$. They also contain spin 1/2 fermions, but we don't need those. Finally, the theory contains massless hypermultiplets too, but they do not affect the solutions of interest to us and can be consistently put to constant values. We will work in this section in units with the 4d Newton constant $G_N \equiv 1$. 

Wrapped D-branes manifest themselves as charged point particles in four dimensions, sourcing the vector multiplet fields. We have both electric (D2/D0) and magnetic (D4/D6) monopole charges. In general, the lattice $L$ of magnetic-electric charges $\Gamma$ carries a fundamental symplectic product, which in a symplectic basis has the canonical form
\begin{equation} \label{symplprod}
\langle \Gamma,\tilde{\Gamma}  \rangle \equiv \Gamma^\Lambda \tilde{\Gamma}_\Lambda-\Gamma_\Lambda \tilde{\Gamma}^{\Lambda}.
\end{equation}
Upper indices denote magnetic and lower indices electric components. In the IIA case at hand, the electric charges are the D0 and D2 charges $Q_0$ and $Q_A$, and the dual magnetic charges are the D6 and D4 charges $N^0$ and $N^A$.\footnote{Usually these are denoted by $P^0$ and $P^A$, but to be consistent with the previous sections we use $N^0$ and $N^A$ instead.} In terms of these, the symplectic product is $\langle (N,Q),(\tilde{N},\tilde{Q}) \rangle = N^0 \tilde{Q}_0 + N^A \tilde{Q}_A - Q_0 \tilde{N}^0 - Q_A \tilde{N}^A$. Integrality of this product is equivalent to Dirac quantization. 

In the weak string coupling regime (\ref{weakcoupling}), the wrapped D-branes are well described as point particles moving in flat $\IR^3$, interacting with each other primarily through the lightest stretched open string modes. The coupling $g_4 \equiv g_s/\sqrt{V_X} \propto \ell_p/\ell_s$ is in a hypermultiplet and can therefore be tuned at will. When it gets larger, excited open string modes become important, until eventually the interactions are better described by massless closed string exchange, i.e.\ graviton, photon and scalar exchange. This is the regime in which the supergravity description becomes valid. When the charges are large, the wrapped D-brane states manifest themselves as weakly curved black hole solutions. 

A single centered BPS (i.e.\ supersymmetric) solution to the equations of motion is necessarily static and spherically symmetric, with a metric of the form
$ds^2 = -e^{2U} dt^2 + e^{-2U} d\vec x^2$, and all fields functions of $r=|\vec x|$ only. The BPS equations of motion take the first order gradient flow form typical for supersymmetric solutions \cite{Ferrara1995,Ferrara1997}:
\begin{align} \label{singflow1}
\dot{U} &= - e^U |Z| ,\\
\dot{t}^A &= - 2 e^U g^{A\bar{B}} \partial_{\bar{B}} |Z|, \label{singflow2}
\end{align}
where $g_{A\bar{B}}$ is the metric on the vector multiplet moduli space, the dot denotes derivation with respect to $\tau \equiv 1/r$, and $Z(\Gamma,t)$ is the \emph{central charge} of the magnetic-electric charge $\Gamma$ in a background with moduli $t^A$. It is a complex function on the vector multiplet moduli space, holomorphic up to a normalization factor, and linear in the charge vector $\Gamma$:
\begin{equation} \label{ZGamma}
 Z(\Gamma;t) = {\scriptsize \frac{1}{|\langle V,\bar{V}\rangle|^{1/2}}} \, \langle \Gamma, V \rangle \, ,
\end{equation}
where
\begin{equation} \label{Vdef}
 V^\Lambda = X^\Lambda \, , \qquad V_\Lambda = \frac{\partial F}{\partial X^\Lambda} \, , \qquad X^A = t^A X^0 \, , \qquad A = 1,\ldots,B_2 \, .
\end{equation}
Here $F(X)$ is the \emph{prepotential} of the $\CN=2$ theory, which determines all couplings and metrics in the 4d action. In general it is a locally defined holomorphic function, homogeneous of degree 2. For our type IIA theory, it takes the form
\begin{equation}
 F(X) = -\frac{1}{6} D_{ABC} X^A X^B X^C + \cdots
\end{equation}
where $D_{ABC}$ was defined in (\ref{DABC}) and the ellipsis denote string worldsheet instanton corrections, which are exponentially suppressed in the K\"ahler moduli $J^A = {\rm Im} \, t^A$ and therefore negligible when the Calabi-Yau is large. Dropping those, (\ref{ZGamma}) boils down to
\begin{equation}
 Z = \frac{1}{\left(\frac{8 \, D_{ABC} J^A J^B J^C}{6}\right)^{1/2}} \, \left(\frac{1}{6} N^0 D_{ABC} t^A t^B t^C - \frac{1}{2} N^A D_{ABC} t^B t^C + Q_A t^A + Q_0 \right) \, .
\end{equation}
The central charge gets its name because it appears as a charge commuting with everything in the $\CN=2$ supersymmetry algebra. Its absolute value equals the lowest mass a particle of charge $\Gamma$ can possibly have in a background specified by $t$. This bound is saturated for BPS states. Its phase determines the supercharges preserved by the BPS state. Two BPS objects are mutually supersymmetric if their central charge phases line up. 

The gradient flow equations (\ref{singflow1}) drive $|Z(\Gamma,t)|$ to its minimal value $|Z_\star|$.\footnote{Although all regular critical points of $|Z|$ are isolated local minima \cite{Moore1998}, in the presence of singularities at finite distance in the moduli space, there can be multiple basins of attraction, but we will ignore this here, as it does not occur in the large volume approximation.} If this minimal value is zero at a nonsingular point in moduli space, no solution exists. If it is nonzero, we get a black hole with near horizon solution $t={\rm const.} = t_\star$, $e^{-2U(r)} = \frac{|Z_\star|^2}{r^2}$, as can be checked directly from the above BPS flow equations. This describes AdS$_2 \times S^2$ with $S^2$ horizon area $A=4\pi|Z_\star|^2$ and therefore Bekenstein-Hawking entropy 
\begin{equation}
 S(\Gamma) = \pi |Z_\star(\Gamma)|^2 \, .
\end{equation}
When the D6-charge $N^0$ is zero, the minimization of $|Z|$ is straightforward and there is a simple closed form expression for $S(\Gamma)$ (still in the large volume approximation) \cite{Shmakova1996}:
\begin{equation} \label{macroS}
 S(N,Q)|_{N^0=0} = 2\pi \sqrt{\frac{\widehat{Q}_0 D_{ABC} N^A N^B N^C}{6}} \, , 
 \qquad \widehat{Q}_0 = Q_0 + \frac{1}{2} D^{AB} Q_A Q_B \, ,
\end{equation}
where $D_{AB} = D_{ABC} N^C$ and $D^{AB}$ is its inverse. Note that to leading order in the large charge limit, this is exactly the same as the microscopic result (\ref{microS})! The subleading term can be reproduced macroscopically as well, from the Wald entropy in the presence of an $R^2$ term \cite{Cardoso1998}. When $N^0$ is not zero there is no general closed form solution for $S(\Gamma)$, except when $Q_A$ is chosen to be proportional to $D_{ABC} N^B N^C$. In that case we define for some fixed $K^A>0$ the charge parameters $(n^0,n,q,q_0)$ by $(N^0,N^A,Q_A,Q_0)=(n^0,n \, K^A,q \, (K^2)_A,q_0 \, K^3)$, where $K^3$ is a short for $D_{ABC} K^A K^B K^C$ and $(K^2)_A$ for $D_{ABC} K^B K^C$. Then we can write 
\begin{equation}
S = \frac{\pi}{3} \, K^3 \,{\sqrt{3\,n^2\,q^2 - 8\,n^0\,q^3
+ 6\,n^3\,q^0 -
        18\,n\,n^0\,q\,{q^0} -
        9\,{{n^0}}^2\,{{q^0}}^2}} \, .
\end{equation}
As we will see below, once the entropy function $S(\Gamma)$ is known, it is easy to write down fully explicit expressions for all fields at all points in space.

\begin{figure}
\begin{center}
\includegraphics[height=7cm]{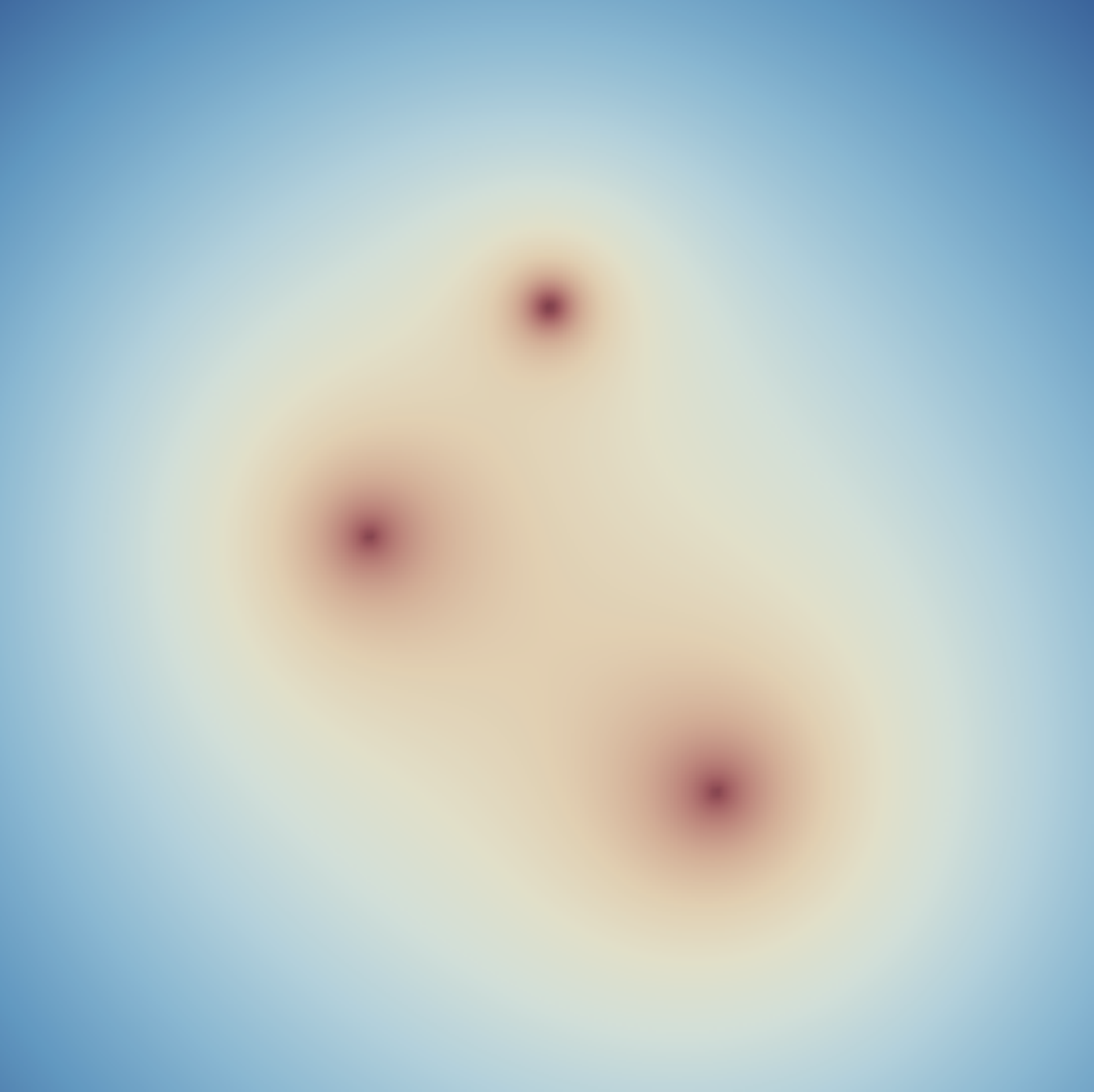}
\end{center}
\caption{\footnotesize Warp factor $e^U$ for some three centered ${\rm D6- \overline{D6} - D4}$ bound state, for a choice of charges that doesn't really matter because this figure is primarily ornamental. 
\label{3holes}}
\end{figure}

Quite remarkably, these theories also have multicentered, supersymmetric, stationary \emph{black hole bound states}, like giant molecules. These are genuine bound states in the sense that the centers are constrained by nontrivial potentials, generated by scalar, electomagnetic and gravitational forces \cite{Denef2000,Cardoso2000,Behrndt1998,Bates2003,Boer2008a,Denef2007a}. They have a metric of the form
\begin{equation}
 ds^2=-e^{2U}(dt+\omega_i dx^i)^2+ e^{-2U}d\vec x^2,
\end{equation}
where $U$ and $\omega$ depend on $\vec x$, and they are fully characterized by harmonic functions $H^\Lambda$, $H_\Lambda$ with sources at the positions $\vec{x}_i$ of the charges $\Gamma_i$:
\begin{equation} \label{harmfu}
 H = \sum_i \frac{\Gamma_i}{|\vec x - {\vec x}_i|} + h \, ,
\end{equation}
where the constant term $h$ is determined by the total charge $\Gamma$ and the asymptotic moduli:
\begin{equation}
\label{hDefinition}
  h=-2\,\mathrm{Im}\bigl(e^{-i \alpha} \tilde{V} \bigr)|_{r=\infty}, \qquad \alpha=\arg Z(\Gamma) \, ,
  \qquad \tilde{V} = V/|\langle V,\bar{V} \rangle|^{1/2} \, ,
\end{equation}
and $V$ is as in (\ref{Vdef}).

As mentioned earlier, once the entropy function $S(\Gamma)$ is known on charge space, the complete solution is known \cite{Bates2003}, simply by substituting $H(\vec x)$ for $\Gamma$ in $S$ and its derivatives: 
\begin{equation} \label{Sigmadef}
e^{-2U} = \frac{1}{\pi} S(H) \, ,
\qquad
{\cal A}^\Lambda=\frac{1}{\pi} \frac{\partial \log S(H)}{\partial H_\Lambda}  \  (dt+\omega)+{\cal A}^\Lambda_{{\rm mon}}, \qquad
t^A=\frac{H^A-\frac{i}{\pi} \frac{\partial S}{\partial H_A}}{H^0- \frac{i}{\pi} \frac{\partial S}{\partial H_0}}.
\end{equation}
The one-form ${\cal A}^\Lambda_{{\rm mon}}$ is  the vector potential for a system of Dirac magnetic monopoles  of charge $N^\Lambda_i$ located at the positions $\vec x_i$. 
The off-diagonal components $\omega$ of the metric are the solutions to
\begin{equation} \label{omegaeq}
  \nabla \times \omega= \langle \nabla H, H\rangle \, ,
\end{equation}  
where $\nabla$ is the flat space gradient. This equation implies an important integrability condition:
 $\nabla \cdot (\nabla \times \omega)=0$ $\Rightarrow$ $\langle \nabla^2 H, H\rangle=0$, from which, using $\nabla^2 \frac{1}{|\vec x|} = -4 \pi \delta^3(x)$, we get for every center $i$ a condition:
\begin{equation}\label{IntegrabilityConditions}
 \sum_j \frac{ \langle \Gamma_i, \Gamma_j \rangle}{|\vec x_i -\vec x_j|}
= -  \langle \Gamma_i, h\rangle = \left. 2 \, {\rm Im} \bigl( e^{-i \alpha} Z(\Gamma_i,t) \bigr) \right|_{r=\infty} \, .
\end{equation}
This imposes constraints on the positions $\vec{x}_i$; solutions are the BPS equilibrium positions of the black holes in their mutual force fields. For the 2-centered case we have in particular
\begin{equation} \label{towcentersep}
 |\vec x_1-\vec x_2| = \frac{\langle \Gamma_1,\Gamma_2 \rangle}{2} \frac{|Z_1+Z_2|}{{\rm Im}(Z_1 \bar{Z}_2)} \, ,
\end{equation}
where $Z_i=Z(\Gamma_i,t)|_{r=\infty}$.
When the right hand side goes from positive to negative through a wall where $Z_1$ and $Z_2$ line up, the equilibrium separation diverges and the bound state decays. This is the supergravity incarnation of the wall crossing phenomenon. 

For completeness we also give the solution to (\ref{omegaeq}) \cite{Bates2003}. For a 2-centered bound state with centers at $(0,0,L)$ and $(0,0,-L)$, up to residual diffeomorphism gauge transformations $t \to t+f(\vec{x})$, $\omega \to \omega - df$:
\begin{equation} \label{omegasol2}
 \omega = \frac{\langle \Gamma_1,\Gamma_2 \rangle}{2 L} \biggl(
 \frac{L^2-r^2}{(L^4+r^4-2L^2r^2\cos 2\theta)^{1/2}}
 + 1 - \cos \theta_1 + \cos \theta_2
 \biggr) \, d\phi \, .
\end{equation}
Here $(r,\phi,\theta)$ are standard spherical coordinates centered at the origin and $\theta_1$, $\theta_2$ are angles of spherical coordinates centered at the two particle positions. The integrability condition is equivalent to the absence of physical singularities along the $z$-axis. Because (\ref{omegaeq}) is linear, the solution for more centers is obtained by superposition. 

The solutions are generically stationary but not static: they have intrinsic angular momentum, given by
\begin{equation}\label{AngularMomentum}
 \vec J=\frac{1}{2}\sum_{i<j}\langle \Gamma_i, \Gamma_j \rangle \, \hat{x}_{ij} \, , \qquad \hat{x}_{ij} = \frac{\vec x_i-\vec x_j}{|\vec x_i-\vec x_j|}.
\end{equation}
This generalizes the well-known electromagnetic field angular momentum sourced by monopole-electron pairs (generated by the $\vec E \times \vec B$ Poynting vector circling around the electron-monopole axis).

\begin{figure}
\begin{center}
\includegraphics[height=6cm]{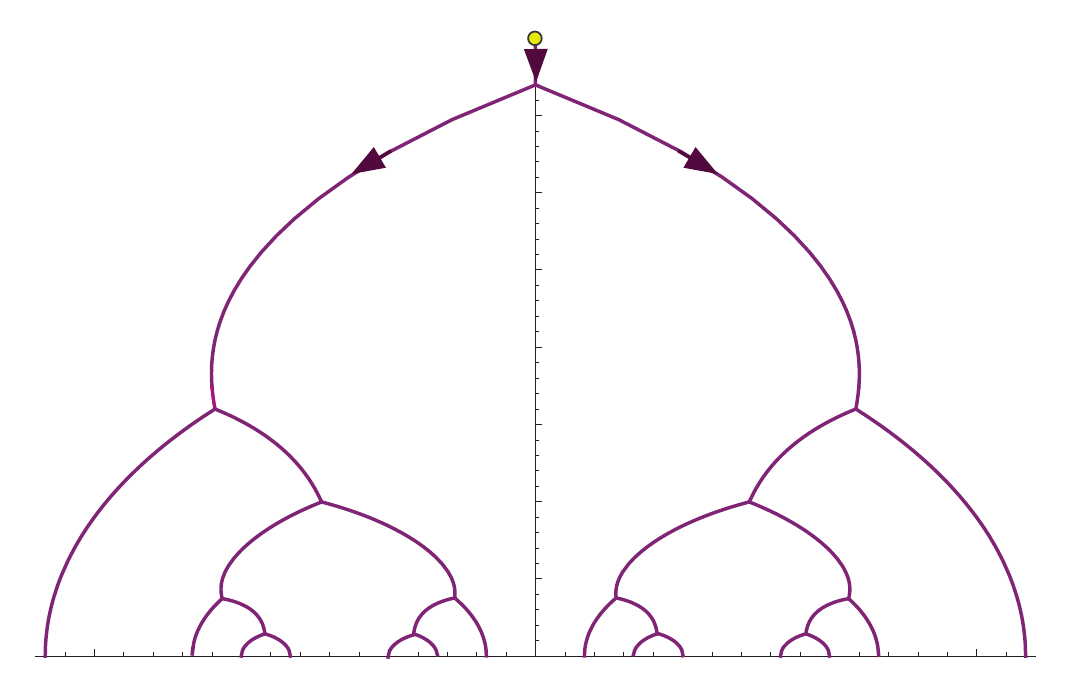}
\end{center}
\caption{\footnotesize (from \cite{Denef2007a}) Attractor flow tree in the $t$-plane for a particular bound state of 7 flux-carrying D6-branes and 7 flux-carrying anti-D6-branes. The flow starts at the yellow dot at the top and goes down. The iterated structure can be thought of as representing clutering of individual centers. In the case at hand all attractor points lie on the boundary of moduli space, which is due the special choice of charges. The symmetric form is also special to our particular choice. The main motivation to pick those charges was esthetics and dramatic effect.  More generic examples can be found in \cite{Denef2007a}.
\label{fractalflow}}
\end{figure}

The integrability conditions are a necessary but not sufficient condition for the existence of solutions. One also needs to check if the entropy function $S(H)$ is positive everywhere. This is in general difficult to do. In \cite{Denef2000,Denef2007a}  the conjecture was formulated that the existence of solutions is equivalent to the existence of the (much easier to analyze) \emph{attractor flow trees}. An attractor flow tree is built out of single attractor flows. The tree starts at the background value of the moduli and terminates at the attractor points of the constituent charges. Each
edge $E$ of an attractor flow tree is given by a single charge attractor flow for some charge $\Gamma_E$. Charge and energy are conserved at the vertices, i.e.\ for each vertex splitting $E \to (E_1,E_2)$,
$\Gamma_E = \Gamma_{E_1} + \Gamma_{E_2}$ and
$|Z(\Gamma_E)|=|Z(\Gamma_{E_1})| + |Z(\Gamma_{E_2})|$. The last
condition is equivalent to requiring the vertices to lie on a line
of marginal stability: $\arg Z(\Gamma_{E_1}) = \arg
Z(\Gamma_{E_2})$. Flow trees can be thought of as ``skeletons'' for the supergravity solutions, or as giving a recipe for adiabatically assembling or disassembling multicentered bound states. This gives a strong physical argument for the conjecture. We refer to section 3.2.2 of \cite{Denef2007a} for a more detailed discussion. 

\subsection{Entropy}

Comparing the macroscopic and microscopic D4-D0 entropies (\ref{macroS}) and (\ref{microS}), we find perfect agreement to leading order when $\hat{Q}_0 \gg N^3$. The subleading correction can be reproduced from the Wald entropy in the presence of an $R^2$ correction term in the 4d effective gravity theory \cite{Cardoso1998}. In fact, in the same regime, an infinite series of corrections matches \cite{Ooguri,Denef2007a,Beasley2006,Boer2006,Gaiotto2007a}

However, as was clear from the discussion in section 4, microcopically things get much more complicated when $\hat{Q}_0$ becomes of the order of $\chi \sim N^3$: the D4-degrees of freedom start getting important and eventually dominate the entropy. On the black hole side things get much more complicated too in exactly the same regime: Multicentered solutions appear, and at some point they start dominating the entropy \cite{Denef2007a}. 

In particular one clearly exits the easy regime where $\hat{Q}_0 \gg N^3$ whenever \emph{all} charges are uniformly scaled up sufficiently. Thus, more precisely, when charges $\Gamma$ are
scaled up as $\Gamma \to \Lambda \Gamma$, there may exist for example two-centered
solutions with  horizon entropy scaling as $\Lambda^3$, while   the
single centered entropy  scales as $\Lambda^2$.

Let us give a  concrete example of this phenomenon, taken from \cite{Denef2007b}. Consider the Calabi-Yau $X = T^2_1 \times T^2_2 \times T^2_3$, a product of three two-tori.  Let $\Sigma$ be the 4-cycle $(T^2_1 \times T^2_2) + (T^2_2 \times T^2_3) + (T^2_3 \times T^2_1)$ and let
$\sigma$ be the 2-cycle $T^2_1+T^2_2+T^2_3$. Then the entropy
function of a charge $\Gamma$ corresponding to wrapping $N^0$ D6-branes on
$X$, $N$ D4-branes on $\Sigma$, $Q$ D2-branes on $\sigma$ and $Q_0$
D0-branes is given by $S(\Gamma) = \pi \sqrt{4 N^3 Q_0 + 3 N^2 Q^2 + 6 N^0 N Q Q_0 - 4 N^0 Q^3 - (N^0
 Q_0)^2}$.
Now consider a total charge
\begin{equation}
\Gamma =
 (N^0,N,Q,Q_0)=\Lambda(0,6,0,-12),
\end{equation}
in a background in which the area of each
$T^2$ equals $v$. Then
for any  $v$, there exists a single centered solution with horizon
entropy 
\begin{equation} \label{singleS}
 S_1 = 72 \sqrt{2} \pi \, \Lambda^2.
\end{equation}
However, there also exists D6-anti-D6 two-centered
black hole bound states, for instance with charges
\begin{equation}
 \Gamma_1 = (1,3 \Lambda,6 \Lambda^2,-6 \Lambda),
 \qquad \Gamma_2 = (-1,3 \Lambda,-6 \Lambda^2,-6 \Lambda).
\end{equation}
provided $v >  \sqrt{18} \Lambda$. The constant terms in the harmonic functions (\ref{harmfu}) are
$h=(0,\frac{1}{\sqrt{2 v}},0,-\sqrt{\frac{v^3}{2}})$, and the
equilibrium separation (\ref{towcentersep}) at large $\Lambda$ is $|\vec x_1 - \vec
x_2| \approx \frac{108 \sqrt{2} \sqrt{v} \, \Lambda}{(v/\Lambda)^2 - 18}$. The resulting metric is well defined, with $S(H)$ real
and positive everywhere. The two centers
have equal horizon entropy, summing up to a total entropy
\begin{equation}
 S_2 \approx 12 \pi \sqrt{3} \, \Lambda^3 \, ,
\end{equation}
which is indeed parametrically larger than the single centered
entropy (\ref{singleS}).

When   $v$ is kept fixed while sending $\Lambda \to \infty$,
the equilibrium separation diverges and then formally goes negative, meaning these
2-centered solutions   disappear from the spectrum. But the
low energy effective quantum mechanics of section \ref{sec:D4susyQM} is supposed to become increasingly reliable in the limit
of large volume $v \to \infty$ (and weak string coupling), so its Witten index for 
given charges must be the total index of all black hole configurations in the limit $v \to \infty$. 
Thus we conclude that at large $\Lambda$, the usual
weakly coupled, weakly curved wrapped D4-branes are not computing the
entropy of a single D4-D2-D0 black hole, but predominantly that of multicentered configurations. The  single center D4 black hole entropy can only be microscopically reproduced from a weak coupling computation in the regime $\hat{Q}_0 \gg N^3$.

This occurs in particular also in the lift to M-theory \cite{Boer2008a}, where a precise holographic correspondence exists between a 1+1 dimensional (0,4) CFT \cite{Maldacena1997} and quantum gravity in AdS$_3 \times S^2$. It was pointed out there that the solution with the highest entropy in the $\Lambda \to \infty$ limit appears to be a configuration in which all entropy carrying charges has been concentrated in one of the D6-charged black holes. 

\subsection{Landscape structure}

In the minimal charge case, namely $\hat{Q}_0 = -\frac{\chi}{24} \sim -\frac{N^3}{24}$, the 4d supergravity solution is a pure D6-anti-D6 bound state, where the D6 branes carry flux but no entropic degrees of freedom \cite{Denef2007a}. Quantization of the centers leads to a spin $j=d/2$ multiplet of supersymmetric ground states with degeneracy $2j+1=d+1$, in agreement with the microscopic picture where this arises as the Euler characteristic of the moduli space $\CM=\ICP^d$. The lifted M-theory solution becomes pure ``spinning'' AdS$_3 \times S^2$ in this case.

Some identifications between the microscopic structure of the D-brane landscape and the macroscopic black hole solutions have been made (where it should be kept in mind of course that the map from D-brane microstates to black hole macroscopic states must be many-to-one, due to the existence of black hole entropy). For example switching on fluxes represented by holomorphic 2-cycles as in section \ref{sec:explconstr} corresponds to ``dressing up'' the D6 or anti-D6 by halos of D2-D0 particles \cite{Denef2007a}, or equivalently in M-theory by adding M2-branes to the north or south poles of the $S^2$ \cite{Boer2006,Gaiotto2007a}. But the full map at higher D0-charge is not well understood. 

Dynamics near zero energy should proceed through tunneling of charged particles in and out of the black hole centers in this picture. This is under study in \cite{Barandes}.

\vskip10mm

\noindent {\bf \Large Acknowledgements} 

\vskip5mm

\noindent I would like to thank the organizers of TASI 2010, Tom Banks, Michael Dine and Subir Sachdev, for giving me the opportunity to teach on these topics. I am very much indebted to Dionysios Anninos, Tarek Anous, Jacob Barandes, Marcus Benna, Hyeyoun Chung, Mike Douglas, Bram Gaasbeek, Hajar Ebrahim, Greg Moore and Andy Strominger, whose insights, collaborations and discussions were crucial in the genesis of these notes. Many of the ideas motivating the theme of these lectures were developed over the past year together with Dionysios Anninos, and we would like to thank Charlies Kitchen for hospitality. I'm grateful to A.P.\ Young and Subir Sachdev for clarifying conversations about spin glasses. Finally, special thanks to Dionysios Anninos, Tarek Anous, Jacob Barandes and Marcus Benna for a careful reading of parts of these notes and their useful suggestions for improvements. This work was supported in part by DOE grant DE-FG02-91ER40654. 

%
%


\bibliography{/Users/frederikdenef/Documents/library}

\end{document}